\RequirePackage{fix-cm}
\documentclass[ba]{imsart}
\pubyear{0000}
\volume{00}
\issue{0}
\doi{0000}
\firstpage{1}
\lastpage{1}
\usepackage{amsthm}
\usepackage{amsmath}
\usepackage{amssymb}
\usepackage{natbib}
\usepackage[colorlinks,citecolor=blue,urlcolor=blue,filecolor=blue,backref=page]{hyperref}
\usepackage{graphicx}
\usepackage{bm}
\usepackage[toc,page]{appendix}
\usepackage[colorlinks,citecolor=blue,urlcolor=blue,filecolor=blue,backref=page]{hyperref}
\usepackage[linesnumbered, algoruled, longend, boxed]{algorithm2e}
\usepackage{multirow}
\usepackage{mathrsfs}
\usepackage{float}
\usepackage{subfigure}
\usepackage{siunitx}
\usepackage{xcolor}

\startlocaldefs
\newcommand{\mmu}{\bm{\mathit{\mu}}}
\newcommand{\mtheta}{\mathbf{\mathit{\theta}}}
\newcommand{\mTheta}{\bm{\theta}}
\newcommand{\mXi}{\bm{\xi}}
\newcommand{\mSigma}{\bm{\Sigma}}

\newcommand{\mPr}{\Pr}
\newcommand{\mL}{\mathcal{L}}

\newcommand{\mpi}{\mathit{\pi}}

\newcommand{\mZ}{\mathcal{Z}}

\newcommand{\mP}{\mathcal{P}}

\newcommand{\mxi}{\mathit{\xi}}

\newcommand{\tmL}{\tilde{\mathcal{L}}}
\newcommand{\tmpi}{\tilde{\mathit{\pi}}}

\newcommand{\mbeta}{\mathit{\beta}}

\newcommand{\calD}{\mathcal{D}}
\newcommand{\calM}{\mathcal{M}}
\newcommand{\calN}{\mathcal{N}}
\newcommand{\calU}{\mathcal{U}}

\newcommand{\pkg}[1]{\texttt{#1}}

\newcommand{\revise}[1]{\textcolor{black}{#1}}

\makeatletter
\newcommand{\mypm}{\mathbin{\mathpalette\@mypm\relax}}
\newcommand{\@mypm}[2]{\ooalign{%
  \raisebox{.3\height}{$#1+$}\cr
  \smash{\raisebox{-.4\height}{$#1-$}}\cr}}
\makeatother

\graphicspath{{img/},{}}
\endlocaldefs
\begin{document}

\begin{frontmatter}
\title{Bayesian posterior repartitioning for nested sampling}

\runtitle{Bayesian posterior repartitioning}

\begin{aug}
\author{\fnms{Xi} \snm{Chen}\thanksref{addr1,addr2}\ead[label=e1]{xc841@bath.ac.uk}\ead[label=e11]{xc253@mrao.cam.ac.uk}},
\author{\fnms{Farhan} \snm{Feroz}\thanksref{addr2}\ead[label=e2]{f.feroz@mrao.cam.ac.uk}}
\and
\author{\fnms{Michael} \snm{Hobson}\thanksref{addr2}%
\ead[label=e3]{mph@mrao.cam.ac.uk}}


\address[addr1]{Department of Computer Science, University of Bath, UK. BA2 7PB. \printead{e1}}

\address[addr2]{Department of Physics, University of Cambridge, UK. CB3 0HE. \\ \printead{e11,e2,e3}}

\end{aug}

\begin{abstract}
Priors in Bayesian analyses often encode informative domain knowledge
that can be useful in making the inference process more efficient.
Occasionally, however, priors may be unrepresentative of the parameter
values for a given dataset, which can result in inefficient parameter
space exploration, or even incorrect inferences, particularly for
nested sampling (NS) algorithms.  Simply broadening the prior in such
cases may be inappropriate or impossible in some applications. Hence our
previous solution to this problem, known as posterior repartitioning
(PR), redefines the prior and likelihood while keeping their product
fixed, so that the posterior inferences and evidence estimates remain
unchanged, but the efficiency of the NS process is significantly
increased. In its most practical form, PR raises the prior to some
power $\beta$, which is introduced as an auxiliary variable that must
be determined on a case-by-case basis, usually by lowering $\beta$
from unity according to some pre-defined `annealing schedule' until
the resulting inferences converge to a consistent solution. Here we
present a very simple yet powerful alternative Bayesian approach, in
which $\beta$ is instead treated as a hyperparameter that is inferred
from the data alongside the original parameters of the problem, and
then marginalised over to obtain the final inference. We show through
numerical examples that this Bayesian PR (BPR) method provides a very
robust, self-adapting and computationally efficient `hands-off'
solution to the problem of unrepresentative priors in Bayesian
inference using NS.  Moreover, unlike the original PR method, we show
that even for representative priors BPR has a negligible
computational overhead relative to standard nesting sampling, which
suggests that it should be used as the default in all NS analyses.
\end{abstract}


\end{frontmatter}

\section{Introduction}
\label{sec:intro}

In recent years, Monte Carlo sampling techniques have been widely used
in Bayesian inference problems both for parameter estimation and model
selection. Nested sampling (NS) \citep{skilling2006nested} is one such
approach that can simultaneously produce samples from the posterior
distribution and estimate the marginal likelihood (or evidence). The
NS algorithm involves drawing samples from a pre-defined prior
distribution, and then evaluating their corresponding likelihoods,
which are dependent on some measurement (or forward) model and the
observed data.  In general, the observed data are fixed and the
measurement model is defined by field experts, with little room for
flexibility. In contrast, the prior that identifies the regions of
interest in the parameter space is typically much more loosely
determined and often defined using simple standard distributions,
together occasionally with physical constraints on the parameters
$\mTheta$ of the problem under consideration.

As we discussed in \cite{chen2018improving}, the NS algorithm can become
very inefficient, or even fail completely in extreme cases, if the
likelihood for a given data set is concentrated far out in the wings
of the assumed prior distribution. This problem can be particularly
damaging in applications where one wishes to perform analyses on many
thousands (or even millions) of different datasets, since those
(typically few) datasets for which the prior is unrepresentative can
absorb a large fraction of the computational resources.

The problem occurs because the NS algorithm begins by drawing a number
of `live' samples from the prior and at each subsequent iteration
replaces the sample having the lowest likelihood with a sample again
drawn from the prior but constrained to have a higher
likelihood. Thus, as the iterations progress, the collection of ‘live
points’ gradually migrates from the prior to regions of high
likelihood.  When the likelihood is concentrated very far out in the
wings of the prior, this process can become very slow, even in the
rare problems where one is able to draw each new sample from the
constrained prior using standard methods (sometimes termed perfect
nested sampling). In practical problems, the issue is yet more pronounced
since algorithms such as MultiNest \citep{feroz2009multinest} and PolyChord
\citep{handley2015polychord} use other methods that may require
several likelihood evaluations before a new sample is
accepted. Depending on the method used, an unrepresentative prior can
thus result in a significant drop in sampling efficiency, thereby
increasing still further the required number of likelihood
evaluations.

In extreme cases, the migration of live points in the NS process can
be very slow, since over many iterations the live points will
typically all lie in a region over which the likelihood is very small
and flat. Indeed, in some such cases, the log-likelihoods of the set
of live points may be indistinguishable to machine precision, so the
‘lowest likelihood’ sample to be discarded will be chosen effectively
at random and, in seeking a replacement sample that is drawn from the
prior but having a larger likelihood, the algorithm is very unlikely
to obtain a sample for which the likelihood value is genuinely larger
to machine precision. These problems can result in the live point set
migrating exceptionally slowly, or becoming essentially stuck, such
that the algorithm (erroneously) terminates before reaching the main
body of the likelihood and therefore fails to produce correct
posterior samples or evidence estimates.

One may, of course, seek to improve the performance of NS in such
cases by increasing the number of live points and/or adjusting the
convergence criterion, so that many more NS iterations are performed,
but there is no guarantee in any given problem that these measures
will be sufficient to prevent premature convergence. Perhaps more
useful is to ensure that there is a greater opportunity at each NS
iteration of drawing candidate replacement points from regions of the
parameter space where the likelihood is larger. This may be achieved
in a variety of ways. In MultiNest, for example, one may reduce the
{\tt efr} parameter to enlarge the volume of the multi-ellipsoidal
bound from which candidate replacement points are
drawn. Alternatively, as in other NS implementations, one may draw
candidate replacement points using either MCMC sampling
\citep{feroz2008multimodalns} or slice-sampling
\citep{handley2015polychord} and increase the number of steps taken
before a candidate point is chosen. All of these approaches may
mitigate the problem to some degree in particular cases, but only at
the cost of a simultaneous dramatic drop in sampling efficiency caused
precisely by the changes made in obtaining candidate replacement
points. Moreover, in more extreme cases, these measures may fail
completely.

Aside from making changes to the implementation of the NS algorithm
itself, one might consider simple solutions such as broadening the
prior range in such cases, but this might not be appropriate or
possible in real-world applications, for example when one wishes to
assume a single standardised prior across the analysis of a large
number of datasets for which the true values of the parameters of
interest may vary.

In \cite{chen2018improving}, we therefore proposed a posterior repartitioning
(PR) method that circumvents the above difficulties. The PR method
exploits the intrinsic degeneracy between the `effective' likelihood
and prior in the formulation of Bayesian inference problems. This is
especially relevant for NS since it differs from other sampling
methods by making use of the likelihood $\mL(\mTheta)$ and prior
$\pi(\mTheta)$ {\em separately} in its exploration of the parameter
space, in that samples are drawn from the prior $\pi(\mTheta)$ such
that they satisfy $\mL(\mTheta) >
L_\ast$, \revise{where $L_\ast$ is some predefined likelihood constraint}. By contrast, Markov chain Monte Carlo (MCMC) sampling methods
or genetic algorithm variants are typically blind to this
separation\footnote{One exception is the propagation of multiple MCMC
  chains, for which it is often advantageous to draw the starting
  point of each chain independently from the prior distribution.}, and
deal solely in terms of the product $\mL(\mTheta)\pi(\mTheta)$, which
is proportional to the posterior $\mP(\mTheta)$. This difference
provides an opportunity in the case of NS to `repartition' the product
$\mL(\mTheta)\pi(\mTheta)$ by defining a new effective likelihood
$\tmL(\mTheta)$ and prior $\tmpi(\mTheta)$ (which is typically
`broader' than the original prior), subject to the condition
$\tmL(\mTheta)\tmpi(\mTheta)= \mL(\mTheta)\pi(\mTheta)$, so that the
(unnormalised) posterior remains unchanged. Thus, in principle, the
inferences obtained are unaffected by the use of the PR method, but,
as we demonstrated in \cite{chen2018improving}, the approach can yield
significant improvements in sampling efficiency and also helps to
avoid the convergence problems that can occur in extreme examples of
unrepresentative priors.

In general, the effective prior $\tmpi(\mTheta)$ can be any
distribution that can be sampled straightforwardly, which
in principle can be arbitrary \citep{Alsing2021}, but should at least
be non-zero everywhere that the original prior is non-zero.  In its
most practical form, however, sometimes termed power posterior
repartitioning (PPR)\footnote{This naming convention, which we shall
  avoid here, should not be confused with the power posterior method
  (also known as thermodynamic integration) for estimating the
  evidence using MCMC sampling; the latter instead involves
  transitioning from the prior to the posterior by powering just the
  likelihood by an inverse temperature.}, the effective prior
$\tmpi(\mTheta)$ is proportional simply to the original prior
$\pi(\mTheta)$ raised to some power $\beta$, which is introduced as an
auxiliary variable. Although we demonstrated the effectiveness of this
approach in \cite{chen2018improving}, one drawback of our original
method is that the auxiliary variable must be determined on a
case-by-case basis, which is typically achieved by lowering $\beta$
from unity, outside the execution of the NS algorithm, according to some
pre-defined `annealing schedule', until the resulting inferences from
successive NS runs converge to a statistically consistent solution for
values below some (positive) threshold $\beta \lesssim
\beta_\ast$. This approach lacks elegance and the repeated NS runs
required can be computationally demanding. Moreover, the final
inference is unavoidably conditioned on the value $\beta_\ast$.

In this paper, we therefore present an alternative, Bayesian approach,
in which $\beta$ is instead treated as a hyperparameter that is
inferred from the data alongside the original parameters $\mTheta$ of
the problem, within a single execution of the NS algorithm.  Although
this approach is conceptually very simple, indeed almost trivial, the
resulting Bayesian PR (BPR) method is considerably more
powerful than our original PR technique in several respects. In
particular, one obtains samples from the joint posterior on
$(\mTheta,\beta)$, which may then be used straightforwardly in a
number of ways. First, one may properly marginalise over $\beta$ in a
Bayesian manner to obtain a final inference on $\mTheta$, rather than
conditioning on a single value of $\beta$ as in the original PR
approach. Moreover, this allows BPR to accommodate
likelihood functions consisting of a number of spatially separated
modes that are located asymmetrically with respect to the prior, and
are therefore characterised by different ranges of $\beta$ values;
this is not possible using the original PR method.  Second, one may
instead marginalise over $\mTheta$ to determine the 1-dimensional
marginal distribution of $\beta$ (\revise{an `effective' posterior that will be introduced in later sections}), which we will show is very useful in diagnosing both the existence and severity of an unrepresentative
prior for a given dataset, and hence for identifying `outlier'
datasets and in refining the inference problem for future
analyses. Finally, since the sampling of the joint posterior on
$(\mTheta,\beta)$ is performed within a single NS run, BPR is much
less computationally demanding than the original PR method, which
requires a separate NS run for each value of $\beta$ in its annealing
schedule. Indeed, since the overhead of introducing just a single
additional hyperparameter $\beta$ to be sampled is negligible for most
practical problems, BPR is typically no more computationally
demanding than a standard NS analysis (i.e., equivalent to setting
$\beta=1$), but automatically safeguards against potentially
inefficient parameter space exploration, or even incorrect inferences,
which may occur in the presence of unrepresentative priors. This
suggests that BPR should, in fact, be used as the
default approach in all NS analyses.

It is worth noting, however, that one may encounter even more extreme
problems than those discussed above, where the likelihood for some
dataset(s) is concentrated outside an assumed prior having compact
support. From a probabilistic point of view, the prior in these
problems has zero probability in the region of the likelihood
distribution. This case, which one might describe as an unsuitable prior, is
not addressed by the BPR method, and is not considered here.

This paper is organised as follows. Section~\ref{Sec:NestedSampling}
briefly introduces the Bayesian inference and the NS
algorithm. In Section~\ref{sec:analytical_example}, we present a simple analytical illustration of the effect of an unrepresentative prior on Monte Carlo sampling in general and NS in particular. Section \ref{Sec:BPRmethod} outlines the original PR
method and then describes the proposed BPR scheme. Section
\ref{Sec:Numeric} demonstrates performance of the BPR method in
some numerical examples, and we present our conclusions in Section
\ref{Sec:Conclu}.

\section{Bayesian inference using nested sampling}
\label{Sec:NestedSampling}

Bayesian inference (see e.g. \citealt{mackay2003}) provides a
consistent framework for estimating unknown parameters $\mTheta$ of
some model by updating any prior knowledge of $\mTheta$ using the
observed data $\calD$ and an assumed measurement process. The complete
inference is embodied in the posterior distribution of $\mTheta$,
which can be expressed using Bayes' theorem as:
\begin{align}
\mPr(\mTheta | \calD, \calM) = \frac{\mPr(\calD | \mTheta, \calM) \mPr(\mTheta | \calM)}{\mPr(\calD | \calM)},
\end{align}
where $\calM$ represents model (or hypothesis) assumption(s), and
adopting a simplifying notation, $\mPr(\mTheta | \calD, \calM) \equiv
\mP(\mTheta)$ is the {\em posterior} probability density, $\mPr(\calD
| \mTheta, \calM) \equiv \mL(\mTheta)$ is the {\em likelihood},
$\mPr(\mTheta | \calM) \equiv \mpi(\mTheta)$ is the {\em prior}
probability density on $\mTheta$ and $\mPr(\calD | \calM) \equiv \mZ$ is
called the {\em evidence} (or marginal likelihood). We then have the
simplified expression:
\begin{align}
\mP(\mTheta) = \frac{\mL(\mTheta) \mpi(\mTheta)}{\mZ},
\label{Eq:Bayes}
\end{align} 
in which the evidence is given by
\begin{align}
\mZ = \int_{\cal R} \mL(\mTheta)  \mpi(\mTheta) d \mTheta, 
\label{Eq:origZ}
\end{align}
where $\cal R$ represents the prior space of $\mTheta$. The evidence
$\mZ$ is often used for model selection. It is the average of the
likelihood over the prior, considering every possible choice of
$\mTheta$, and thus is not a function of the parameters $\mTheta$. The
constant $\mZ$ is usually ignored in parameter estimation, since the
posterior $\mP(\mTheta)$ is proportional to the product of likelihood
$\mL(\mTheta)$ and prior $\mpi(\mTheta)$.

Nested sampling \citep{skilling2006nested} explores the posterior
distribution in a sequential manner using a fixed number of $N_{\rm
  live}$ `live samples' of the parameters $\mTheta$ at each iteration
of the process\footnote{\revise{The method has recently been extended to so-called dynamic nested sampling \citep{Higson2018}, which allows the number of live samples to vary as the NS interations proceed, but we will not use this variant here.}}. \revise{Pseudo-code of standard NS algorithm is described in Supplementary Materials.} Among the various implementations of the NS algorithm \citep{chopin2007contemplating, feroz2009multinest, brewer2011diffusive, handley2015polychord, baldock2017constant, Higson2018, buchner2019collaborative, speagle2020dynesty, buchner2021nested}, two widely used packages are MultiNest \citep{feroz2009multinest,feroz2013importance}
and PolyChord \citep{handley2015polychord}. MultiNest draws the new
sample at each iteration using rejection sampling from within a
multi-ellipsoid bound approximation to the iso-likelihood surface
defined by the discarded point; the bound is constructed from the
live points present at that iteration. PolyChord draws the new sample at
each iteration using a number of successive slice-sampling steps taken
in random directions, which is particularly well suited to
higher-dimensional problems.  Please see \cite{feroz2009multinest} and
\cite{handley2015polychord} for more details.

\revise{\section{Analytical illustration of Monte Carlo sampling with an unrepresentative prior}\label{sec:analytical_example}}

\revise{We begin by considering a simple intuitive example that illustrates the effect of an unrepresentative prior on  Monte Carlo sampling algorithms in general and on NS in particular. We consider the case where both the likelihood and prior distributions are one-dimensional Gaussians  defined by $\mathcal{L}(\mtheta)=\mathcal{N}(\mtheta;0,1)$ and $\mathcal{\pi}(\mtheta)= \mathcal{N}(\mtheta;\mu_{\mpi},1)$,  respectively, where $\mu_{\mpi}$ is the mean of prior and $\mTheta$ is rewritten as $\mtheta$ for the one dimensional example. In this case, the fractional prior volume $X$ (which ranges between $[0,1]$ for NS) occupied by the region for which the likelihood exceeds the level $\mathcal{L}=\lambda$ is given by
\begin{align}
    &X(\lambda) = \int_{-|\mtheta(\lambda)|}^{|\mtheta(\lambda)|}\mathcal{\pi}(\mtheta)\,d\mtheta = \frac{1}{2}\left[\texttt{erf}\left(\frac{|\mtheta(\lambda)|-\mu_{\mpi}}{\sqrt{2}}\right) - \texttt{erf}\left(-\frac{|\mtheta(\lambda)|+\mu_{\mpi}}{\sqrt{2}}\right)\right]. \label{Eq: bigX_l}
\end{align}
where $\texttt{erf}$ is the error function and $\mtheta(\lambda) = \pm \sqrt{-\log(2\pi) - 2\log(\lambda)}$ is the value of random variable $\mtheta$ where the likelihood value is $\lambda$. Also, the fractional evidence contained in the complement region where the likelihood is below the level $\mathcal{L}=\lambda$ is:
\begin{align}
    \mathcal{Z}(X=X(\lambda)) &= \left(\int_{-\infty}^{\infty} - \int_{-|\mtheta(\lambda)|}^{|\mtheta(\lambda)|}\right)\mathcal{L}(\mtheta)\mathcal{\pi}(\mtheta)d\mtheta \nonumber \\
&= \frac{1}{2\sqrt{\pi}}\exp\left(-\frac{\mu_{\mpi}^2}{4}\right)\left[1-\frac{1}{2}\left\{\texttt{erf}\left(\frac{\mu_{\mpi}}{2}+|\mtheta(\lambda)|\right) - \texttt{erf}\left(\frac{\mu_{\mpi}}{2}-|\mtheta(\lambda)|\right)\right\}\right].\label{Eq: evi_Xt}
\end{align}}

\begin{figure*}[!ht]
\centering
\subfigure[$\text{log}\mathcal{L}$ against $\text{log}X$]{
\includegraphics[width = 0.48\linewidth]{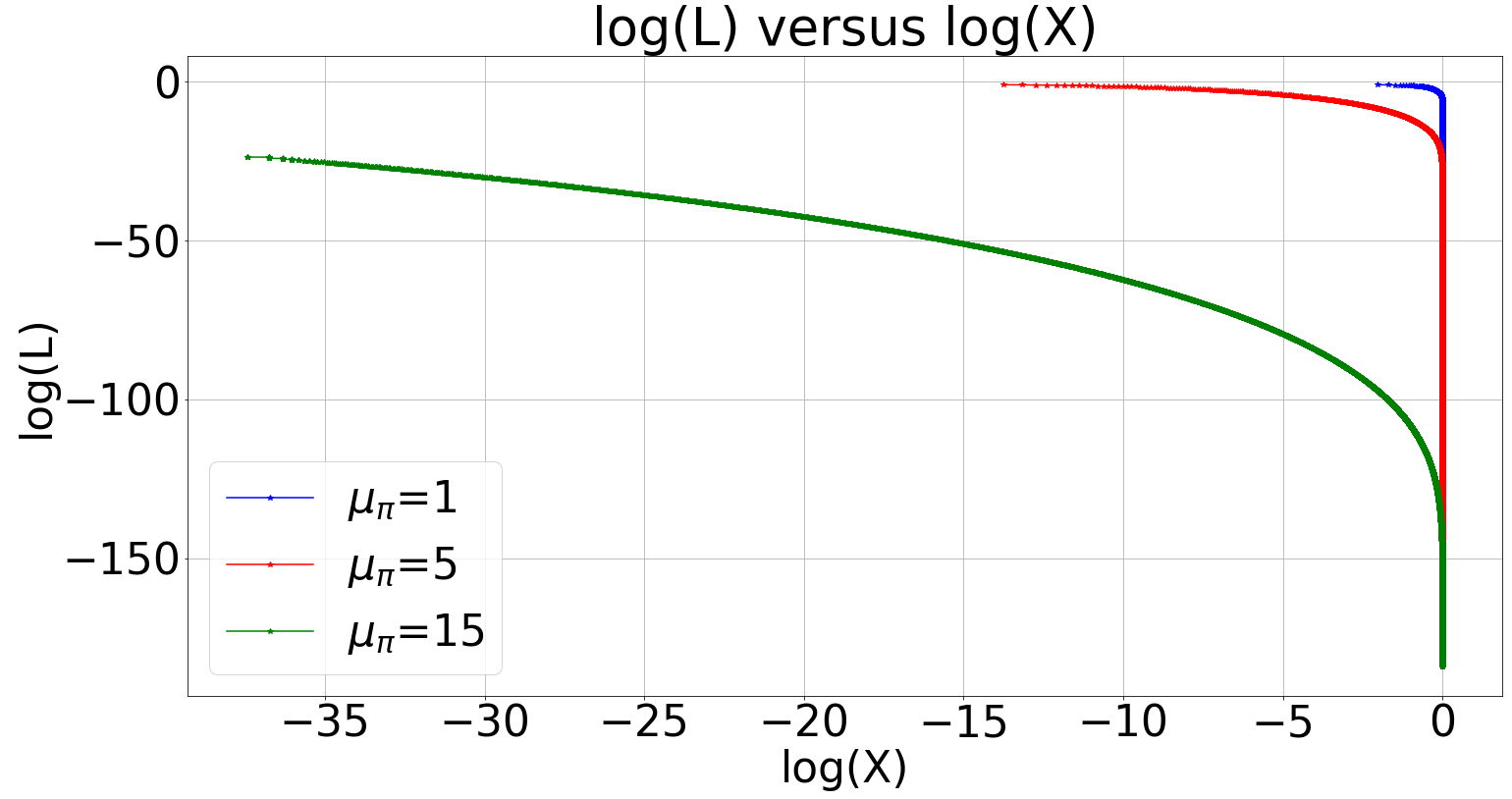}}
\subfigure[$\text{log}\mathcal{Z}(X)$ against $\text{log}X$]{
\includegraphics[width = 0.48\linewidth]{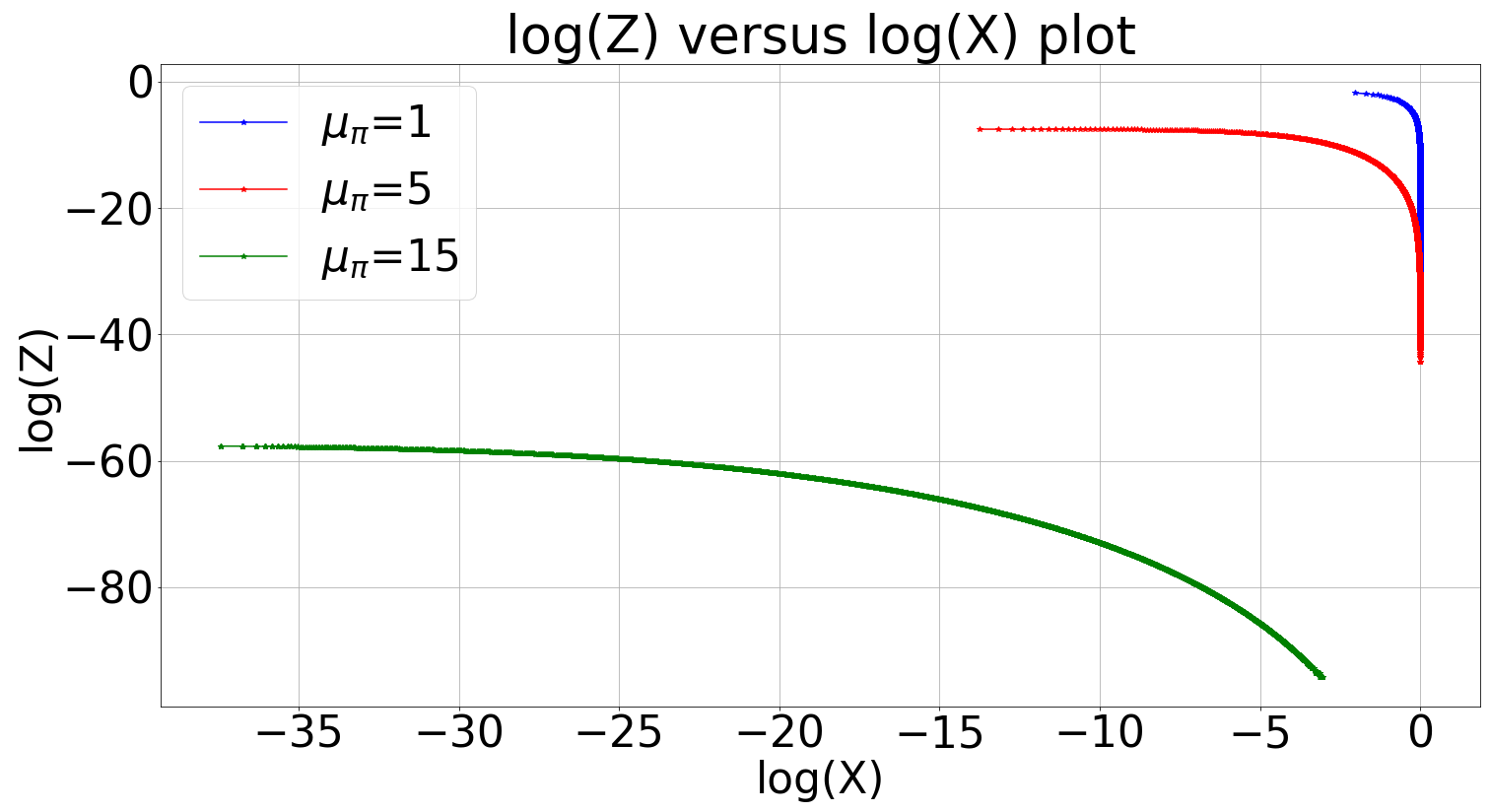}}
\caption{\revise{Relationship between likelihood $\mathcal{L}(X)$ and evidence $\mathcal{Z}(X)$ (up to prior volume $X$) against the prior volume $X$ for the toy examples discussed in Sec. \ref{sec:analytical_example}. The three curves in each figure represent the priors with mean $\mu_{\mpi}$ equal to $1, 5$ and $15$, respectively.}}
\label{fig:ExtraDemo}
\end{figure*}

\revise{We consider three cases for which the mean $\mu_{\mpi}$ of the Gaussian prior equal to $1$, $5$ and $15$, respectively. In the first case, the prior and likelihood are consistent,  while the other two cases correspond to increasingly unrepresentative priors. Figure \ref{fig:ExtraDemo} shows (a) the behavior of the likelihood $\mathcal{L}$ and (b) the evidence $\mathcal{Z}$ as a function of the prior volume $X$, all on a logarithmic scale.}

\revise{This figure shows that both $\mathcal{L}(X) \approx 0$ and $\mathcal{Z}(X) \approx 0$ (i.e., large negative values on a log-scale) in the case of an extreme unrepresentative with $\mu=15$ (green curve) for all values of $X$ except those very close to zero (or $\text{log}X \approx -\infty$). Also both $\log\mathcal{L}(X)$ and $\log\mathcal{Z}(X)$ increase more gradually with decreasing $\text{log}X$ as $\mu_{\mpi}$ increases (and the prior becomes more unrepresentative). Therefore, for higher $\mu_{\mpi}$ it becomes increasingly  difficult for any Monte Carlo algorithm in general and NS in particular to proceed as the likelihood is almost zero over most of the prior space and increases very gradually as the NS algorithm proceeds to regions contained within iso-likelihood contours at progressively higher likelihood levels, resulting in premature convergence. This also explains the results shown in Figure \ref{fig:1DVarNlike}(a) in the Supplementary Material, where standard NS experiences a sudden failure as the prior becomes increasingly unrepresentative.}

\section{Bayesian posterior repartitioning}
\label{Sec:BPRmethod}

In previous research, we first considered the possibility of improving the robustness and efficiency of NS by exploiting the intrinsic degeneracy between the `effective' likelihood and prior in the formulation of Bayesian
inference problems in \cite{feroz2009multinest},
and then further developed the idea as the original PR method in
\cite{chen2018improving}.

As mentioned in the Introduction, the central idea is to `repartition'
the product of the likelihood and prior, such that
\begin{align}
\mL(\mTheta) \mpi(\mTheta) = \tmL(\mTheta) \tmpi(\mTheta),
\label{Eq:equality}
\end{align}
where $\tmL(\mTheta)$ and $\tmpi(\mTheta)$ are the newly-defined
effective (or modified) likelihood and prior, respectively.  As a
result, the (unnormalised) posterior remains unchanged and hence, in
principle, any parameter inferences are unaffected. Moreover, if
$\tmpi(\mTheta)$ is normalised, then the evidence also remains
unchanged. In general, $\tmpi(\mTheta)$ may be any distribution that
can be straightforwardly sampled, and in principle can be arbitrary
with sufficient effort \citep{Alsing2021}. For example, there is no
requirement for the modified prior to be centred at the same parameter
value as the original prior. One could, therefore, choose a modified
prior that broadens and/or shifts the original one, or a modified
prior that has a completely different form from the original. In this
generalised setting, however, the modified prior should at least have
the same support as the original prior.

\begin{figure}[t]
\centering \includegraphics[width = 0.6\linewidth]{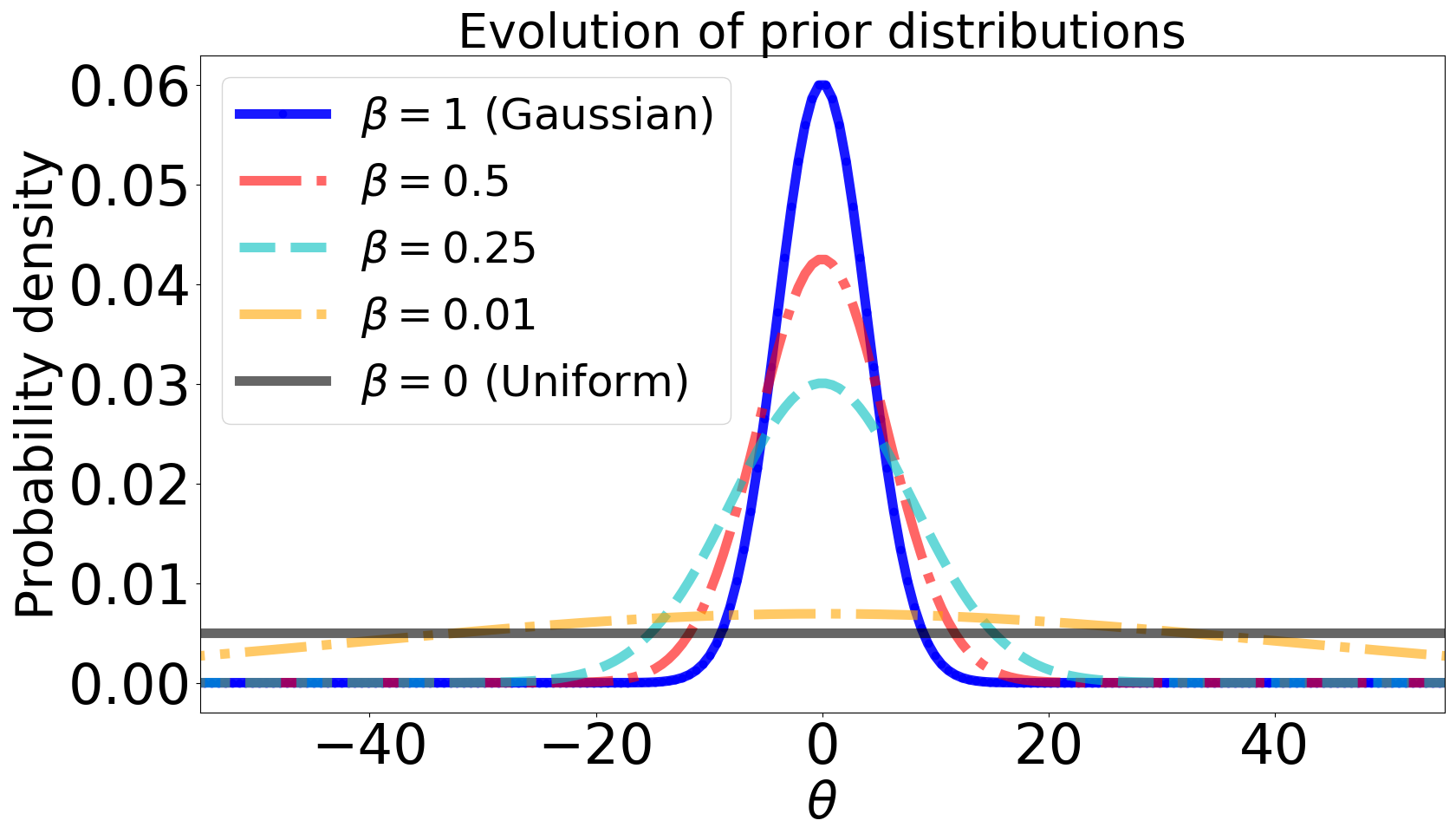}
\caption{One-dimensional prior evolution for $\mbeta \in [0,1]$. The
  original prior is a Gaussian distribution with $\sigma_\pi=4$
  (truncated in the range $[-50,50]$) when $\mbeta=1$ (solid blue curve), and is an uniform distribution when $\mbeta=0$ (solid black curve). The remaining three curves correspond to $\mbeta=0.5$ (dot-dashed red curve), $0.25$ (dashed light blue curve), $0.01$ (dot-dashed yellow curve),
  respectively.}
\label{fig:priorEvo}
\end{figure}

In practice, rather than introducing a completely new prior
distribution into the problem, a convenient choice (which we
previously termed power PR)
is simply to take $\tmpi(\mTheta)$ to be the original prior
$\mpi(\mTheta)$ raised to some (real) power $\beta$, and then
renormalised to unit volume, such that
\begin{eqnarray}
\tmpi(\mTheta) &=& \frac{\mpi(\mTheta)^{\mbeta}}{\mZ_\pi(\beta)},
\label{Eq:proportion} \\
\tmL(\mTheta) &=& \mL(\mTheta)
\mpi(\mTheta)^{(1-\mbeta)}\mZ_\pi(\beta), \label{eqn:newlike}
\end{eqnarray}
where $\mbeta \in [0,1]$ and $\mZ_\pi(\beta) \equiv \int
\mpi(\mTheta)^{\mbeta} d\mTheta$ is the normalisation constant of the
modified prior.  By altering the value of $\mbeta$, the modified prior
varies across a range of distributions between the original prior
($\mbeta=1$) and the uniform distribution ($\mbeta=0$). This is
illustrated in Figure \ref{fig:priorEvo} for five specific $\mbeta$
values in a one-dimensional case, where the original prior is a
Gaussian with zero mean and standard deviation $\sigma_\pi=4$ and the
normalisation depends on the assumed support $[-50,50]$ of the unknown
parameter $\mTheta$.  The $\beta=0$ limit clearly yields a uniform
modified prior $\tmpi(\mTheta) \sim \calU({\cal R})$ over the allowed
region ${\cal R}$ of the parameter space, which is an
important special case in that it maximises the dispersion of the
initial live point set across the prior space, but consequently can
result in very inefficient sampling. It is worth noting that in the
special case in which the original prior is uniform $\mpi(\mTheta)
\sim \calU({\cal R})$, the power PR method is insensitive to the value of
$\beta$ and defaults to standard NS. \revise{In principle, one may extend the upper limit on $\beta$ to exceed unity; this produces a new prior distribution that is sharper than the original, which may prove useful if the latter is over-dispersed. The power PR method does, however, have the limitation that one must be able to sample from the resulting modified prior and evaluate the normalising constant $\mathcal{Z}_{\pi}(\beta)$. If necessary, the latter may be estimated numerically, most efficiently in a separate NS run, for each value of $\beta$ used, but this may increase the computational costs significantly. Nonetheless, this issue may potentially be mitigated by pre-calculating $\mathcal{Z}_{\pi}(\beta)$ for a set of $\beta$ values and storing them in a look-up table, from which interpolation can be performed to approximate $\mathcal{Z}_{\pi}(\beta)$ for other $\beta$ values.} 
\label{Sec:powerPR}

A clear drawback of the original PR method, however, is that an
appropriate value of the auxiliary variable $\beta$ must be determined
on a case-by-case basis, and this can depend sensitively on the nature
of the original prior and likelihood, as well as on the dimensionality
of the problem under consideration. In \cite{chen2018improving}, we
therefore suggested an approach in which $\beta$ is gradually lowered
from unity (outside of the NS algorithm) according to some `annealing
schedule', until the resulting inferences from successive NS runs
converge to a statistically consistent solution, which typically
occurs for $\beta$ values below some (positive) threshold, $\beta
\lesssim \beta_\ast$. Although we demonstrated in
\cite{chen2018improving} that this method is robust and effective, it
has the disadvantages that there is a significant computational
overhead associated with the multiple NS runs required and that the
final inference of the parameters of interest $\mTheta$ is conditioned
on the adopted value $\beta=\beta_\ast$.

We therefore propose an alternative approach here, which we consider
to be considerably more elegant, whereby $\beta$ is instead treated as a
hyperparameter that is inferred in a fully Bayesian manner alongside
the original parameters $\mTheta$, within a single run of the NS
algorithm. Although this approach is admittedly conceptually very
simple, we view this as a virtue, in particular because the resulting
Bayesian PR (BPR) method is considerably more powerful and flexible
than the original PR technique in a number of important ways. The BPR method is based on defining the joint posterior
\begin{align}
    \tilde{\mP}(\mTheta, \mbeta)  \propto \tilde{\mL}(\mTheta, \mbeta) 
\tilde{\mpi}(\mTheta,\mbeta) = \tilde{\mL}(\mTheta, \mbeta) 
\tilde{\mpi}(\mTheta | \mbeta) \mpi(\mbeta),
\label{eqn:joint}
\end{align}
where $\mpi(\mbeta)$ denotes the assumed prior on the hyperparameter
$\mbeta$, and $\tilde{\mpi}(\mTheta | \mbeta)$ and
$\tilde{\mL}(\mTheta, \mbeta)$ have precisely the forms
(\ref{Eq:proportion}) and (\ref{eqn:newlike}), respectively. Since
$\beta$ lies naturally in the range $[0,1]$, we define $\mpi(\mbeta)$
to be the uniform prior over this interval, although other choices may
be accommodated if there is a strong motivation to adopt an
alternative form in a particular problem. \revise{For example, if one wishes to extend the upper limit on $\beta$ to exceed unity to accommodate an over-dispersed original prior $\mpi(\mTheta)$ on the parameters, a natural prior on $\mbeta$ is
$\mpi(\mbeta)=e^{-\mbeta}$ in the range $[0, \infty]$, which is the maximum-entropy distribution for a non-negative quantity for which one knows only that its expectation value is unity.} 
In any case, from the relations
(\ref{Eq:equality}) and (\ref{Eq:Bayes}), one sees that (by construction)
\begin{align}
    \tilde{\mP}(\mTheta, \mbeta)  \propto \mP(\mTheta)\pi(\beta).
\label{eq:modpost}
\end{align}
Moreover, the corresponding evidence is given by
\begin{align}
\tilde{\mathcal{Z}} &= \int\!\!\!\int \tilde{\mL}(\mTheta, \mbeta) 
\tilde{\mpi}(\mTheta,\mbeta)\,d\mTheta\,d\beta \\
&= 
\int \mL(\mTheta)\mpi(\mTheta)\,d\mTheta\,\int \pi(\beta)\,d\beta =
\mathcal{Z},
\label{eq:modevid}
\end{align}
so that the proportionality in (\ref{eq:modpost}) can be replaced by
an equality.

In principle, a single NS run thus provides samples from the full
joint posterior distribution (\ref{eq:modpost}), which can be used
straightforwardly to obtain inferences on the original parameters
$\mTheta$ in a properly Bayesian manner by marginalising over $\beta$, since
\begin{align}
\tilde{\mP}(\mTheta) = \int\tilde{\mP}(\mTheta, \mbeta)\,d\beta =
\int {\mP}(\mTheta)\mpi(\mbeta)\,d\beta =      {\mP}(\mTheta).
\end{align}
Hence, the overall computational burden is much reduced compared to
the original PR method, since one does not require the latter's
multiple NS runs, namely one for each value of $\beta$ in its
annealing schedule, or more if runs with the same value of $\beta$
(particularly for large $\beta$ values) are liable to exhibit
significant variation in their results because of the failure of the
NS algorithm, thereby complicating the process of determining
convergence of results as the annealing schedule proceeds.  Moreover, the final
inference on the parameters $\mTheta$ in BPR is not conditioned on a
particular value of $\beta$.  As we will show in examples in Section~\ref{Sec:Numeric}, marginalising over $\beta$ also allows BPR to accommodate likelihood functions consisting of multiple spatially separated modes that are located asymmetrically with respect to the prior, and which are therefore characterised by different ranges of $\beta$ values; this is not possible using the
original PR method.

A further advantage of BPR is that one may instead choose to
marginalise over the original parameters $\mTheta$ to obtain the 1-D
\revise{`effective'} posterior on $\beta$. This should in principle simply recover the prior on
$\mbeta$, since
\begin{align}
\tilde{\mP}(\mbeta) = \int\tilde{\mP}(\mTheta, \mbeta)\,d\mTheta =
\int {\mP}(\mTheta)\mpi(\mbeta)\,d\mTheta =  \pi(\mbeta).
\end{align}
Thus, if the NS procedure has correctly sampled from the joint
posterior distribution $\tilde{\mP}(\mTheta, \mbeta)$, one has neither
gained nor lost anything by introducing the hyperparameter $\beta$,
apart perhaps from a slight increase in the computational burden as a
result of increasing by one the dimensionality of the space to be
sampled.

In practice, however, the situation is more subtle. For illustration,
let us consider the case where the original prior $\pi(\mTheta) =
\tilde{\mpi}(\mTheta,1)$ is extremely unrepresentative of the dataset
under analysis, so that the original likelihood
$\mL(\mTheta)=\tilde{\mL}(\mTheta,1)$ is concentrated very far into
the wings of $\pi(\mTheta)$.  In the limit $N_{\rm live} \to \infty$,
the NS algorithm would nonetheless converge correctly, yielding
samples from the posterior (\ref{eq:modpost}) and an estimate of the
evidence (\ref{eq:modevid}). For finite $N_{\rm live}$, however, live
points drawn from $\tilde{\mpi}(\mTheta,\mbeta)$ at any NS iteration
will typically have very low likelihood $\tilde{\mL}(\mTheta,\mbeta)$
for values of $\beta$ above some limiting threshold $\beta \gtrsim
\beta_+$ (which will depend on $N_{\rm live}$), since the chance of
these points lying within the main body of the likelihood
$\tilde{\mL}(\mTheta,\mbeta)$ is vanishingly small. Depending on the
precise nature of the original prior $\pi(\mTheta)$ and likelihood
$\mL(\mTheta)$, a similar phenomenon may also occur for values of
$\beta$ below some other limiting threshold $\beta \lesssim \beta_-$
(which will also depend on $N_{\rm live}$), since in this case the
modified prior may have a support that far exceeds that of the
likelihood.

Thus, in the presence of unrepresentative priors where the NS
algorithm may become very inefficient or even fail, one may obtain
samples that are drawn not from (\ref{eq:modpost}), but instead from
some `effective' posterior
\begin{align}
    \tilde{\mP}_{\rm eff}(\mTheta, \mbeta) \propto \mP(\mTheta)
    \tilde{\mP}(\beta),
\label{eq:effpost}
\end{align}
where the (unnormalised) marginal posterior $\tilde{\mP}(\beta)$ is
non-zero only in some range $\sim[\beta_-,\beta_+]$. The form of this
marginal posterior, in particular its extent in $\beta$, will be
determined by how the NS algorithm fails for particular (extreme)
values of $\beta$, which is in turn dependent on the particular NS
implementation used and the associated control parameters (including
generic NS parameters such as $N_{\rm live}$), as well as the nature
of the original prior $\pi(\mTheta)$ and likelihood
$\mL(\mTheta)$. One would, however, expect there to be a range of
$\beta$ values for which the NS algorithm does not fail, and so
$\tilde{\mP}(\beta)$ should coincide with $\pi(\beta)$ in this
range. Beyond these very general considerations, there is a paucity of
further theoretical arguments from which to predict the form of
$\tilde{\mP}(\beta)$.

Nonetheless, by marginalising (\ref{eq:effpost}) over $\beta$ one
still obtains the posterior $\mP(\mTheta)$ on the original
parameters. Conversely, marginalising over $\mTheta$ one obtains
$\tilde{\mP}(\beta)$ (in practice as a histogram constructed from the
equally weighted posterior samples) and may hence observe its form and
determine the values $\beta_-$ and $\beta_+$. In particular, the value
of $\beta_+$ is useful in diagnosing both the existence and severity
of an unrepresentative prior for a given dataset, and thereby
identifying the dataset as an `outlier'. Here we will take $\beta_-$
and $\beta_+$ simply as the smallest and largest $\beta$ values,
respectively, in the set of equally weighted posterior
samples. Alternatively, one could apply some percentile thresholds
(e.g. 1\% and 99\%) to the $\beta$ marginal, but in practice this
leads to very similar values of $\beta_-$ and $\beta_+$. In any case,
observing the form of $\tilde{\mP}(\beta)$ obtained in the BPR method
provides a far more robust and computationally efficient approach to
identifying the presence and severity of unrepresentative priors than
determining and interpreting the value $\beta_\ast$ in the original PR
method below which the inferences converge.

Finally, in the presence of unrepresentative priors, the NS process
will not yield an estimate of the evidence (\ref{eq:modevid}), but
rather the `effective' evidence
\begin{align}
\tilde{\mathcal{Z}}_{\rm eff} \approx
\tilde{\mathcal{Z}}\int\tilde{\mP}(\mbeta)\,d\mbeta =
\mathcal{Z}\int\tilde{\mP}(\mbeta)\,d\mbeta.
\label{eq:zeff}
\end{align}
One may, however, estimate the required evidence
$\mathcal{Z}$, provided one can evaluate the factor
$\int\tilde{\mP}(\mbeta)\,d\mbeta$. Fortunately, from the discussion
above, this may be achieved by first scaling the distribution
$\tilde{\mP}(\mbeta)$, represented as a histogram of equally-weighted
posterior samples, such that the bin containing the largest number of
samples has the volume $\int_{\beta_{\rm l}}^{\beta_{\rm
    u}}\pi(\beta)\,d\beta$, where $\beta_{\rm l}$ and $\beta_{\rm u}$
are the lower and upper limits of the bin, respectively. The factor
$\int\tilde{\mP}(\mbeta)\,d\mbeta$ may then be calculated by summing
up the volumes of all the bins in the resulting scaled histogram. \revise{This summation is equivalent to the simplest form of numerical quadrature, namely the rectangle rule, which adopts a polynomial of degree zero (a constant function) to pass through points $\textstyle \left({ (\beta_{\rm l}^{b}+\beta_{\rm u}^{b}) / 2}, h(\beta_{\rm l}^{b}+\beta_{\rm u}^{b}) / 2 \right)$, where $\beta_{\rm l}^{b}$ and $\beta_{\rm u}^{b}$ are the lower and upper limits of the $b$th bin, and $h(\cdot)$ indicates the height of the bin. In principle, one may estimate the error in the quadrature and propagate this through to $\tilde{\mathcal{Z}}_{\rm eff}$ using Eqn. (\ref{eq:zeff}), but we do not perform this here.}

\section{Numerical examples}
\label{Sec:Numeric}

We now illustrate the performance of the BPR method in some numerical
examples. Since we extensively studied the behaviour of our original
PR method in a wide range of problems in \cite{chen2018improving},
including comparisons with other sampling algorithms such as Markov
Chain Monte Carlo (MCMC) and importance sampling, we focus here
primarily on the performance of BPR in the canonical case where the
likelihood and prior on the original variables $\mTheta$ are both
Gaussian (although very mismatched in some examples), and hence so too
is the posterior. In particular, we begin by considering the
univariate case, before moving on to a bivariate example for which we
consider both circularly-symmetric and asymmetric priors, where the
latter may have zero, positive or negative correlation coefficient,
respectively.  \revise{To demonstrate the method further, however, we also
consider several 2-dimensional non-Gaussian likelihoods: multi-modal likelihoods consisting of an
equal mixture of four identical but spatially separated
Gaussians, and a unimodal Laplacian likelihood (the latter example is presented in the Supplementary Material)}. Finally, we investigate higher-dimensional unimodal
Gaussian likelihood examples, up to 10 dimensions. In these further
examples, we consider only circularly-symmetric Gaussian priors for
the sake of brevity.

In all our numerical examples, we use the NS package MultiNest
\citep{feroz2009multinest}, with algorithm parameter settings similar
to those used in \cite{chen2018improving} for the study of our
original PR method. Specifically, we set the number of live points
$N_{\rm live}=100$, the desired sampling efficiency parameter ${\tt
  efr} = 0.8$, which equals the ratio of the remaining prior volume to
the volume of the multi-ellipsoid bound enclosing the active point set
at each NS iteration, and the convergence tolerance parameter ${\tt
  tol} = 0.5$, which corresponds to acceptable uncertainty on the
estimated log-evidence (please refer to \cite{feroz2009multinest} for
more details).

\subsection{Univariate Gaussian likelihood}\label{sec:univariate_gaussian_likelihood}

We begin by considering a simple one-dimensional estimation problem,
for which the data consist of $N$ independent measurements $M =
\{m_{1}, \cdots, m_{n}, \cdots, m_{N}\}$ of an unknown parameter
$\mtheta$, such that
\begin{align}
m_{n} = \mtheta + \mxi, \label{Eq:Simpexp}
\end{align}
where $\mxi$ denotes Gaussian noise $\mxi \sim \calN (\mu_{\mxi},
\sigma_{\mxi}^2)$. The likelihood is thus given by
\begin{equation}
\mL(\mtheta) = \prod_{n=1}^{N} 
\left\{ \frac{1}{\sqrt{2 \pi \sigma_{\mxi}^2}} \exp \left[-\frac{(m_n - \revise{(\mtheta + \mu_{\mxi})})^2}{2 \sigma_{\mxi}^2}\right] \right\},
\label{eqn:likelihood}
\end{equation}
which has a Gaussian form. In particular, we set $\mu_{\mxi} = 0$,
$\sigma_{\mxi} = 1$ and the number of measurements as $N=20$, so that
the likelihood is a Gaussian centred around the true value of $\theta$
with a standard deviation of $\sim 1/\sqrt{20} \approx 0.22$. The
prior distribution $\mpi(\mtheta)$ is also assumed to be Gaussian
$\mtheta \sim \calN (\mu_{\mpi}, \sigma_{\mpi}^2)$, where we set
$\mu_{\mpi} = 0$ and $\sigma_{\mpi}=4$. Therefore one expects {\em a
  priori} that the true value of $\mtheta$ will lie in the range
$[-12,12]$ with greater than $99.5\%$ probability. Finally, we assume
the prior distribution of the auxiliary factor $\mbeta$ to be the unit
uniform distribution $\mbeta \sim \calU[0,1]$.

\begin{figure*}[!ht]
\centering
\subfigure[unrepresentative prior, $\mtheta_\ast = 50$]{
\includegraphics[width = 0.48\linewidth]{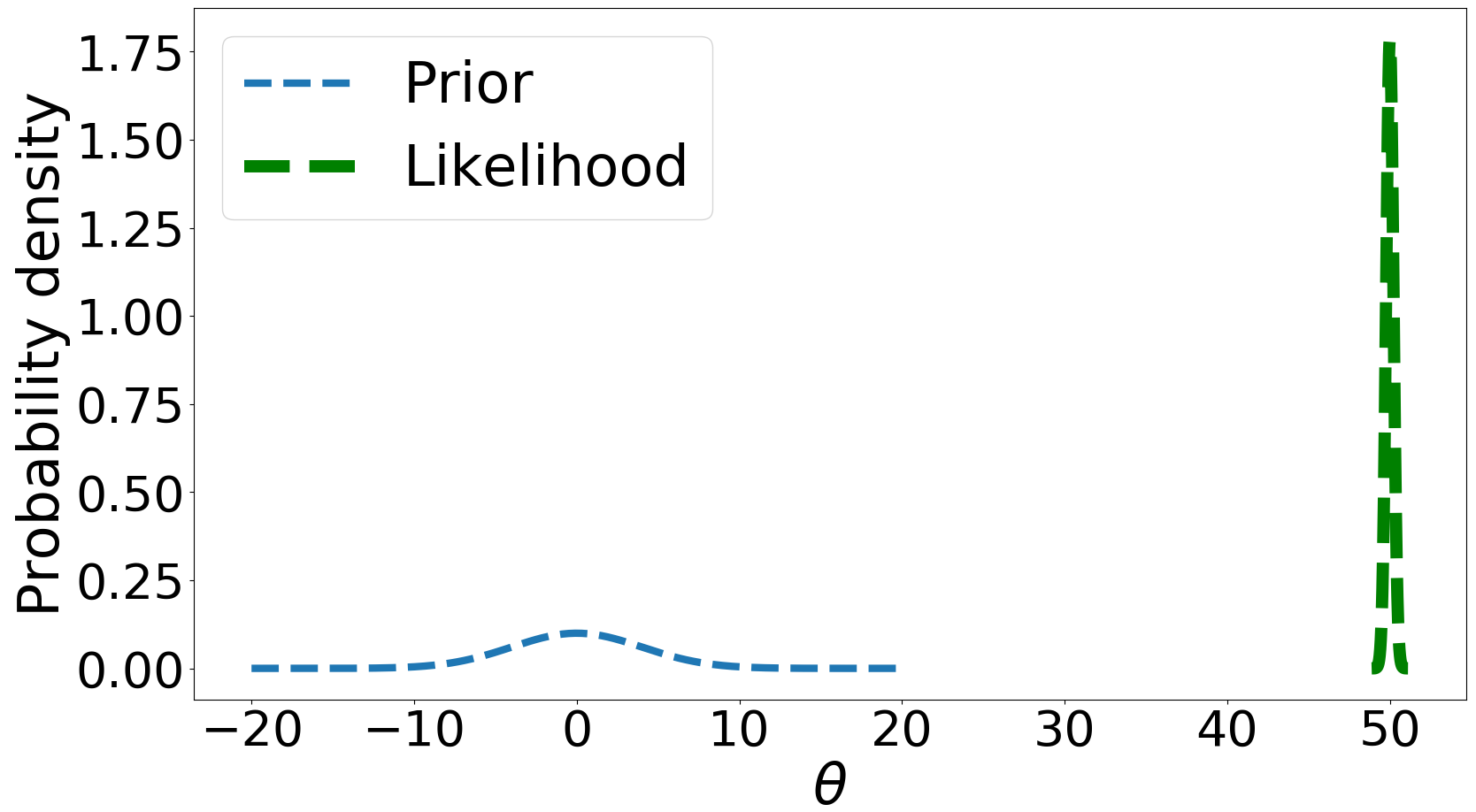}}
\subfigure[estimated posterior by MultiNest, $\mtheta_\ast = 50$]{
\includegraphics[width = 0.48\linewidth]{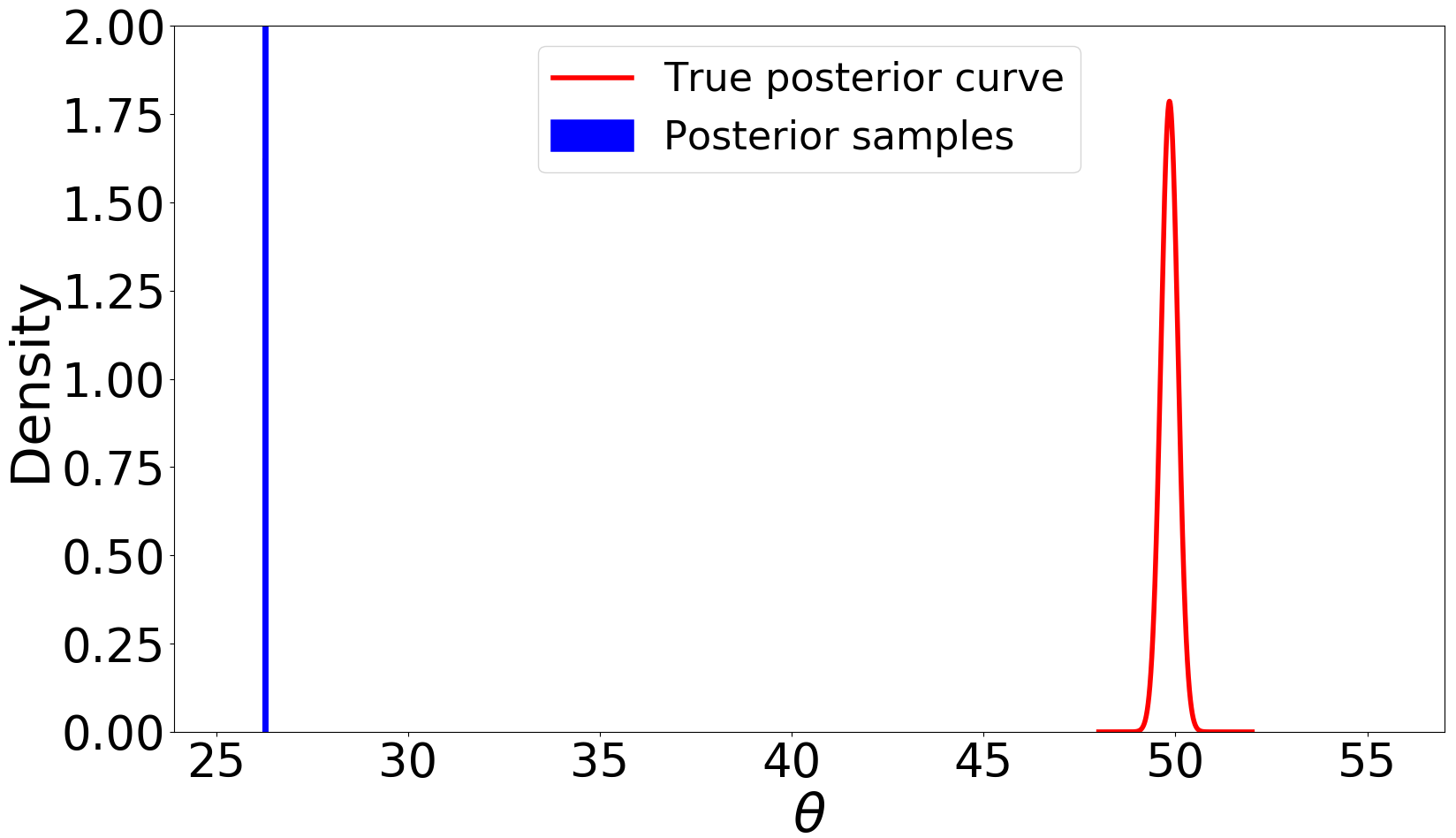}}
\caption{\revise{Univariate Gaussian likelihood examples illustrating
  the unrepresentative case and the posterior samples obtained by standard MultiNest (blue histogram line). The light blue and green dashed curves denote the prior
  and the likelihood, respectively. The red curve represents the true Gaussian posterior.}}
\label{fig:prior1}
\end{figure*}

We consider datasets corresponding to a series of true values $\mtheta_{\ast}$ ranging between 5 and 50. \revise{Figure \ref{fig:prior1}(a) shows the limiting example for $\mtheta_{\ast} = 50$, where the
likelihood lies very far into the wings of the prior and may thus be considered unrepresentative for this data set. Figure \ref{fig:prior1}(b) shows the posterior samples obtained using standard MultiNest without PR, together with the corresponding true Gaussian posterior. The behaviour in the limiting case is characteristic of a catastrophic failure mode of the NS algorithm in practice for extreme unrepresentative priors. This was first discussed in \cite{chen2018improving} and here we take a step further to illustrate and discuss this behaviour in Figure \ref{fig:1DVarNlike} and Section \ref{AP: UG} of the Supplementary Materials. Moreover, a discussion on failure probability of standard NS is also presented using the same univariate example as follows. }

\revise{\subsubsection{Failure probability of NS with univariate Gaussian likelihood}}
\label{AP: NS failure}
\revise{We calculate the failure probability of standard NS for the case where prior is a univariate Gaussian with mean $\mu_{\pi}$ and standard deviation $\sigma_{\pi}$. The likelihood is also a univariate Gaussian with mean $\mu_{\xi}$ and standard deviation $\sigma_{\xi}$ with $\mu_{\xi} > \mu_{\pi}$. We consider failure to be the case where there are no live points within $c\sigma_{\xi}$ of the likelihood mean $\mu_{\xi}$ after $I$ iterations. One could use $c = 3$ here. The failure will very likely happen before $I$ iterations due to all live points being the same within machine precision, so we can take the failure probability estimate $p_{\rm fail}$ obtained here as its lower bound.}

\revise{Before the first iteration, for the failure to happen, all $N_{\rm live}$ points have to be at values less than $(\mu_{\xi} - c\sigma_{\xi})$ which has probability: 
\begin{equation}
    p_{\rm fail, 0} = \Phi(\mu_{\xi} - c\sigma_{\xi}; \mu_{\pi}, \sigma_{\pi})^{N_{\rm live}},
\end{equation}
where $\Phi(x; \mu, \sigma)$ is the cumulative distribution function of Normal distribution and $p_{\rm fail, 0}$ denotes the failure probability estimate at the $i=0$ iteration. }

\revise{At each subsequent iteration $i$, if the remaining prior volume is $v_{\rm i}$ and the point with minimum likelihood value is at position $y_{i}$, then we have:
\begin{equation}
    1 - \Phi(y_{\rm i};  \mu_{\pi}, \sigma_{\pi}) = v_{\rm i},
\label{min_like_point}
\end{equation}
We can then calculate the probability of failure at iteration $i$, $p_{\rm fail, i}$ as:
\begin{equation}
\begin{aligned}
    p_{\rm fail, i} & = \int_0^1 \frac{P(v_{\rm i}) \left(\Phi(\mu_{\xi} - c\sigma_{\xi}; \mu_{\pi}, \sigma_{\pi}) - \Phi(y_{\rm i}; \mu_{\pi}, \sigma_{\pi}) \right)}{v_{\rm i}} dv_{\rm i},\\
    & = \int_0^1 \frac{P(v_{\rm i}) \left(\Phi(\mu_{\xi} - c\sigma_{\xi}; \mu_{\pi}, \sigma_{\pi}) - 1 + v_{\rm i} \right)}{v_{\rm i}} dv_{\rm i},
\end{aligned}
\label{p_fail_i}
\end{equation}
where the integral is over the distribution $P(v_{\rm i})$ of remaining prior volume $v_{\rm i}$ at iteration $i$ which itself equals to $v_{\rm i} = \prod_{\rm j=1}^{\rm i} t_{\rm j}$,
where $P(t) = N_{\rm live}t^{N_{\rm live} - 1}$ \citep{feroz2009multinest} i.e. the distribution of the maximum of $N_{\rm live}$ uniformly distributed random variables and therefore $v_{\rm i}$ is the product of beta distributed random variables. 
It isn't possible to get an analytical solution for the integral given in Eqn. \ref{p_fail_i}. However, we can get an approximation by assuming $v_{\rm i} \approx \mathop{\mathbb{E}}[v_{\rm i}]$,
where $\mathop{\mathbb{E}}[v_{\rm i}] = \exp(-i / N_{\rm live})$ \citep{feroz2009multinest}. Using this approximation in Eqn. \ref{p_fail_i} we get:
\begin{equation}
    p_{\rm fail, i} \approx \frac{\Phi(\mu_{\xi} - c\sigma_{\xi}; \mu_{\pi}, \sigma_{\pi}) - 1 + \mathop{\mathbb{E}}[v_{\rm i}]}{\mathop{\mathbb{E}}[v_{\rm i}]}.
\end{equation}
We can now approximate the total failure probability $p_{\rm fail}$ as:
\begin{equation}
\begin{aligned}
    p_{\rm fail} & \approx p_{\rm fail, 0} \prod_{\rm i=1}^{\rm I} p_{\rm fail, i}\\
    & = \Phi(\mu_{\xi} - c\sigma_{\xi}; \mu_{\pi}, \sigma_{\pi})^{N_{\rm live}} \prod_{\rm i=1}^{\rm I} \frac{\Phi(\mu_{\xi} - c\sigma_{\xi}; \mu_{\pi}, \sigma_{\pi}) - 1 + \mathop{\mathbb{E}}[v_{\rm i}]}{\mathop{\mathbb{E}}[v_{\rm i}]}.
\end{aligned}
\label{prob_fail}
\end{equation}}

\begin{figure}[t]
\centering \includegraphics[width = 0.6\linewidth]{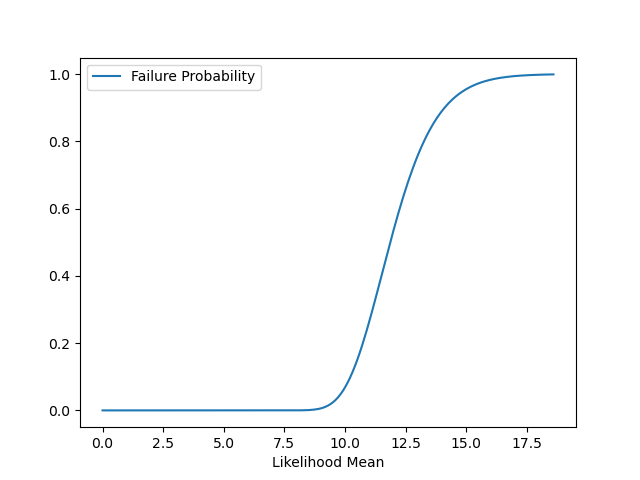}
\caption{\revise{Failure probability of standard NS for the univariate Gaussian example as a function of mean of the likelihood distribution $\mu_{\xi}$ (which is equal to $\theta_{*}$ in the rest of this section).}}
\label{fig:p_fail}
\end{figure}

\begin{figure*}[!ht]
\centering
\includegraphics[width = 0.6\linewidth]{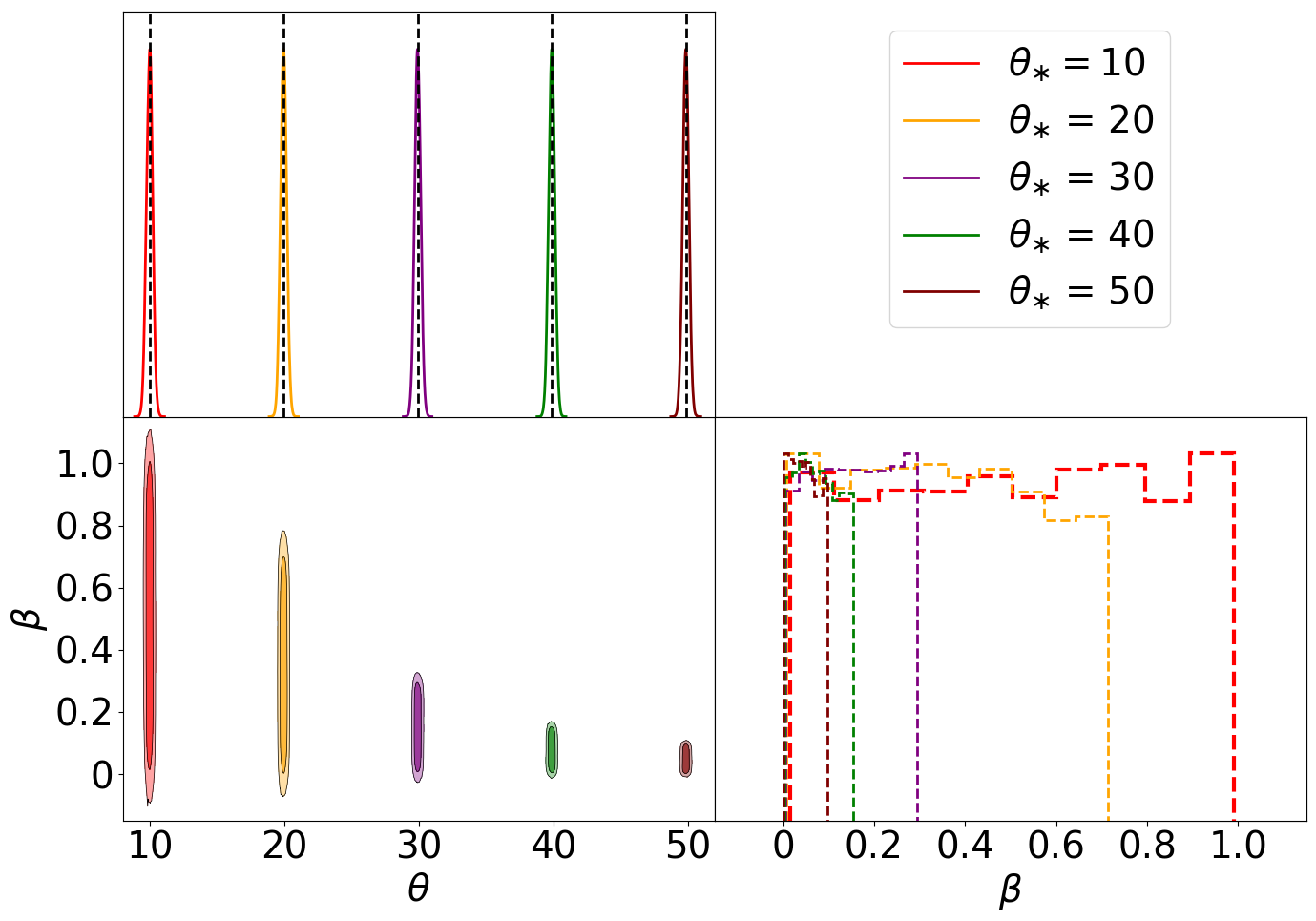}
\caption{A `corner plot' showing the joint posterior distribution on
  $(\theta,\beta)$ obtained by MultiNest using BPR, together with
  its marginals, for the univariate Gaussian likelihood example with a
  range of true values $\theta_\ast$. The vertical dashed black lines
  in the top left panel indicate the mean of the
  true Gaussian posterior in each case. }
\label{fig:1DComb}
\end{figure*}

\revise{We can use $p_{\rm fail}$ given above to explain the failure of standard NS in the univariate example discussed in this Section, i.e., $c = 3, I = 100, N_{\rm live} = 100, \mu_{\pi} = 0, \sigma_{\pi} = 4, \sigma_{\xi} = 0.22$. We can now calculate the failure probability $p_{\rm fail}$ for different values of $\mu_{\xi}$ (which is equal to $\theta_{*}$ in Figure  \ref{fig:1DComb}). As shown in Figure~\ref{fig:p_fail}, the failure probability increases rapidly for $\mu_{\xi} > 10$ and therefore, we would expect $\beta < 1$ for $\mu_{\xi} > 10$ with BPR which is indeed what we see in Figure \ref{fig:1DComb}.}

The above illustrated issue of NS may be straightforwardly addressed by applying the BPR method. Figure~\ref{fig:1DComb} shows the resulting MultiNest (with BPR) joint posterior on $(\theta,\beta)$ and its marginals for a selection
of true values $\theta_\ast$ in the range $[5,50]$ (\revise{the $\mtheta_{\ast} = 50$ case is plotted in brown}). One sees that the joint posterior is precisely of the form expected in Eqn. (\ref{eq:effpost}), in that it is the product of two independent
distributions on $\theta$ and $\beta$, respectively.  One also
observes the evolution of the marginal distribution on $\mbeta$. For
each value of $\theta_\ast$, this is consistent with a top-hat
distribution in the range $[0,\beta_+]$, where $\beta_+$ gradually decreases as
$\theta_\ast$ increases. For $\theta_\ast \lesssim 10$, one sees that
$\beta_+ \approx 1$, so one recovers the original uniform prior
distribution $\mbeta \sim \calU[0,1]$, indicating that the original
prior $\pi(\theta)$ is representative for these data sets.  For
\revise{$\theta_* \gtrsim 10$}, however, the value of $\beta_+$ gradually
decreases from unity as $\theta_\ast$ increases, which indicates that
the original prior is increasingly unrepresentative for
these data sets.

Turning to the marginals on $\mtheta$ in Figure~\ref{fig:1DComb}, one
sees that they are correctly centred on the mean of the corresponding
true Gaussian posterior for every value of $\mtheta_\ast$. Moreover,
the widths of the marginals on $\mtheta$ are equal for each value of
$\mtheta_\ast$ and consistent with the width of the true Gaussian
posterior, hence showing that BPR yields the correct
inferences on $\mtheta$, independent of the value of $\mtheta_\ast$,
in a fully automated manner, without any need for tuning. \revise{Quantitative results on both the parameter estimation accuracy on
$\mtheta$ and the evidence calculation are given in the Supplementary Material Section \ref{AP: UG}}.

\subsection{Bivariate Gaussian likelihood}
\label{sec:bgl}

We now move to a bivariate example, for which we consider both
circularly-symmetric and asymmetric priors, where the latter may have
zero, positive or negative correlation coefficient, respectively.  In
particular, consider a vectorised version of Eqn.
\eqref{Eq:Simpexp} from the univariate Gaussian likelihood example,
with some $K = 2$ dimensional parameter vector
$\mTheta=(\mtheta_1,\mtheta_2)^{\top}$, such that
\begin{align}
\mathbf{m}_n = \mTheta + \mXi, \label{Eq:Biexp}
\end{align}
where $\mathbf{m} = (m_1, m_2)^{\top}$, and
$\mXi=(\mxi_1,\mxi_2)^{\top}$ denotes the two dimensional Gaussian
noise $\mXi \sim \calN (\mmu_{\mXi}, \mSigma_{\mXi})$ with mean
$\mmu_{\mXi}$ and covariance $\mSigma_{\mXi}$. The prior distribution
is also Gaussian, $\mTheta \sim \calN (\mmu_{\mtheta},
\mSigma_{\mtheta})$. We assume unbiased measurements and priors
centered on the origin, such that
$\mmu_{\mXi} = (0,0)^{\top} = \mmu_{\mtheta}$, and parameterise the
noise and prior covariances matrices by
\begin{align*}
    \mSigma_{\mXi} = \begin{bmatrix} 
                        \sigma_{\mxi_1}^2 &  \rho_{\mxi}\sigma_{\mxi_1}\sigma_{\mxi_2} \\
                        \rho_{\mxi}\sigma_{\mxi_2}\sigma_{\mxi_1} & \sigma_{\mxi_2}^2 
                     \end{bmatrix}, \\
                     \qquad
    \mSigma_{\mtheta} = \begin{bmatrix} 
                    \sigma_{\mtheta_1}^2 & \rho_{\mtheta}\sigma_{\mtheta_1}\sigma_{\mtheta_2} \\
                   \rho_{\mtheta}\sigma_{\mtheta_2}\sigma_{\mtheta_1} & \sigma_{\mtheta_2}^2 
                 \end{bmatrix}, 
\end{align*}
where $\rho_{\mxi}$ and $\rho_{\mtheta}$ are the standard correlation
coefficients in each case. 

In this example, we take the opposite approach to that used in the
univariate example, in that we retain the same likelihood function for
each case considered and instead vary the form of the assumed prior,
although in all cases the prior is centred on the origin of the
parameter space. In particular, we assume throughout the true value
$\mTheta_{\ast} = (40,40)^{\top}$, uncorrelated measurement noise
with unit standard deviation, such that $\rho_{\mxi}=0$ and
$\sigma_{\mxi_1}=\sigma_{\mxi_2}=1$, and that the number of
measurements is just $N = 1$, as this yields a circularly-symmetric
bivariate Gaussian likelihood distribution of unit standard deviation,
which is convenient for our investigations. We present results only
for the BPR method, since the standard NS approach fails in all the
cases considered.

\subsubsection{Uncorrelated priors}

\begin{figure}[!ht]
\centering
\subfigure[uncorrelated prior]{
\includegraphics[width = 0.47\linewidth]{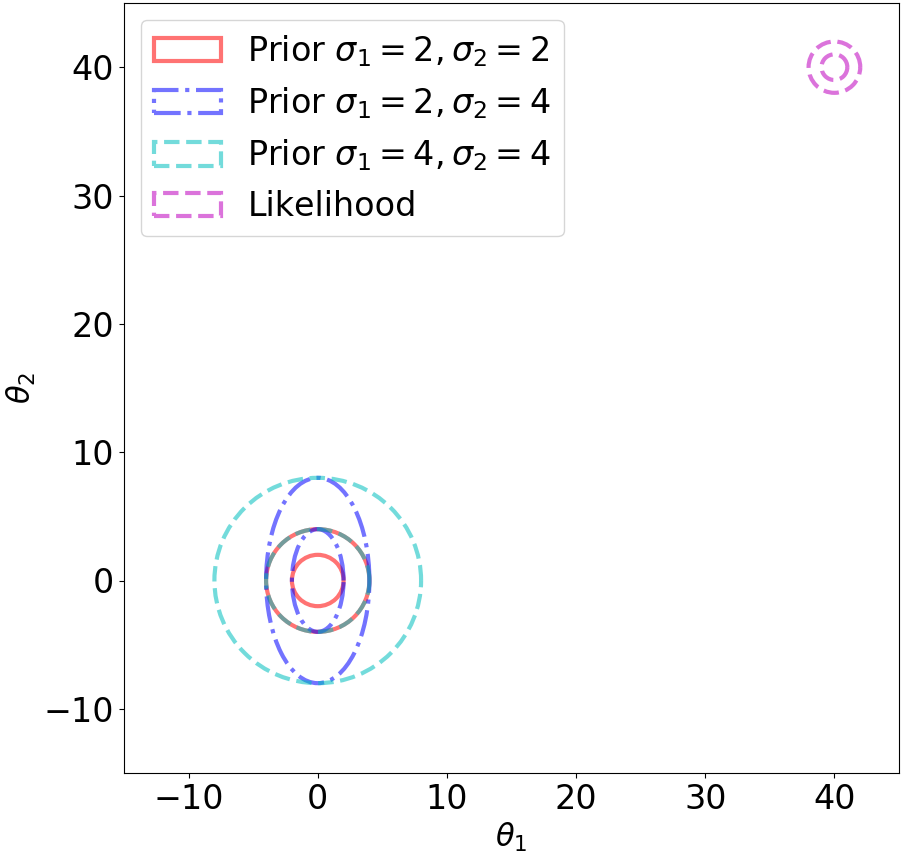}}
\subfigure[correlated prior]{
\includegraphics[width = 0.48\linewidth]{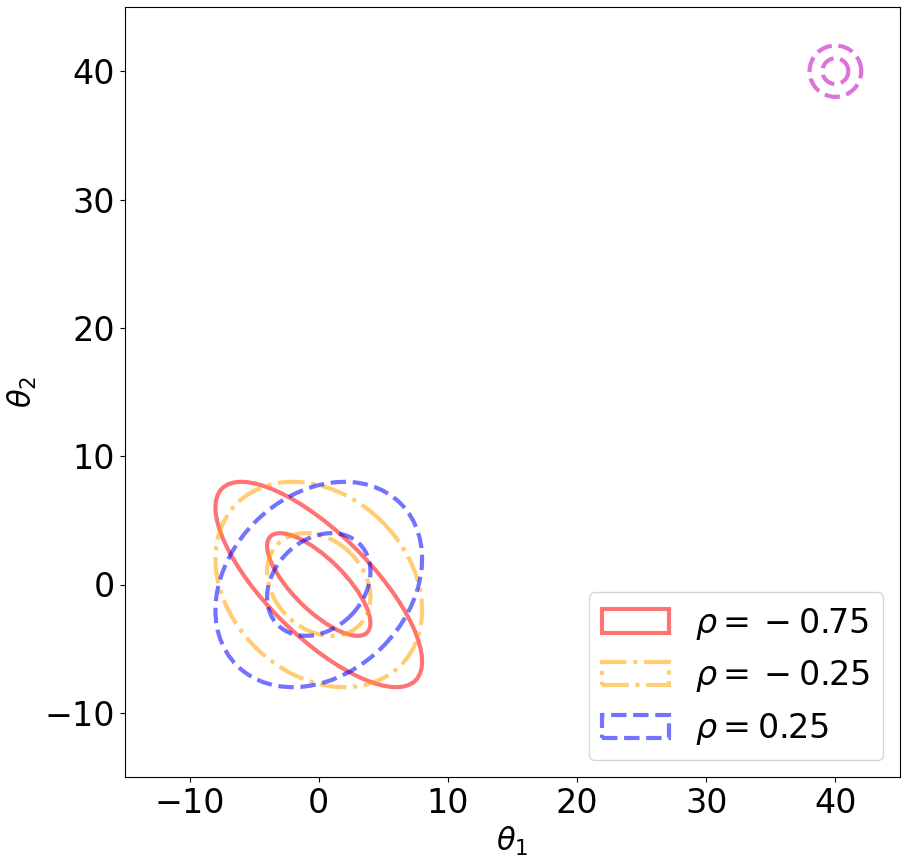}}
\caption{\revise{Illustration of the three test cases in the bivariate
  Gaussian likelihood example with uncorrelated/correlated priors. In the left panel, the cyan dashed, red solid and blue dot-dashed lines denote the Gaussan priors centred on the origin,
  with $\{\sigma_{\mtheta_1},\sigma_{\mtheta_2}\}$ equals to $\{4, 4\}$, $\{2, 2\}$, and $\{2, 4\}$, respectively. The right panel shows correlated prior with $\{4, 4\}$ and correlation coefficients
  $\rho_{\theta} = \{-0.75, -0.25, 0.25\}$, respectively, denoted in a rainbow colour order. The pink dashed lines in the upper right corner denote the likelihood distribution. Each distribution contains two coloured contours corresponding to the $2\sigma$ ($68\%$) and $3\sigma$ ($95\%$) iso-probability levels.}}
\label{fig:2DillBoth}
\end{figure}

We begin by considering uncorrelated priors, for which $\rho_{\mtheta}
= 0$.  In particular, we consider the two circularly-symmetric cases where 
$\{\sigma_{\mtheta_1}$, $\sigma_{\mtheta_2}\}$ equals to $\{4, 4\}$ and
$\{2, 2\}$, and the intermediate non-circularly-symmetric case $\{2, 4\}$. These priors and the likelihood are plotted in Figure~\ref{fig:2DillBoth}(a), which illustrates that
all the priors are unrepresentative.

\begin{figure}[!ht]
\centering
\includegraphics[width = 0.6\linewidth]{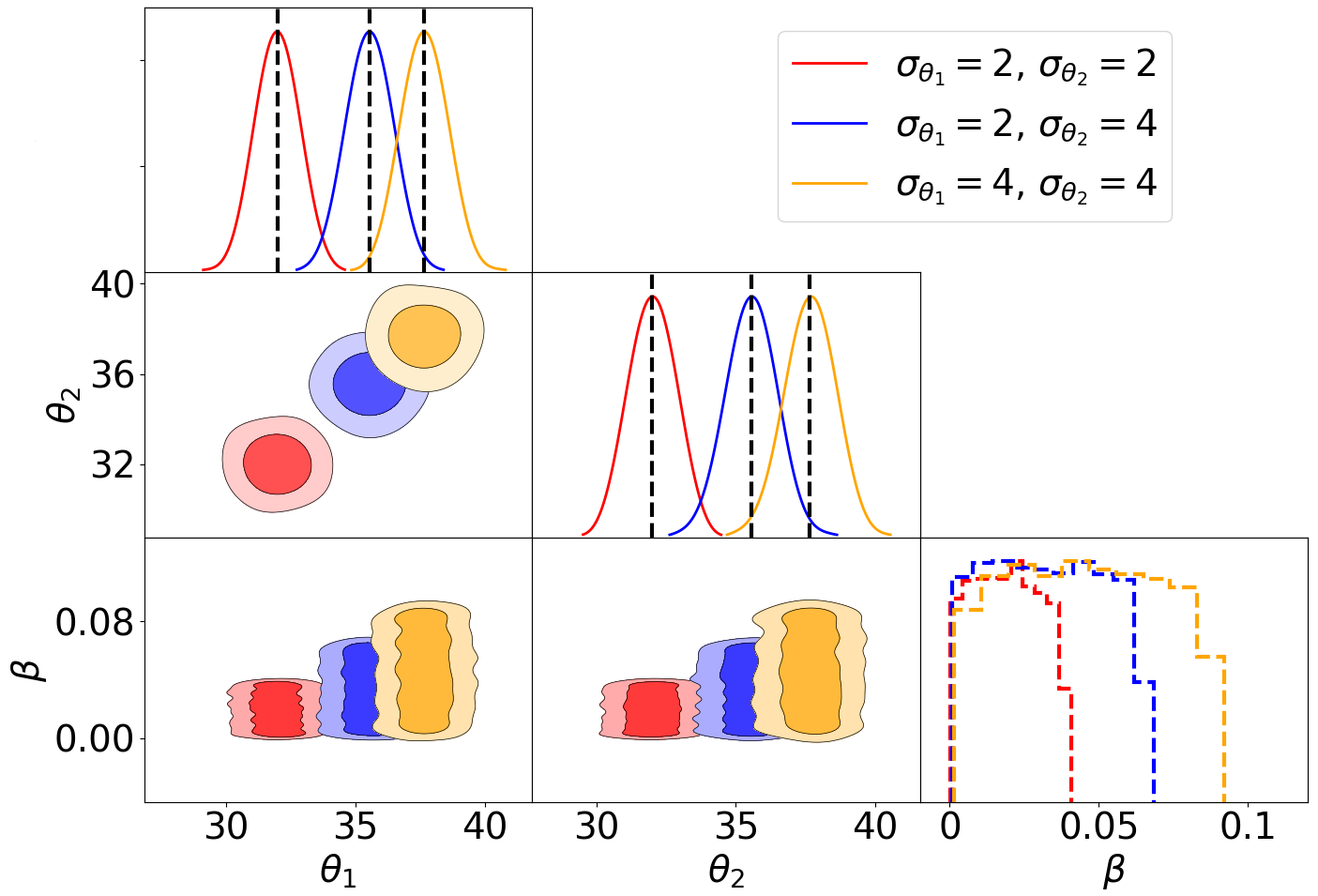}
\caption{A `corner plot' showing the 1-dimensional and 2-dimensional
  marginals of the joint posterior distribution on
  $(\theta_1,\theta_2,\beta)$ obtained by MultiNest using BPR for
  the bivariate Gaussian likelihood example with the likelihood and
  priors illustrated in Figure~\ref{fig:2DillBoth}(a). The vertical
  dashed black lines indicate the mean of the true
  Gaussian posterior in each case.}
\label{fig:2D40}
\end{figure}

Figure~\ref{fig:2D40} is a `corner plot' showing the 1-dimensional and
2-dimensional marginal distributions of the joint `effective'
posterior on $(\theta_1,\theta_2,\beta)$ for each of the three priors
described in Figure~\ref{fig:2DillBoth}(a). In each case, the joint posterior again has the form
of the product of two independent distributions on
$(\theta_1,\theta_2)$ and $\beta$, respectively, and is consistent
with a marginal on $\beta$ having the approximate form of a top-hat
distribution in the range $[0,\beta_+]$. One also observes the
expected evolution of this marginal across the three test cases,
whereby $\beta_+$ gradually decreases as the likelihood is
concentrated progressively further into the wings of the corresponding
prior. Since $\beta_+ \lesssim 0.1$ in all three cases, one may
confirm that each prior is indeed unrepresentative, and one may use
the values of $\beta_+$ for each case to rank their severity.

Turning to the marginals on $\mtheta_1$ and $\theta_2$ in
Figure~\ref{fig:2D40}, one sees that they are correctly centred on the
mean of the corresponding true Gaussian posterior for each
case. Moreover, the widths of the marginals are stable across the
different priors and consistent with the width of the true Gaussian
posterior in each case; the RMSE of the $\theta_\ast$ estimate
$\approx 0.05$ in all cases. Thus, as in the univariate Gaussian
likelihood example, the BPR method yields the correct inferences on
$(\mtheta_1,\theta_2)$ without any need for fine tuning.

\begin{table}
\caption{Comparison of the mean and standard deviation of the
  estimated log-evidence over 10 realisations of the data in the
  bivariate Gaussian likelihood example, obtained using the BPR method for a selection of uncorrelated priors. The `true' value of the evidence in each case is also given,
  as estimated using standard quadrature techniques.}
\smallskip
\centering
\begin{tabular}{c|rrr}
\hline
Prior & True & BPR & BPR \\
 &  & mean & s.d. \\
\hline
$\{\sigma_{\mtheta_1} = 4,\sigma_{\mtheta_2} = 4\}$ & $-98.79$ &
$-98.70$ 
& $0.88$ \\
$\{\sigma_{\mtheta_1} = 2,\sigma_{\mtheta_2} = 4\}$ & $-182.01$ &
$-182.01$ 
& $1.24$ \\
$\{\sigma_{\mtheta_1} = 2,\sigma_{\mtheta_2} = 2\}$ & $-325.76$ &
$-325.66$ & $1.49$ \\
\hline
\end{tabular}
\label{tab:2DUncorrEvid}
\end{table}

One may also verify that the BPR method yields accurate evidence
estimates in the above cases. Table~\ref{tab:2DUncorrEvid} lists the
mean and standard deviation of the estimated log-evidence over 10
realisations of the data for each of the three uncorrelated priors
considered; it also lists the corresponding true evidences, estimated
using standard quadrature techniques. One sees that in each case the
log-evidence estimates obtained using BPR, after making the
correction in (\ref{eq:zeff}), are all consistent with the true value,
and have a standard deviation that remains stable for all the priors
considered.

\subsubsection{Correlated priors}

We now consider priors with fixed standard deviations
$\{\sigma_{\mtheta_1} = 4$, $\sigma_{\mtheta_2} = 4\}$, but three typical
correlation coefficients $\rho_{\mtheta} = \{-0.75,-0.25, 0.25\}$. The three prior distributions are
illustrated in Figure \ref{fig:2DillBoth}(b), together with the likelihood, from which one can see that all the priors are again unrepresentative.

\begin{figure}[!ht]
\centering
\includegraphics[width = 0.8\linewidth]{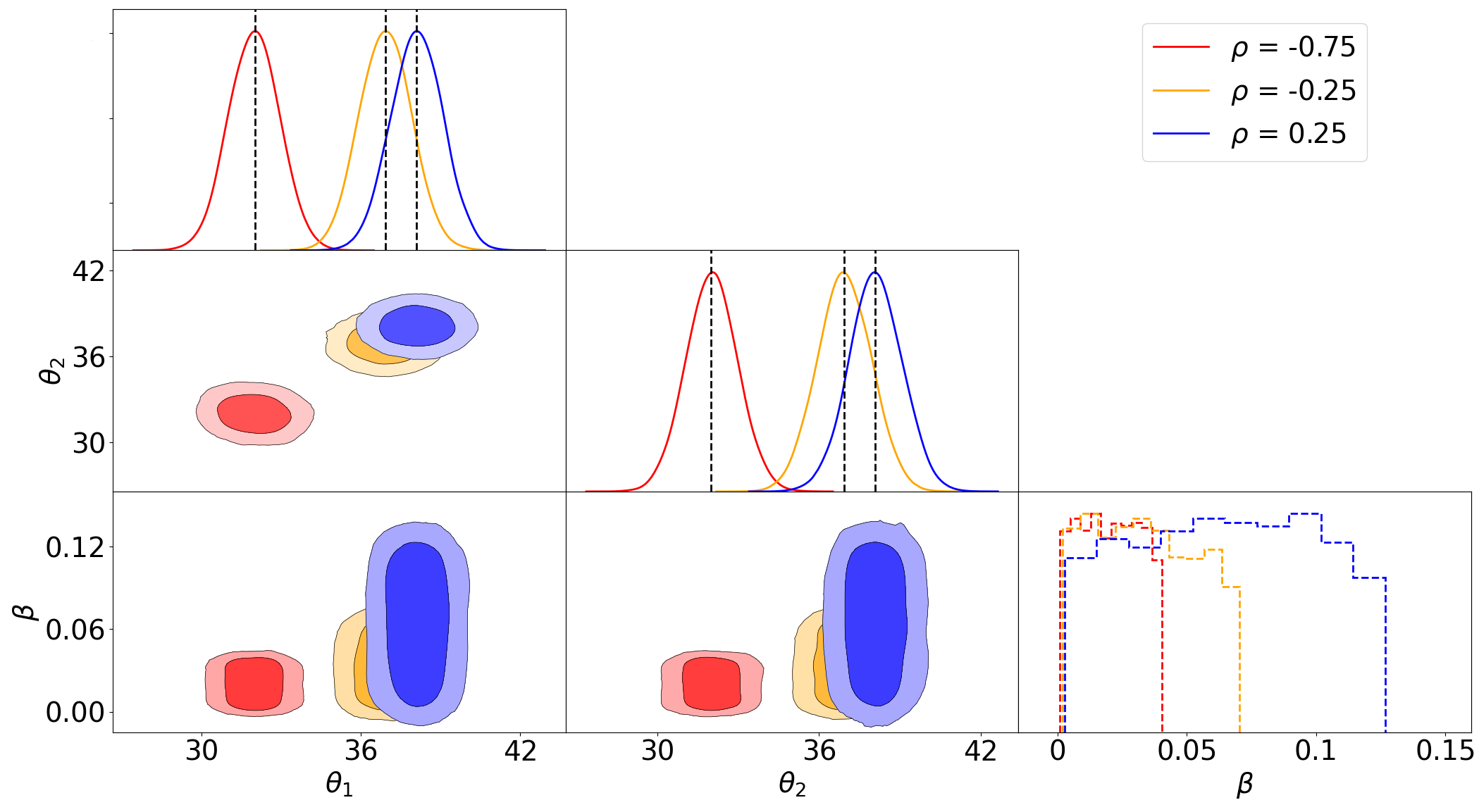}
\caption{A `corner plot' showing the 1-dimensional and 2-dimensional
  marginals of the joint posterior distribution on
  $(\theta_1,\theta_2,\beta)$ obtained by MultiNest using BPR for
  the bivariate Gaussian likelihood example with the likelihood and priors illustrated in
  Figure~\ref{fig:2DillBoth}(b). The vertical dashed black lines
  indicate the mean of the true Gaussian posterior.}
\label{fig:2DCorre}
\end{figure}

Figure~\ref{fig:2DCorre} is a `corner plot' showing the 1-dimensional
and 2-dimensional marginal distributions of the joint `effective'
posterior on $(\theta_1,\theta_2,\beta)$ for each of the three priors
described, using the same colour order as in
Figure~\ref{fig:2DillBoth}(b). As in previous examples, the joint
posterior in each case has the expected form of the product of two in
independent distributions on $(\theta_1,\theta_2)$ and $\beta$,
respectively, and the marginal on $\beta$ is consistent with an
approximately top-hat distribution in the range $[0,\beta_+]$. The
evolution of this marginal from $\rho_{\theta} = 0.25$ (blue) to
$\rho_{\theta} = -0.75$ (red) is as expected, with $\beta_+$ gradually
decreasing as the likelihood is concentrated further into the wings of
the prior distribution, as the latter `rotates' away from the peak of
the likelihood.  Moreover, since $\beta_+ \lesssim 0.15$ in every
cases, one may conclude that all the priors are indeed
unrepresentative and again use the $\beta_+$ values to rank
their severity.

The marginals on $\mtheta_1$ and $\theta_2$ in
Figure~\ref{fig:2DCorre} are again correctly centred on the
corresponding true Gaussian posterior for each case, the means of which
are indicated by the vertical black dashed lines. The widths of the
marginals are stable across the range of priors and consistent with
the width of the true Gaussian posterior in each case; the RMSE of the
$\theta_\ast$ estimate $\approx 0.06$ in all cases. Hence the
inferences on the parameters $(\mtheta_1,\theta_2)$ using BPR are again accurate and robust in each case, without any tuning.

\begin{table}
\caption{Comparison of the mean and standard deviation of the
  estimated log-evidence over 10 realisations of the data in the
  bivariate Gaussian likelihood example, obtained using the BPR
  method for a selection of priors with $\{\sigma_{\mtheta_1} = 4$,
  $\sigma_{\mtheta_2} = 4\}$ and correlation coefficient
  $\rho_{\mtheta}$. The `true' value of the evidence in each case is
  also given, as estimated using standard quadrature techniques.}
\smallskip
\begin{tabular}{c|rrc}
\hline
$\rho_{\mtheta}$ & True & BPR mean & BPR s.d.  \\
\hline
$-0.75$ & $-324.39$ & $-324.32$ & $1.27$ \\
$-0.25$&  $-127.72$ & $-127.65$ & $0.85$ \\
$0.25$&  $-80.83$ & $-80.75$ & $0.98$ \\
\hline
\end{tabular}
\label{tab:2DCorrEvid}
\end{table}

One may again verify that the BPR method yields accurate evidence
estimates in the above cases. Table~\ref{tab:2DCorrEvid} lists the
mean and standard deviation of the estimated log-evidence values over
10 realisations of the data for each of the three correlated priors
considered, and compares them with the corresponding true evidences
estimated using standard quadrature techniques. As for the
uncorrelated priors, the log-evidence estimates are all consistent
with the true value, and have a stable standard deviation across all
the priors considered.

\subsection{Bivariate multi-modal likelihoods}
\label{Sec: BiMulti}
\begin{figure*}
\centering
\subfigure[4-mode symmetric arrangement]{
\includegraphics[width = 0.45\linewidth]{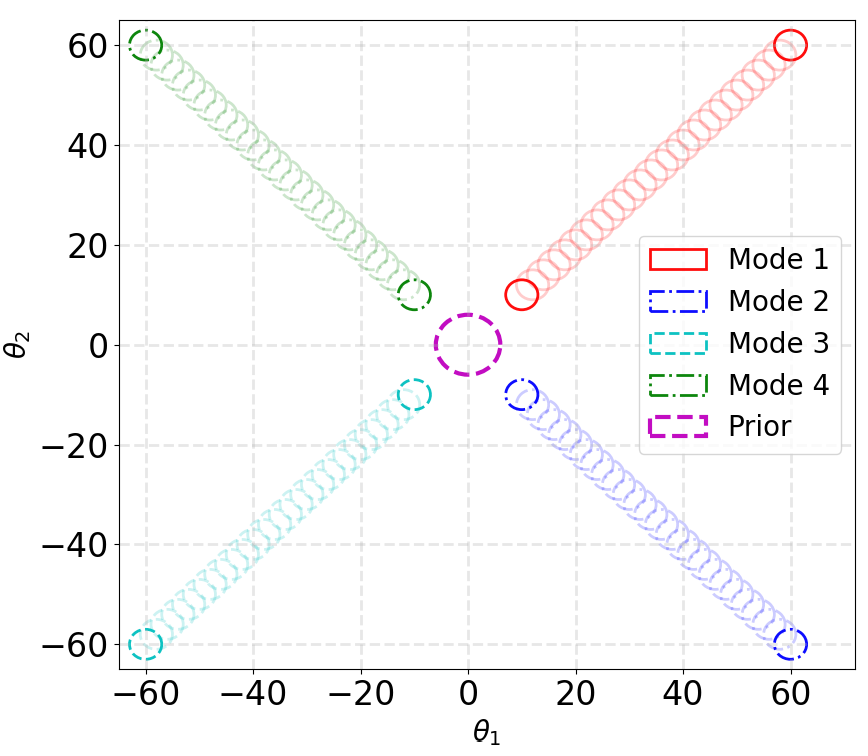}}\qquad
\subfigure[4-mode asymmetric arrangement]{
\includegraphics[width = 0.45\linewidth]{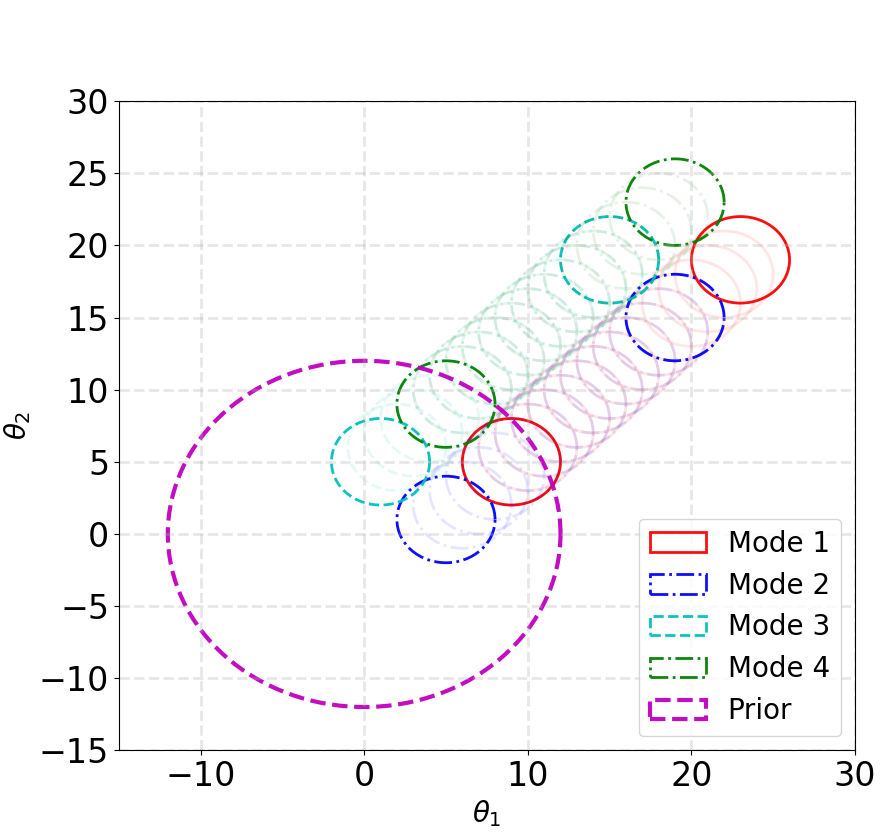}}
\caption{Illustration of the bivariate multimodal
    likelihoods considered, each of which consists of an equal mixture
    of four identical spatially-separated Gaussians for: (a) the
    symmetric arrangement; and (b) the asymmetric arrangement
    (relative to the prior). In each case, the dashed purple circle
    (centred at the origin) represents the 3-sigma iso-probability
    contour of the prior, and the other smaller circles represent the
    3-sigma iso-probability contour of each of the four modes.} 
\label{fig:4modesIllust}
\end{figure*}

We now consider bivariate multi-modal likelihoods consisting of an
equal mixture of four identical but spatially separated Gaussians.
Once again, for the sake of brevity, we consider only an uncorrelated
Gaussian prior centred on the origin, with $\sigma_{\theta_1}=4$ and
$\sigma_{\theta_2} = 4$. Nonetheless, we do consider two different
arrangements of the 4 modes of the likelihood relative to the centre
of the prior. In particular we consider the {\em symmetric} and {\em
  asymmetric} arrangements illustrated in Figure
\ref{fig:4modesIllust}, in which the circles denotes the 3-sigma
iso-probability contour of each mode of the likelihood and the prior,
respectively. As illustrated, for each arrangement, we consider a
range of likelihoods for which the centres of the modes lie at varying
distances from the origin. For the symmetric arrangement, each mode
lies at the same distance from the origin, with positions ranging from
$|\theta_1|=|\theta_2|=5$ to $|\theta_1|=|\theta_2|=60$. For the asymmetric
arrangement, as shown in Figure \ref{fig:4modesIllust}(b), the centres of the 4 modes are placed a distance of $4$ units from their geometric centre, the position of which ranges from $\theta_1=\theta_2=5$ to $19$. From Figure \ref{fig:4modesIllust}, one sees that for all the cases considered
in the symmetric arrangement the prior is unrepresentative, whereas
this is not true for some of the cases considered 
in the asymmetric arrangement. All
other settings are identical to those used in the analysis of the
bivariate Gaussian likelihood in Section~\ref{sec:bgl}. 

\subsubsection{Symmetric arrangement}
\begin{figure*}[!t]
\centering
\subfigure[$\log (\beta)$, varying mode distance]{
\includegraphics[width = 0.45\linewidth]{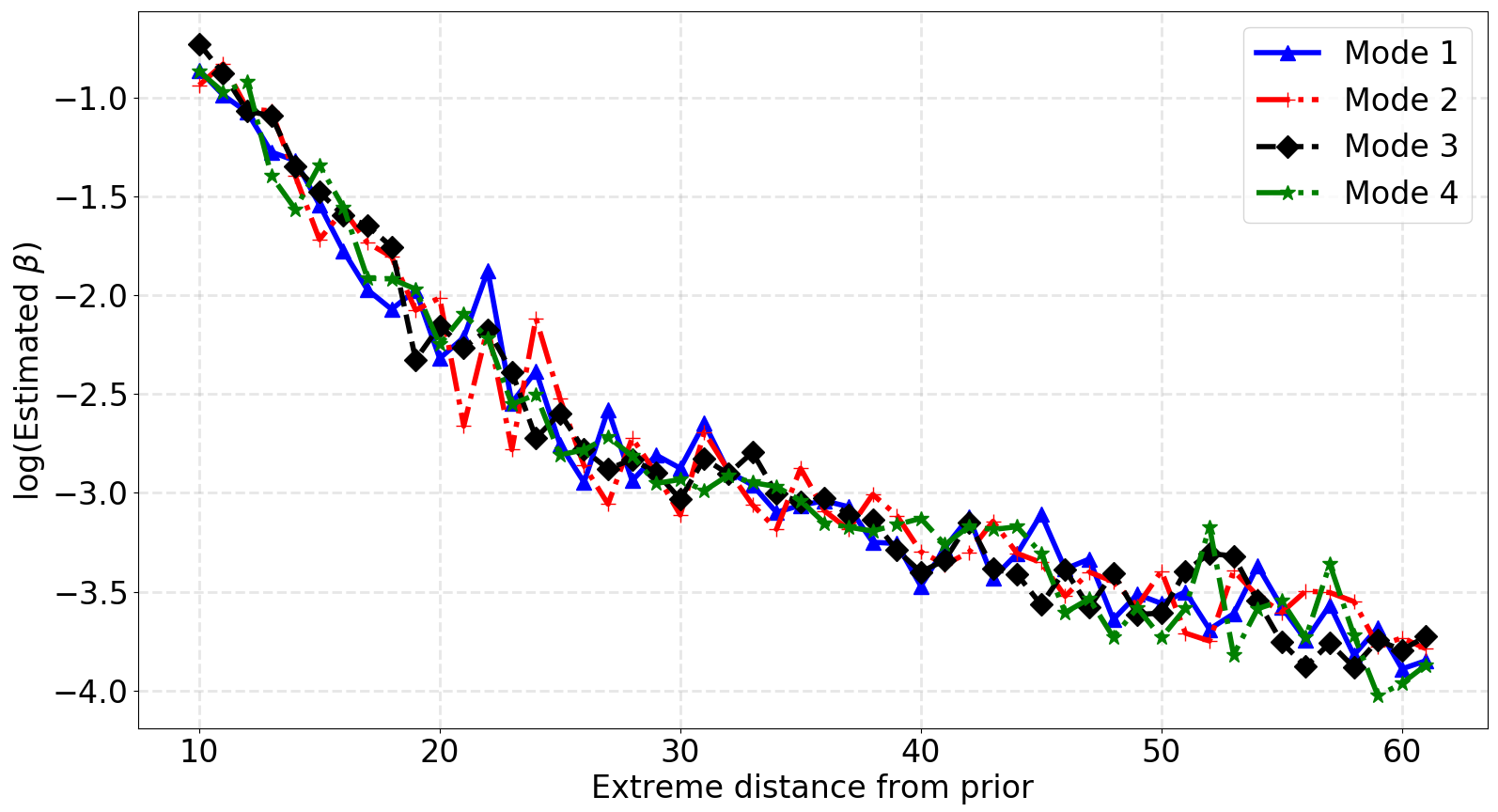}}
\subfigure[RMSE, varying mode distance]{
\includegraphics[width = 0.45\linewidth]{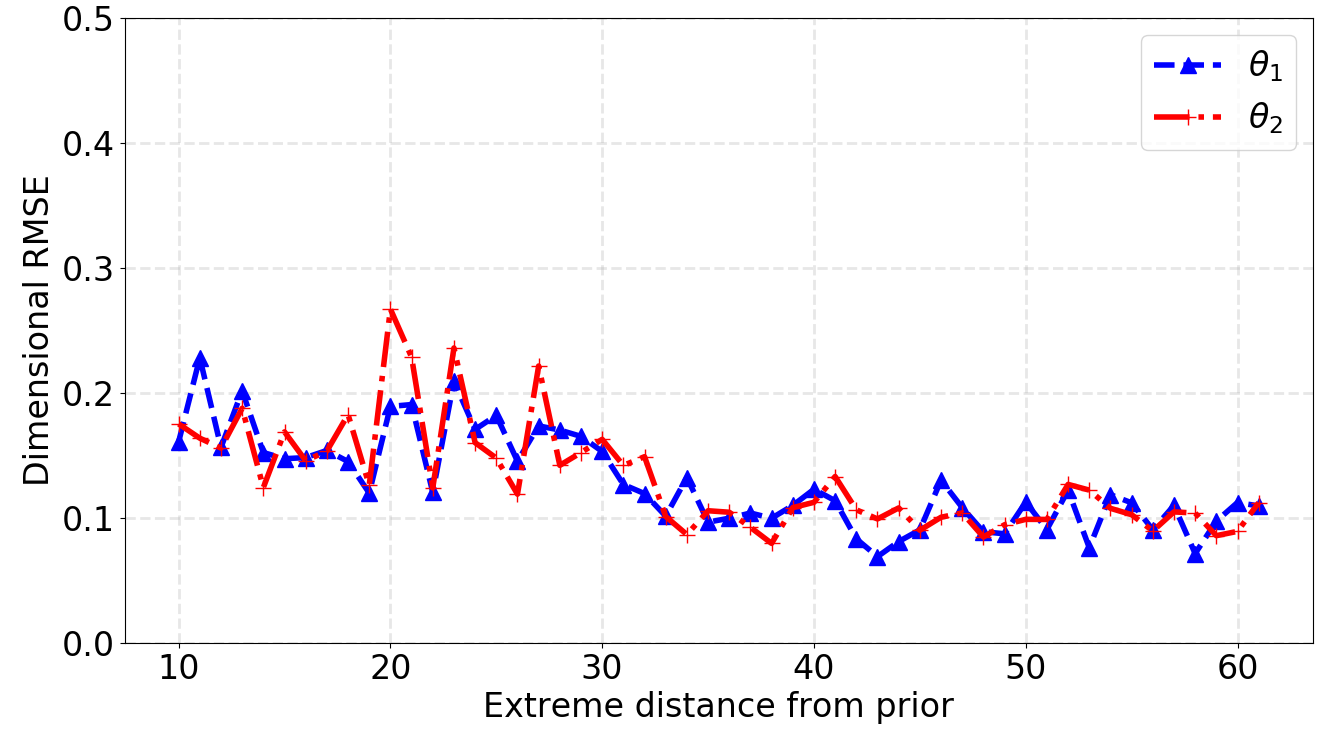}}
\\
\subfigure[$\log (\beta)$, varying $N_{\rm live}$]{
\includegraphics[width = 0.45\linewidth]{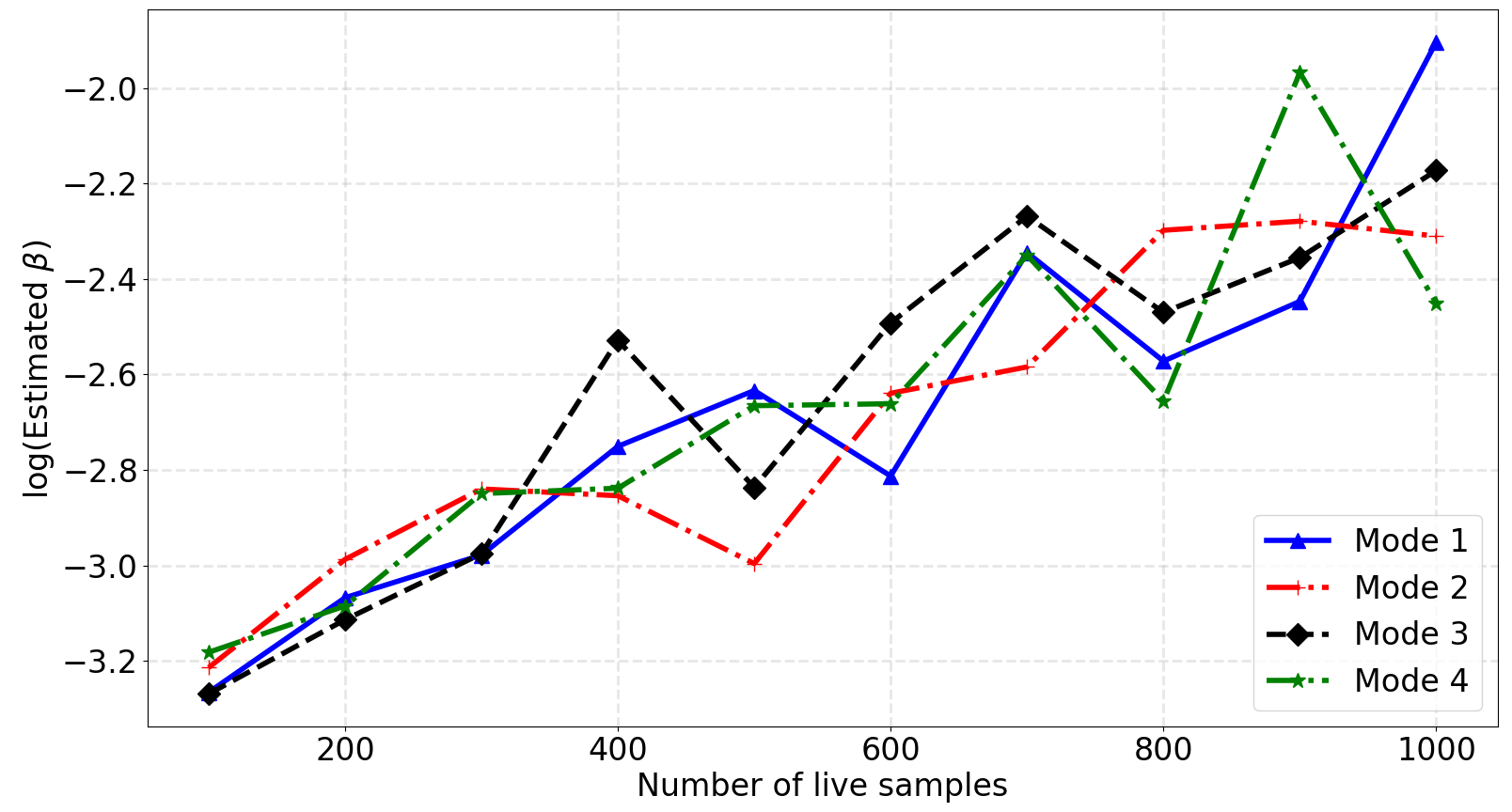}}
\subfigure[RMSE, varying $N_{\rm live}$]{
\includegraphics[width = 0.45\linewidth]{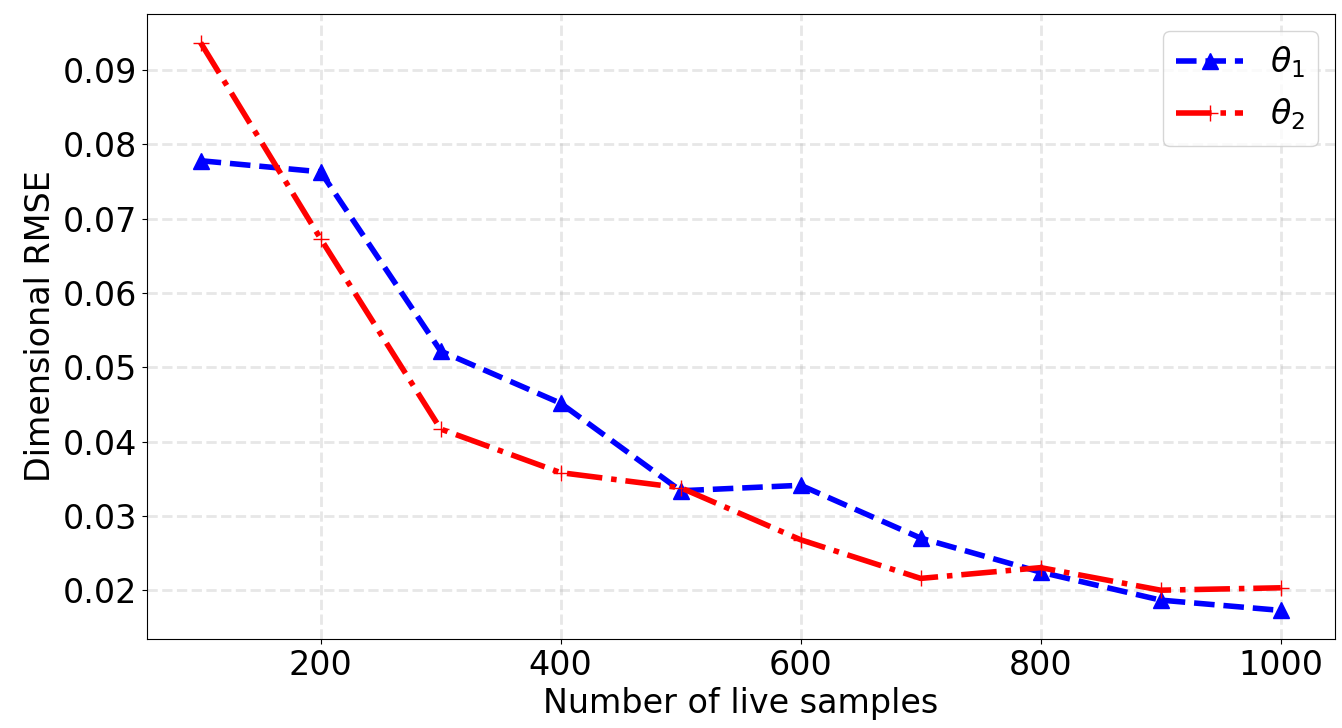}}
\caption{Performance of the BPR method applied to
      multimodal likelihoods consisting of an equal mixture of 4
      identical but spatially-separated Gaussians in the {\em symmetric
      arrangement} illustrated in Figure~\ref{fig:4modesIllust}(a).
      The upper row shows the effect of increasing
      the (equal) distance of the 4 mode centres from the origin for a
      fixed number of live points $N_{\rm live} = 100$. The lower row shows the effect of increasing the number of
      live points $N_{\rm live}$ for fixed modes centres at
      $|\theta_1| = |\theta_2| = 40$.  Panels (a) and (c) show
      $\log(\hat{\beta})$ obtained from posterior samples associated
      with each of the four modes, separately. Panels (b) and (d)
show the RMSE in the estimation of
  $\theta_1$ and $\theta_2$, averaged across the 4 modes. }
\label{fig:4modesGaussian}
\end{figure*}

Figure~\ref{fig:4modesGaussian} summarises the
  performance of the BPR method in the symmetric arrangements
  illustrated in Figure~\ref{fig:4modesIllust}(a). The upper row of
  panels (a)-(b) shows the effect of increasing the (equal) distance
  of the 4 mode centres from the origin for a fixed number of live
  points $N_{\rm live} = 100$. The lower row of panels (c)-(d) shows
  the effect of increasing the number of live points $N_{\rm live}$ for
  fixed modes centres at $|\theta_1| = |\theta_2| = 40$.

In the left-hand column, i.e. panels (a) and (c), we
  plot $\log(\hat{\beta})$ (to accommodate the large dynamic range in
  the estimates of $\hat{\beta}$) obtained from the mean of the posterior samples
  associated with each of the four modes, separately. In panel (a),
  the value of $\log(\hat{\beta})$ is consistent across the four
  modes, as one might expect since they are symmetrically arranged
  relative to the centre of the prior located at the origin. Moreover,
  as the (equal) distance of each mode from the origin increases, then
  the value of $\log(\hat{\beta})$ decreases monotonically (to within
  the statistical uncertainties of the nested sampling process). In
  panel (c), one again sees consistency on the values of $\hat{\beta}$
  across the four modes, but that $\log(\hat{\beta})$ increases
  quasi-monotonically as $N_{\rm live}$ increases. This is expected
  since the failure of the nested sampling algorithm, and the
  consequent need for the posterior repartitioning characterised by
  low values of $\beta$, is reduced as $N_{\rm live}$ increases.

In the right-hand column of Figure~\ref{fig:4modesGaussian}, i.e. panels
(b) and (d), we plot the RMSE in the estimation of $\theta_1$ and
$\theta_2$ for each mode, but averaged across the 4 modes. In panel
(b), one sees that the RMSE for $\theta_1$ and $\theta_2$ are
consistent, as expected, and that the values remain stable in the
range $\sim$ 0.1--0.2 as the (equal) distance of the 4 modes from the
origin varies. Panel (d) shows that the RMSE for $\theta_1$ and
$\theta_2$ are again consistent and that these values decrease
monotonically as $N_{\rm live}$ increases, as expected.



\subsubsection{Asymmetric arrangement}

\begin{figure*}[!ht]
\centering

\subfigure[$\log (\beta)$, varying mode distance]{
\includegraphics[width = 0.47\linewidth]{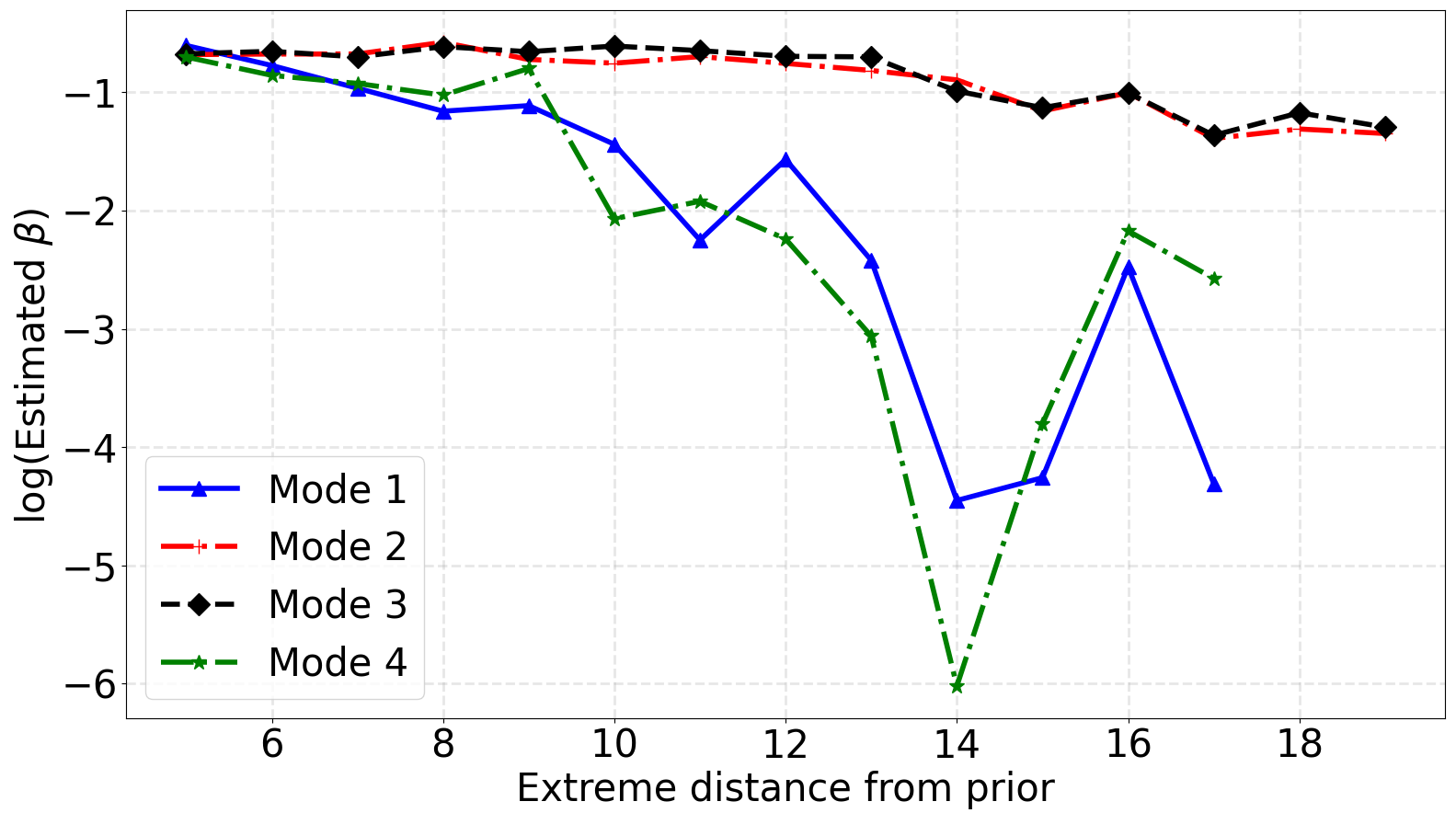}}
\subfigure[$\log (\beta)$, varying $N_{\rm live}$]{
\includegraphics[width = 0.49\linewidth]{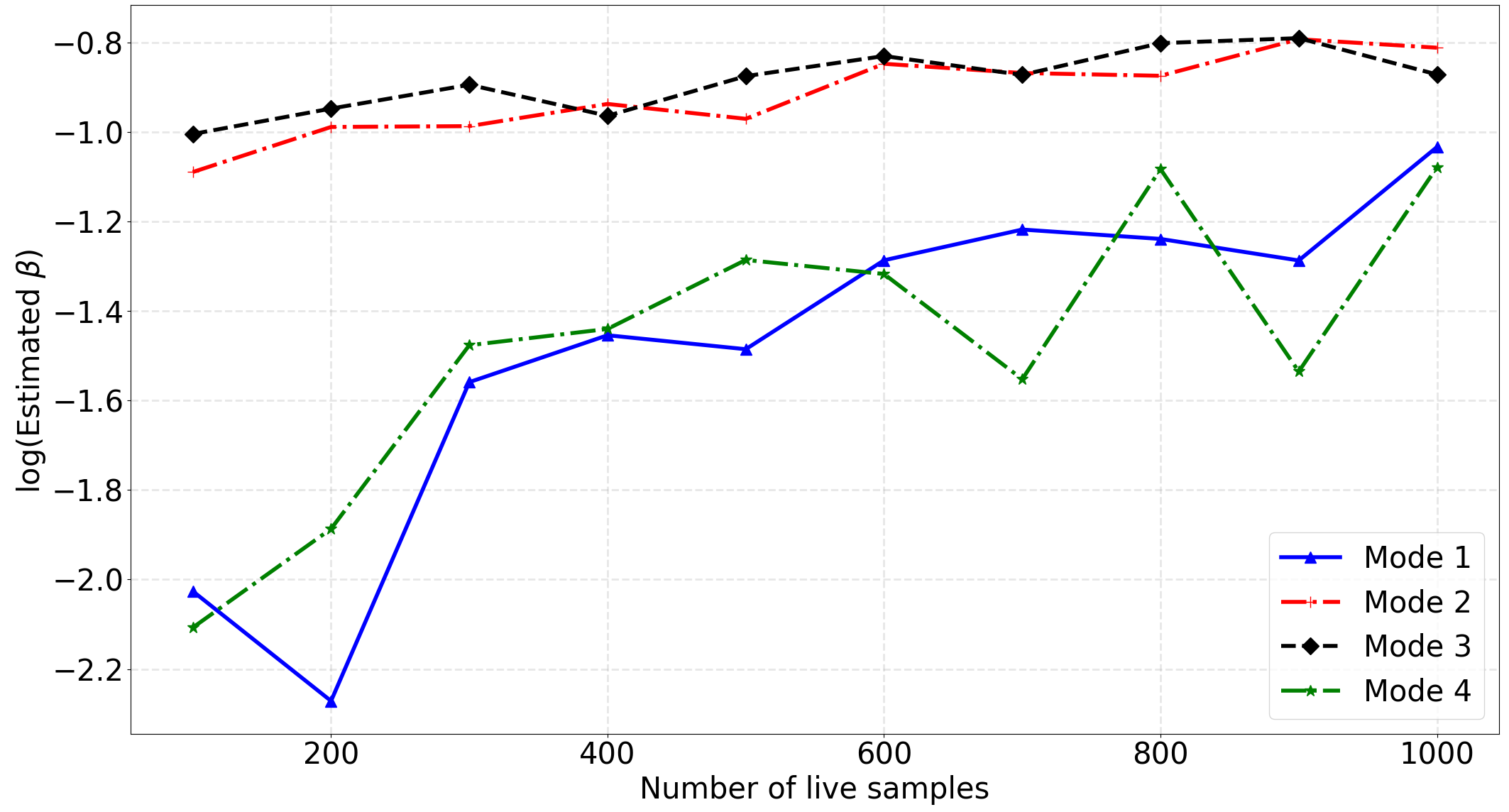}}\\
\subfigure[RMSE, varying mode distance]{
\includegraphics[width = 0.47\linewidth]{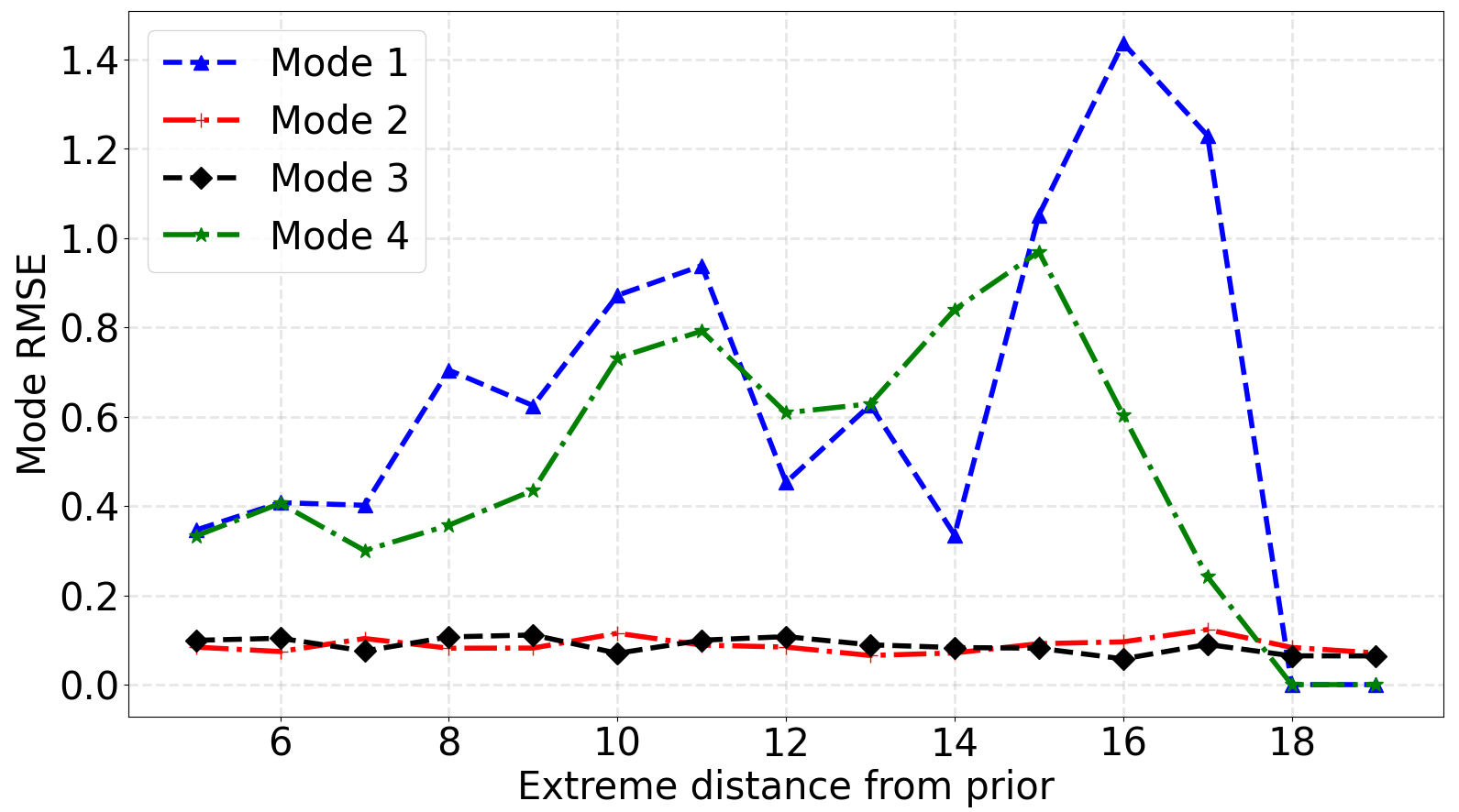}}
\subfigure[RMSE, varying $N_{\rm live}$]{
\includegraphics[width = 0.49\linewidth]{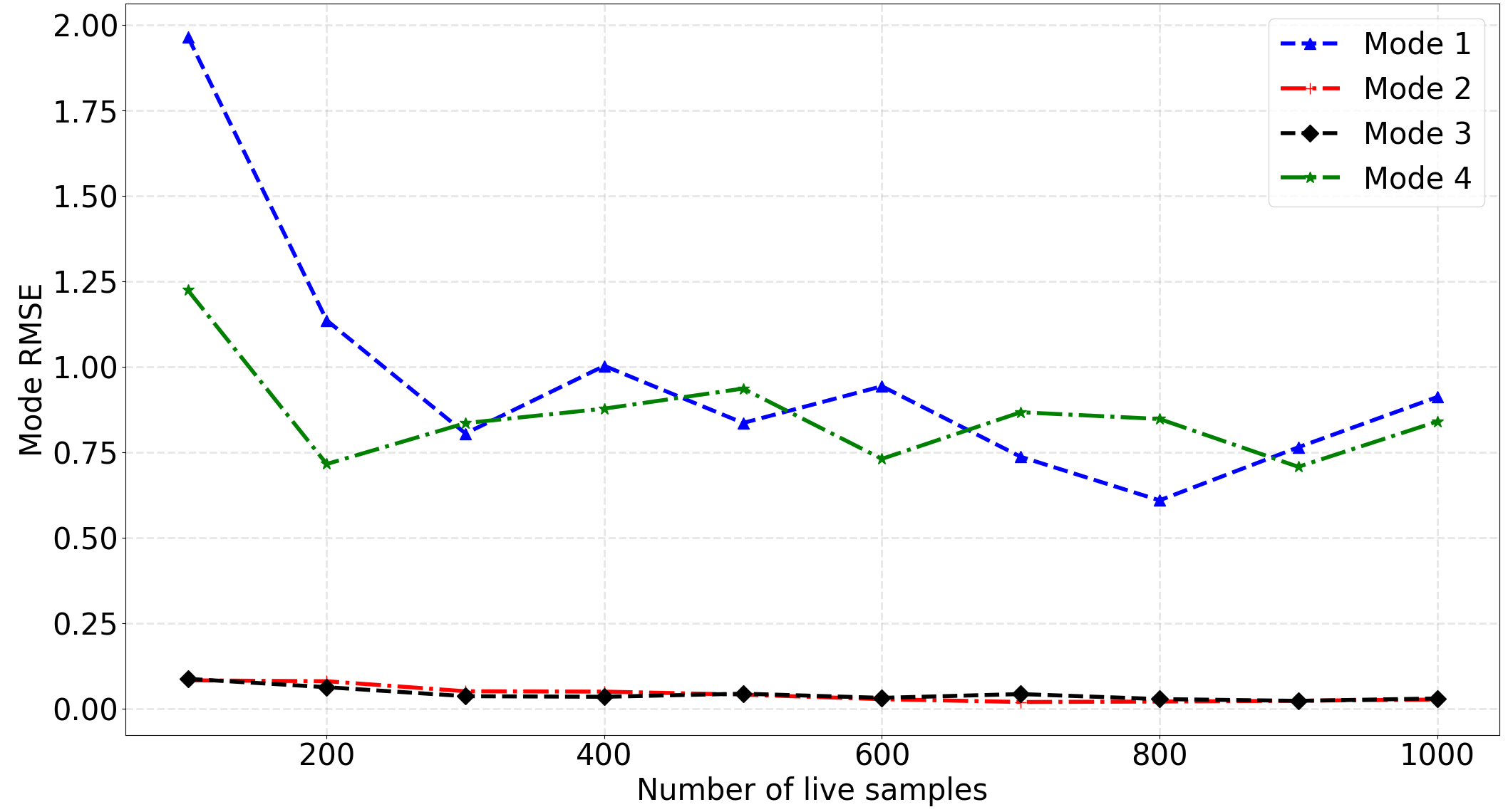}}
\caption{{As for Fig~\ref{fig:4modesGaussian}, but for the {\em
      asymmetric arrangement} illustrated in
    Figure~\ref{fig:4modesIllust}(b). In this case, the panels (a) and (c) show the effects as a function of the distance of
    the geometric centre of the 4 modes from the origin for $N_{\rm
      live}=100$, and the panels (b) and (d) show the effects of increasing $N_{\rm live}$ for fixed mode centres, of which the geometric centre is $15$.
    Also, the panels (c) and (d) show the average RMSE in the
    estimation of $\theta_1$ and $\theta_2$ for each mode separately.}}
\label{fig:4modesGaussianNonSym}
\end{figure*}

Figure~\ref{fig:4modesGaussianNonSym} summarises the performance of
the BPR method in the asymmetric arrangements illustrated in
Figure~\ref{fig:4modesIllust}(b), and the quantities displayed in each
panel are the same as those in Figure~\ref{fig:4modesGaussian}. The
only differences here are that the panels (a) and (c) show
the effects as a function of the distance of the geometric centre of the
4 modes from the origin, and that panels (b) and (d) show the RMSE for
each mode separately. \revise{Further heatmap illustrations for this example can be found in Figure~\ref{fig:4modesGaussianNonSymHeatmap} in the Supplementary Material.} This asymmetric modes scenario presents quite an extreme test of the BPR method, but equally illustrates some
further key advantages over our original PR method presented in
\cite{chen2018improving}.

\noindent \revise{\textbf{A unique feature of BPR in asymmetric multi-modal scenarios:}} In the upper row of panels (a) and (b), one sees that for each mode the behaviour of $\log(\hat{\beta})$ as a function of
distance from the origin and $N_{\rm live}$, respectively, is similar
to that found for the symmetric arrangement, and for the same reasons
as discussed above. By contrast, however, for the asymmetric arrangement
the values of $\log(\hat{\beta})$ are not the same across the 4 modes
of the posterior. Instead, the $\log(\hat{\beta})$ values are very
similar for modes 2 and 3, and broadly consistent for modes 1 and 4,
with the values for the latter pair of modes being much smaller than
those for the former pair. By inspecting
Figure~\ref{fig:4modesIllust}(b), one sees that this is to be
expected, since modes 2 and 3 lie closer (but equidistant) from the
centre of the prior at the origin, whereas modes 1 and 4 lie further
away (but again equidistant). Thus, the prior is more unrepresentative
for modes 1 and 4, which thus require a lower value of
$\log(\hat{\beta})$. 

This demonstrates an important feature of the
BPR method, which is not shared by the original PR method, in that
different regions of the posterior can be characterised by {\em
  different} ranges of $\beta$ values, depending on how far into the
wings of the prior the corresponding regions of likelihood lie, and
this is accommodated in a fully automated manner. From panel (a), it
is also worth noting that the difference between $\log(\hat{\beta})$
values for modes $(2,3)$ and modes $(1,4)$ is maximal for intermediate
distances from the origin. This again makes sense as it corresponds to
the point at which there is the largest difference between how
representative the prior is for these two sets of modes, since the
prior is changing most rapidly between them: for smaller distances
(near the peak of the prior) or larger distances (in the wings of the
prior), there is less change in the prior between the two sets of
modes.

In panels (c) and (d) of Figure~\ref{fig:4modesGaussianNonSym}, one
first sees from panel (c) that the average RMSE in the estimation of
$\theta_1$ and $\theta_2$, is low and insensitive to the distance from
the origin for modes 2 and 3, which are closer the origin.  By
contrast, the RMSE is large and more volatile for modes 1 and 4, which
lie further into the wings of the prior, and rises slowly with
distance from the origin. One sees, however, that the RMSE drops to
zero at very large distances;
this occurs because, for the default setting of $N_{\rm live}=100$,
one obtains no samples in modes 1 and 4. This therefore defines the
limit of the applicability of the BPR method in this example,
although the issue is resolved by increasing $N_{\rm live}$. In panel
(d), one sees that the RMSE is low and insensitive to $N_{\rm live}$
for modes 2 and 3, whereas modes 1 and 4 exhibit a larger and more
volatile RMSE, albeit decreasing rapidly up to $N_{\rm live} \approx
300$ and being relatively insensitive to $N_{\rm live}$ thereafter.

\subsection{Higher-dimensional Gaussian likelihoods}

Finally, we extend the bivariate Gaussian likelihood example in
Section~\ref{sec:bgl} to higher dimensions from 3D
to 10D.  We begin by adopting an analogous likelihood, corresponding
to $N=1$ data points and centered on $\mtheta_{\ast} = 40$ in all
dimensions, but restrict our analysis to the case of a single
circular-symmetric Gaussian prior, with $\sigma_{\mtheta} = 4$ in all
dimensions. Once again, we consider only the BPR results as the
standard NS method fails completely in this case. All results
presented are based on 10 realisations of the data.

\begin{figure*}[!ht]
\centering
\subfigure[Estimate of $\mbeta_+$]{
\includegraphics[width = 0.46\linewidth]{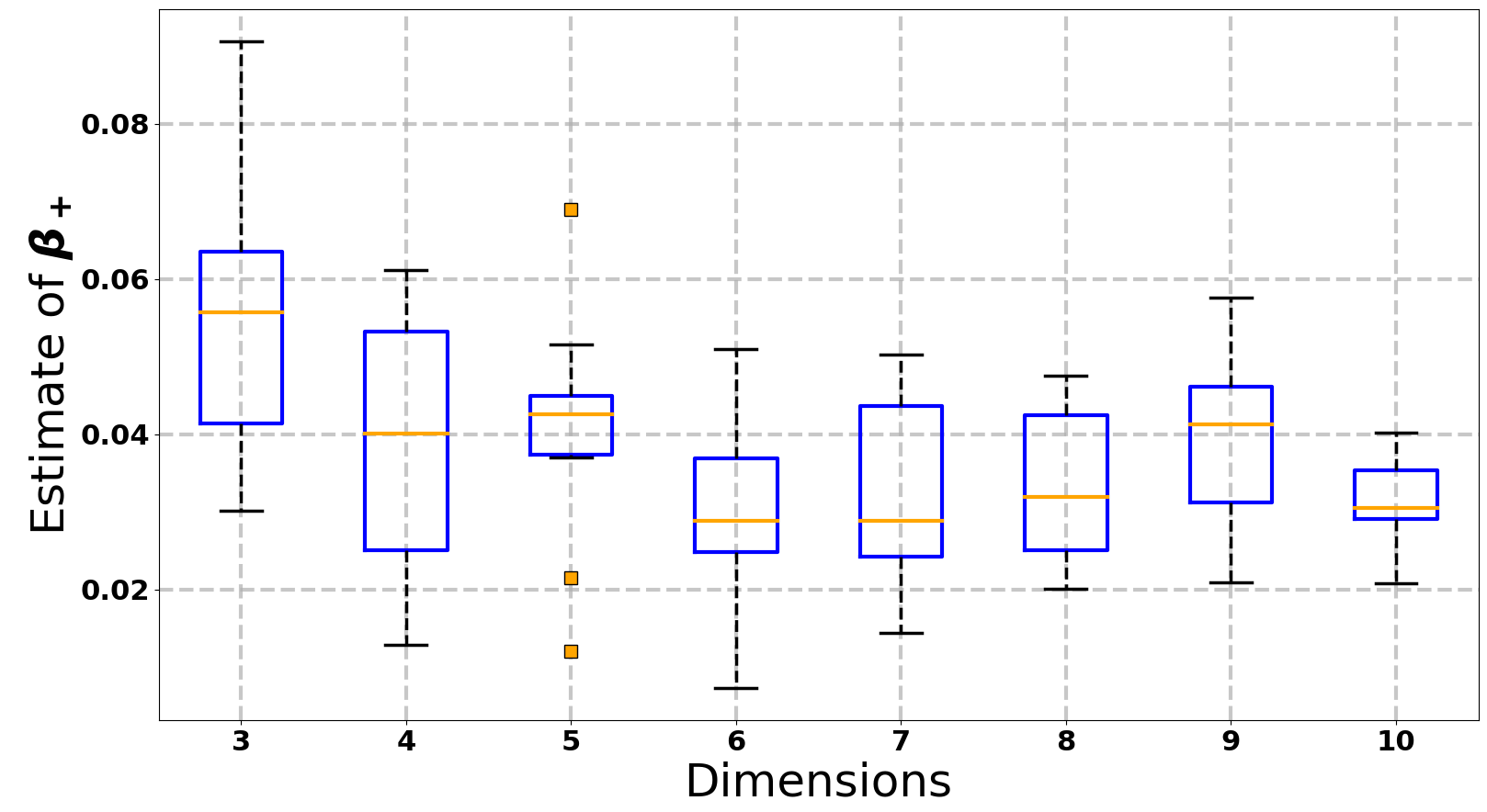}}
\subfigure[RMSE of $\mtheta$ estimate]{
\includegraphics[width = 0.46\linewidth]{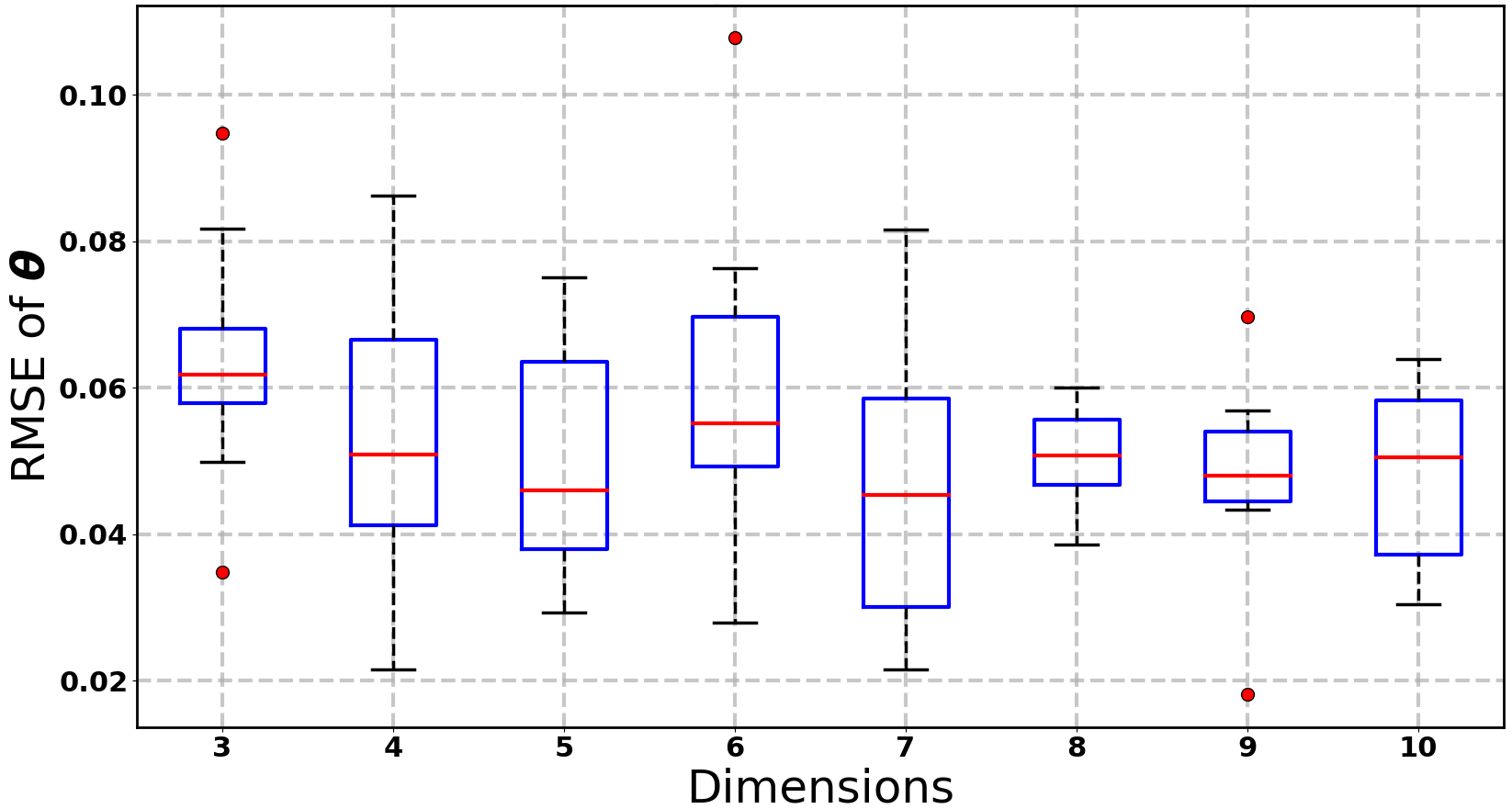}}\\
\subfigure[Error of log-evidence estimates]{
\includegraphics[width = 0.46\linewidth]{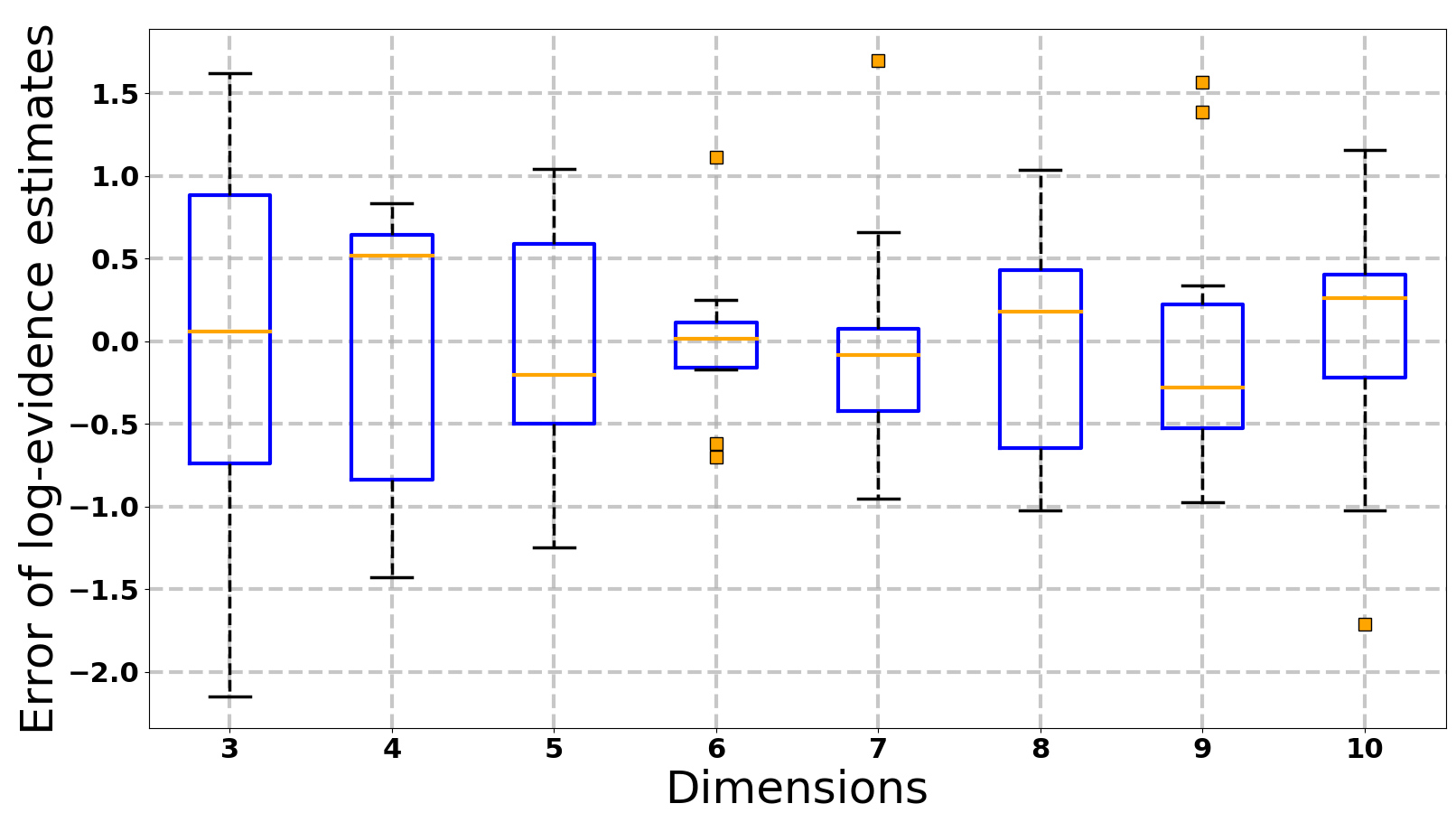}}
\subfigure[Number of likelihood evaluations]{
\includegraphics[width = 0.46\linewidth]{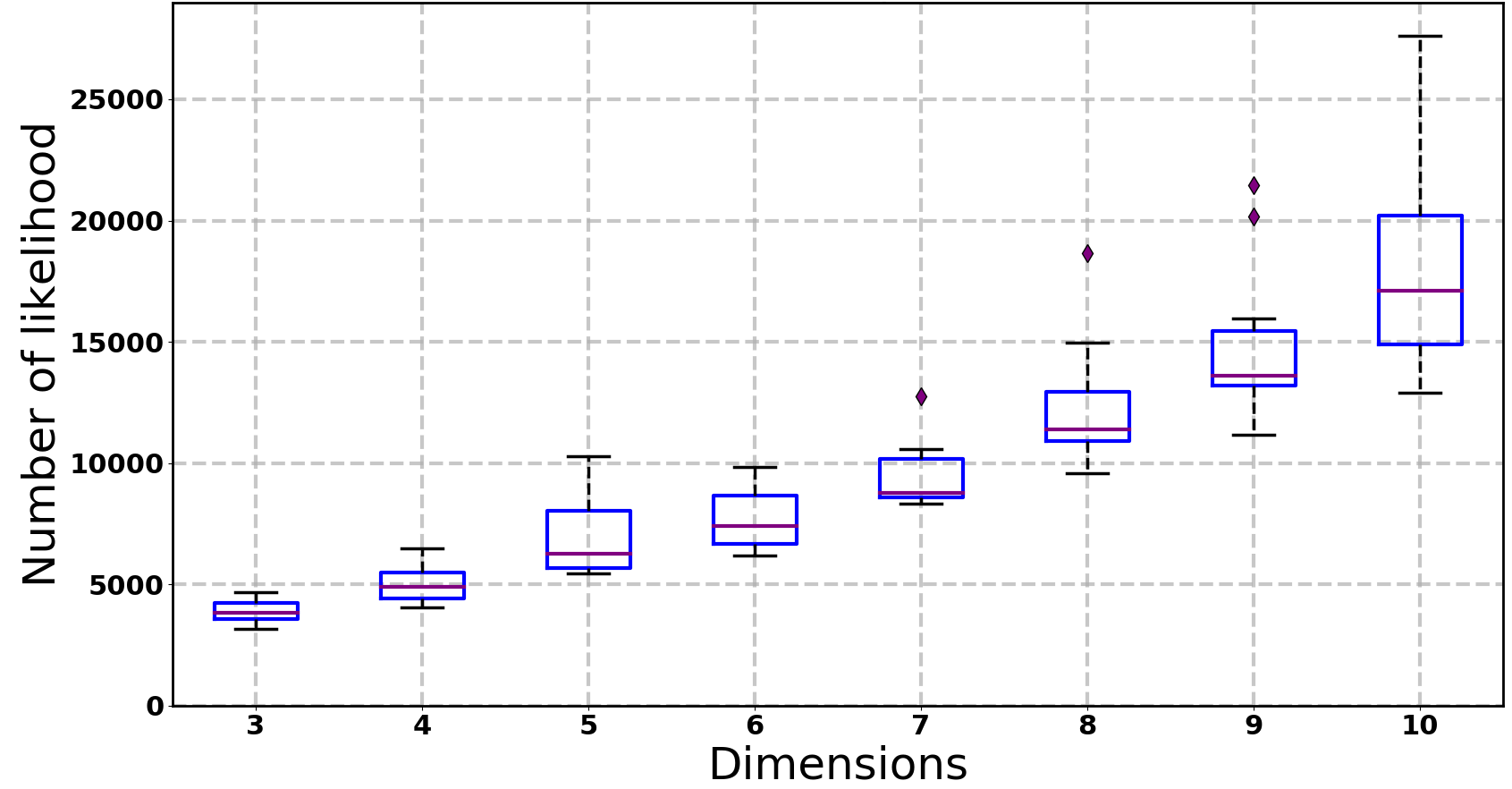}}
\caption{\revise{Boxplots for estimated parameter values
    using the BPR method applied to the higher-dimensional examples with
  $\mtheta_{\ast} = 40$, from 3D to 10D. The error bars denote the minimum and
  maximum values, whereas the boxes indicate the 25th to 75th
  quantiles. The orange/red line in the box represents the median value over 10 realisations of the data and the points
  denote outliers.}}
\label{fig:highD40}
\end{figure*}

We first consider the constraints obtained on the parameters
$(\mTheta,\beta)$. Rather than presenting corner plots of the
posterior in these higher dimensionalities, Figure~\ref{fig:highD40}(a)-(b)
instead shows boxplots of the estimated value of $\beta_+$ (the 99\%
point of the unnormalised marginal posterior $\tilde{P}(\beta)$, as
discussed in Section~\ref{Sec:BPRmethod}) and the RMSE of the
estimate of $\theta_\ast$ (averaged across all dimensions), as a
function of the dimensionality of the problem. The estimated $\beta_+$
values are broadly consistent across different dimensionalities, with
a mean of $\approx 0.05$ and standard deviation $\approx 0.01$, with
only a slight trend to smaller values as the dimensionality
increases. This insensitivity to dimensionality is a result of the
(effective) joint posterior having the product form
(\ref{eq:effpost}). Similarly, the RMSE of the $\theta_*$ estimate is
also seen to be broadly stable across the range of dimensionalities,
with a mean value $\approx 0.05$, which is consistent with the
accuracy obtained in the bivariate example, and standard deviation
$\approx 0.015$.

Turning to the accuracy of the evidence estimates,
Figure~\ref{fig:highD40}(c) shows a boxplot of the error in the
log-evidence, over 10 realisations of the data, as a function of
dimensionality. Once again, one sees that the distributions are
relatively stable across different dimensionalities, with errors
typically lying within one log-unit, which is consistent with the
accuracies achieved in the bivariate example.

Finally, we consider the number of likelihood evaluations $N_{\rm
  like}$ required for MultiNest to converge. Figure
\ref{fig:highD40}(d) shows a boxplot of $N_{\rm like}$ over 10
realisations of the data, as a function of dimensionality. The mean
value of $N_{\rm like}$ rises from $\sim 3800$ at 3D to $\sim 18100$
at 10D. This trend is consistent with that reported using the original
PR method in \cite{chen2018improving}, and follows roughly an ${\cal O}
(n \log n)$ increase with dimensionality. The range of
$N_{\rm like}$ values over the 10 realisations is also
seen to widen slightly as the number of dimensions increases, but this
is a relatively minor effect, at least up to 10D.


\section{Conclusions}
\label{Sec:Conclu}

We have demonstrated that one may straightforwardly automate our
previous prior repartitioning method for improving the robustness and
efficiency of nested sampling in the presence of an unrepresentative
prior. This is achieved by taking an explicitly Bayesian approach that
treats the auxiliary parameter $\beta$ in the power PR approach as a
hyper-parameter that is estimated alongside the parameters of interest
$\mTheta$ of the problem under consideration. Since this estimation
process is performed within a single run of the nested sampling
algorithm, the BPR method provides a substantial reduction in the
computational requirements relative to the annealing schedule approach
adopted in the original PR method. In addition to retaining all the
advantages of the original scheme in providing reliable parameter
constraints and evidence estimates, the BPR method adapts
automatically to each problem and thus requires no tuning
whatsoever. Indeed, by treating $\beta$ as a hyper-parameter, one may
use its resulting marginal to determine both the presence and the
severity of an unrepresentative prior. We illustrate these properties
in a range of numerical examples, from 1D to 10D, some exhibiting
multi-modal likelihoods, with a selection of unrepresentative priors
with varying degrees of severity. In particular, for multi-modal
likelihoods, we show that different regions of the posterior may be
characterised by different values of $\beta$, depending on how far
into the wings of the prior the corresponding regions of likelihood
lie. Moreover, this is accommodated in a fully automated manner and is
an important feature of the BPR method that is not shared by our
original PR method. 

Perhaps most interestingly, we show that there is negligible
computational overhead relative to standard nested sampling when the
BPR method is used in cases where the prior is representative, and
that it offers significant advantages in terms of efficiency and
robustness even if the prior is only marginally unrepresentative. This
suggests that the BPR should always be used in the application of
nested sampling to Bayesian inference problems, irrespective of
whether one suspects that the assumed priors may be unrepresentative
for some data sets. Moreover, BPR might usefully be
integrated directly into nested sampling algorithms and/or data
analysis pipelines. \revise{It should be recalled, however, that the computational cost of BPR may be considerably increased if the normalising constant of the modified prior cannot be calculated analytically and so must be estimated numerically.}

\begin{acknowledgement}
The authors thank Will Handley and Lukas Hergt for their
support with the powerful Python visualisation package \pkg{anesthetic}
\citep{anesthetic}.
\end{acknowledgement}

\bibliographystyle{ba}      
\bibliography{bibFile}   

\newpage
\section{Supplementary Material}
\subsection{Nested sampling algorithm}
\label{AP: NS}

Algorithm \ref{alg:NS} describes the pseudo-code for the standard NS algorithm. $X_i$ denotes the expected fraction of the
prior volume lying within the isolikelihood contour $\mL(\mTheta) =
\mL_i$ at iteration $i$. In general, the fractional prior volume $X$
lying above some likelihood threshold $\lambda$ is defined as
\begin{align}
X(\lambda) = \int_{\mL(\mTheta)> \lambda} \mpi(\mTheta) d \mTheta,
\label{eq:priorvoldef}
\end{align}
where $\lambda$ gradually rises from the minimum to the maximum of
$\mL(\mTheta)$ as the NS iterations proceed. The NS algorithm provides
an estimate of the evidence \eqref{Eq:origZ} by casting it as a
one-dimensional integral with respect to fractional prior volume $X$:
\begin{align}
\mZ = \int_{0}^{1} \mL(X) d X,
\label{Eq:evidence}
\end{align}
subject to the existence of the inverse $\mL(X)$ 
to (\ref{eq:priorvoldef}).
Finally, {\tt{tol}} denotes a pre-defined
convergence criterion, which affects the number of NS iterations. 

\begin{algorithm}[!ht]
\tcp{Nested sampling initialisation}
At iteration $i=0$:\\ 
Draw $ N_{\rm live}$ samples $\{\mTheta_n\}_{n=1}^{N_{\rm live}}$ from
prior $\pi(\mTheta)$. \\
Initialise evidence $Z=0$ and prior volume $X_0 = 1$. \\
\tcp{NS iterations}
\For{$i=1, 2, \cdots, I$}{
    \For{$n=1, 2, \cdots, N_{\rm live}$}{
        	$\bullet$ Compute likelihood $\mL(\mTheta_{n})$.\\
        }
	$\bullet$ Find the lowest likelihood and save in $\mL_i$. \\
	$\bullet$ Calculate weight $w_i = \frac{1}{2}(X_{i-1} -
    X_{i+1})$, where $X_i = \exp(-i/N_{\rm live})$ \\
    $\bullet$ Increment $Z$ by $\mL_i w_i$. \\
	$\bullet$ Draw a new sample $\mTheta$ to replace one with
    likelihood $\mL_i$, subject to $\mL(\mTheta) > \mL_i$.\\ 
    $\bullet$ Stop if $\mbox{max}\{\mL(\mTheta_n)\}X_i <
    {\tt{tol}}\times Z$. 
	}
Increment $Z$ by $\sum_{n=1}^{N_{\rm live}}\mL(\mTheta_n)X_{I}  / N_{\rm live}$. \\
Assign the sample replaced at iteration $i$ the importance weight $p_i=L_iw_i/Z$.\\
\caption{Nested sampling algorithm}
\label{alg:NS}
\end{algorithm}

\subsection{A BPR example with univariate Gaussian likelihood}
\label{AP: Example}

In this section we further illustrate BPR using the same settings of the univariate Gaussian likelihood in Section 4.1. 

In practice, BPR can be easily obtained by modifying the likelihood function (for each example) in standard MultiNest. For the 1D Gaussian example, we have to set the hyper-parameter `nDim' in MultiNest to 2, where the first variable is $\theta$ and the second one is $\beta$ to be estimated.  

$\pi(\beta)$ is an uniform distribution $\mbeta \sim \calU[0,1]$, and the original prior $\pi(\mtheta)$ is a standard Gaussian:
\begin{equation}
\pi(\mtheta) = (2 \pi \sigma_{\mpi}^2)^{\frac{1}{2}} \exp \left( -\frac{(\mtheta - \mu_{\mpi})^2}{2 \sigma_{\mpi}^2} \right)
\end{equation}

\begin{algorithm}[!h]
\tcp{Initialisation}
At iteration $i=0$:\\ 
Initialise evidence $Z=0$ and prior volume $X_0 = 1$. \\
Draw $ N_{\rm live}$ samples $\{\beta_n\}_{n=1}^{N_{\rm live}}$ from
prior $\pi(\beta)$. \\
Draw $\{\mTheta_n\}_{n=1}^{N_{\rm live}}$ samples from prior $\tilde{\pi}(\mTheta | \beta_n)$, and compute the constant $C$. \\

\tcp{BPR iterations}
\For{$i=1, 2, \cdots, I$}{
    \For{$n=1, 2, \cdots, N_{\rm live}$}{
        	$\bullet$ Compute likelihood $\tmL(\mTheta_n, \beta_n)$.\\
        }
	$\bullet$ Find the lowest likelihood and save in $\mL_i$. \\
	$\bullet$ Calculate weight $w_i = \frac{1}{2}(X_{i-1} -
    X_{i+1})$, where $X_i = \exp(-i/N_{\rm live})$ \\
    $\bullet$ Increment $Z$ by $\mL_i w_i$. \\
	$\bullet$ Draw a new sample $\mTheta$ to replace one with
    likelihood $\mL_i$, subject to $\mL(\mTheta) > \mL_i$.\\ 
    $\bullet$ Stop if $\mbox{max}\{\mL(\mTheta_n)\}X_i <
    {\tt{tol}}\times Z$. 
	}
Increment $Z$ by $\sum_{n=1}^{N_{\rm live}}\mL(\mTheta_n)X_{I}  / N_{\rm live}$. \\
Assign the sample replaced at iteration $i$ the importance weight $p_i=L_iw_i/Z$.\\
\caption{BPR 1D Gaussian example}
\label{alg:BPR}
\end{algorithm}

The modified prior and likelihood can then be derived:
\begin{eqnarray}
\tmpi(\mTheta | \beta) &=& \frac{\mpi(\mTheta)^{\mbeta}}{\mZ_\pi(\beta)}, \\
\tmL(\mTheta, \beta) &=& \mL(\mTheta) \mpi(\mTheta)^{(1-\mbeta)}\mZ_\pi(\beta),  \label{eq:moLike}
\end{eqnarray} 
We can easily obtain an unnormalised powered distribution:
\begin{align}
    \mpi(\mTheta)^{\mbeta} = (2\pi)^{\frac{\beta}{2}}  \sigma_{\mpi}^{\beta} \exp \left( -\frac{(\mtheta - \mu_{\mpi})^2}{2 (\frac{\sigma_{\mpi}}{\sqrt{\beta}})^2} \right),
\end{align}
which is equivalent to designing a new Gaussian distribution $\calN(\mu_{\mpi}, (\frac{\sigma_{\mpi}}{\sqrt{\beta}})^2)$ by multiplying a constant $C = (2 \pi)^{\frac{1-\beta}{2}} \sigma_{\mpi}^{1-\beta} \beta^{- \frac{1}{2}}$. $\mZ_\pi(\beta)$ is the normalising constant and can be easily computed by finite summation from drawn samples. Therefore, the newly designed Gaussian distribution forms the modified prior $\tmpi(\mTheta | \beta)$.  

The pseudo code is shown in Algorithm \ref{alg:BPR}. In each iteration for the $n$th live point, BPR will draw a $\beta_n$, and convert sample $\mtheta_n$ (which is originally drawn from the unit hypercube) to the modified Gaussian $\tmpi(\mTheta | \beta)$. $\mpi(\mTheta)^{(1-\mbeta)}$ can also be easily computed following the same rule and finally the modified log-likelihood $\log \tmL(\mTheta, \beta)$ can be computed by summing up the logarithm of the terms in Eqn. \eqref{eq:moLike}.

\subsection{More results for the univariate Gaussian likelihood example}
\label{AP: UG}

\begin{figure*}[!h]
\centering
\subfigure[Log$_{10}$(RMSE) of $\hat{\mtheta}$]{
\includegraphics[width = 0.48\linewidth]{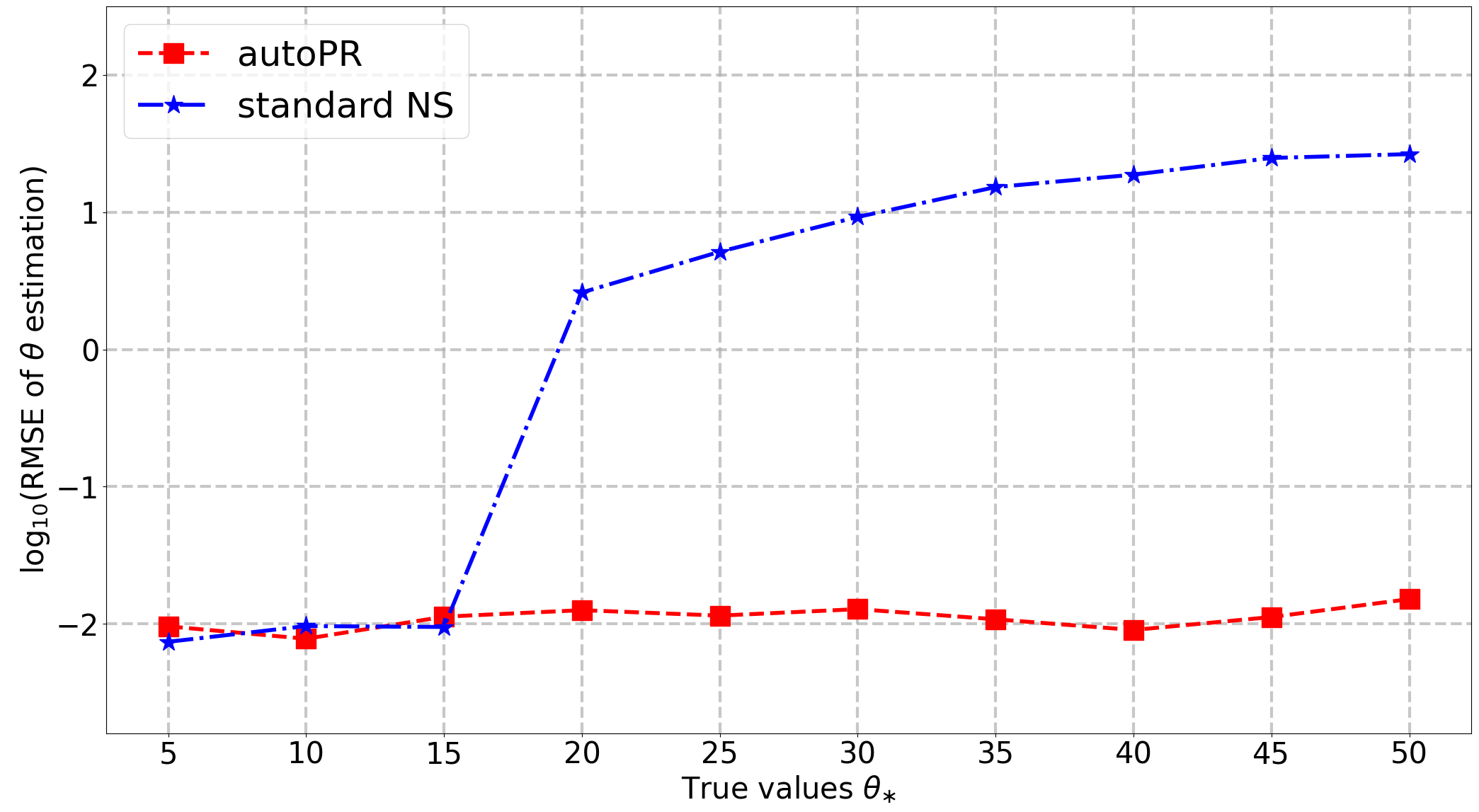}}
\subfigure[Number of likelihood evaluations]{
\includegraphics[width = 0.48\linewidth]{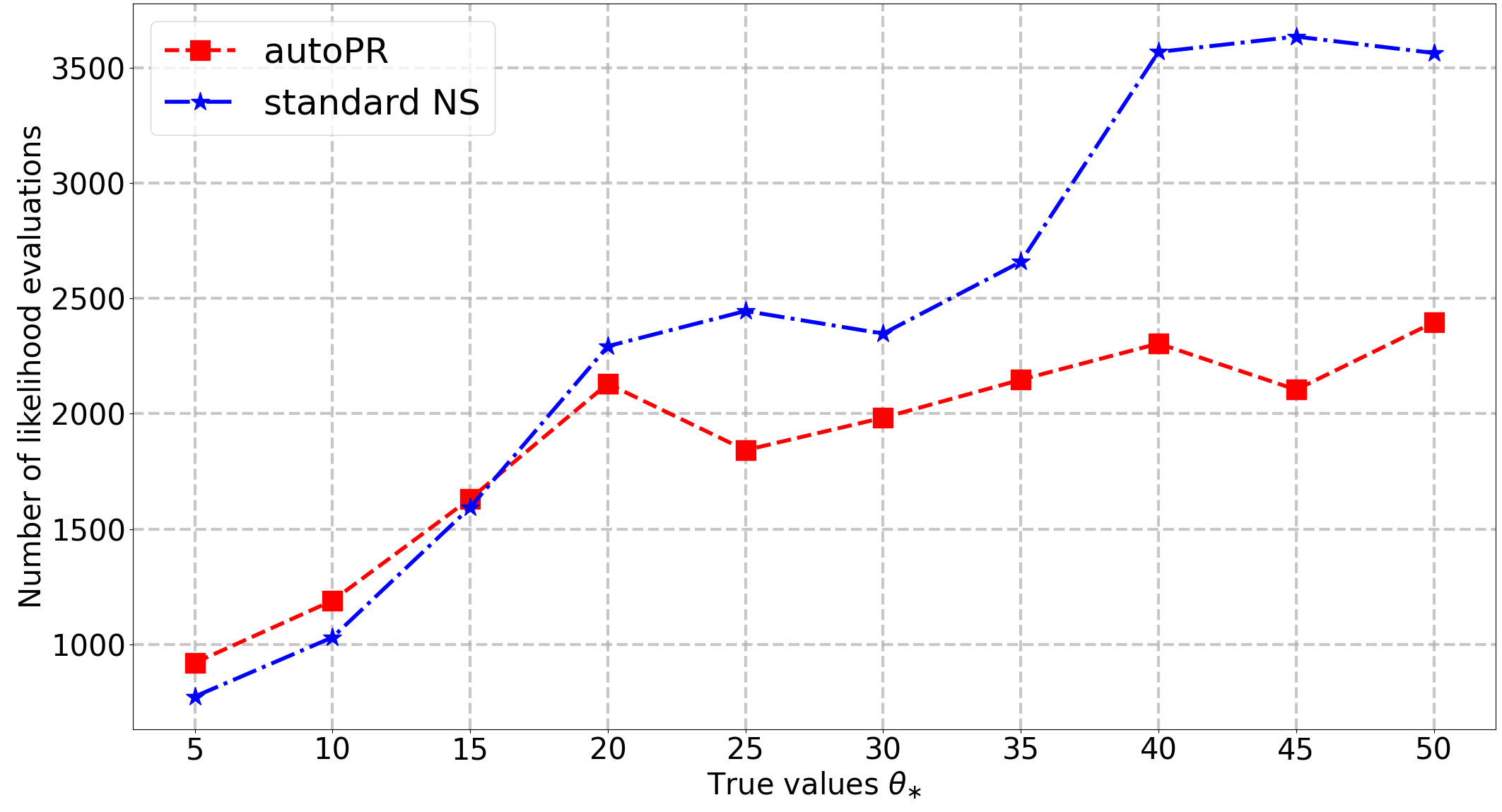}}
\caption{Algorithm performance comparison in the univariate Gaussian
  likelihood example over 10 realisations of the data for each value
  of $\mtheta_\ast$. The left panel shows Log$_{10}$(RMSE) of the
  estimate $\hat{\mtheta}$, and the right panel shows the mean number
  of likelihood evaluations, both of which are obtained using the
  standard NS approach (blue star points) and the BPR method (red
  square points).}
\label{fig:1DVarNlike}
\end{figure*}

Figure~\ref{fig:1DVarNlike} shows the (logarithm of the) root mean squared error (RMSE) of the estimate
$\hat{\mtheta}$ over 10 realisations of the data for each value of $\mtheta_\ast$, for both the standard NS approach and the BPR method. one sees that the RMSE of $\hat{\mtheta}$ for the standard NS approach rapidly increases for $\theta_\ast \gtrsim 20$, as the prior
$\pi(\theta)$ becomes increasingly unrepresentative and the NS
algorithm begins to fail, but that the RMSE of the BPR method
remains stable as $\theta_\ast$ increases, indicating robust parameter
estimation in all cases. Moreover, for $\theta_\ast \lesssim 15$, the
RMSE of BPR is consistent with that of the standard NS
approach, demonstrating that there is no reduction in estimation
accuracy associated with the BPR method in analysing datasets for
which the prior is representative. This is an important conclusion,
since it suggests that one sacrifices nothing in terms of parameter
estimation accuracy by {\em always} using the BPR method, which
also has the advantage that one stands to gain considerably from its
use when one (unexpectedly) encounters a dataset for which the prior
is unrepresentative. 

\revise{As we discussed in Section~\ref{sec:intro}, the failure of standard NS, which occurs in this example around $\mtheta_{\ast} = 15$ to $20$, is associated with the issue of likelihood values being indistinguishable to machine precision. 
For double precision arithmetic, the smallest
number that can be represented is $\sim 10^{-308} \sim e^{-709}$. For a Gaussian distribution $\sim \exp(-d^2/2)$, where $d$ is Mahalanobis distance, this corresponds to $d \sim 37$. In this example, the Gaussian likelihood has width $\sim 0.22$, so that the `basin of attraction' of the likelihood, which is defined by where
likelihoods become zero to machine precision corresponds to a distance $\sim 37 \times 0.22 \approx 8$ units from its centre. Turning to the Gaussian prior centred on zero, one would expect one of the $N_{\rm live} =100$ samples to lie beyond a distance of $\sim 2.5\sigma_\pi = 10$. Thus, one expects one of the samples from the prior to lie within the basin of attraction of the likelihood provided its centre does not lie beyond $\sim ~ 10 + 8 = 18$ units, which agrees well with the numerical results. This simple argument should be refined to take into account that, in total, one draws $N_{\rm like}$ samples from the prior during the full NS run, which is typically considerably larger than $N_{\rm live}$, and depends on the convergence criterion used. Even if $N_{\rm like} \sim 10^4$, however, one would expect one of the samples to lie beyond a distance of only $\sim 3.5\sigma_\pi = 14$, so that the standard NS should fail when the likelihood centre lies beyond $\sim 14+8 = 22$ units, which is again consistent with the numerical results.}

To demonstrate further the practicality of always using the BPR
method, Figure \ref{fig:1DVarNlike}(b) shows the mean number of likelihood
evaluations performed by MultiNest over 10 realisations as a function
of $\mtheta_\ast$, for both the standard NS and BPR methods. One
sees that both methods require similar numbers of likelihood
evaluations for $\mtheta_\ast \lesssim 20$, demonstrating that there
is very little additional computational overhead resulting from
extending the parameter space (in this case doubling its
dimensionality) to include the extra hyperparameter $\beta$. Moreover,
for larger values of $\mtheta_\ast$, the number of likelihood
evaluations required by the standard NS approach continues to increase
with $\mtheta_\ast$, whereas the number required by the BPR method
remains roughly constant. This again suggests that one should use the
BPR method in all applications of NS.

One may also compare the accuracy of the standard NS algorithm and the
BPR method in estimating evidences. Table \ref{tab:1DEvid} lists
the mean and standard deviation of the estimated log-evidence over 10
realisations of the data for each method, for each considered value of
$\theta_\ast$. The table also lists the `true' value of the evidence
in each case, estimated using standard quadrature techniques.  One
sees that for the standard NS approach, the log-evidence estimates
quickly become increasingly biased and volatile as $\theta_\ast$
increases. By contrast, the log-evidence
estimates obtained using BPR, after making the correction in
(\ref{eq:zeff}), are all consistent with the true
value, and have a standard deviation that remains stable for all values
of $\mtheta_\ast$. This once again suggests that one should {\em
  always} using the BPR method, since the required accuracy in
evidence estimation is achieved even for datasets for which the prior
is representative.

\begin{table*}
\caption{Comparison of the mean and standard deviation of the
  estimated log-evidence over 10 realisations of the data, obtained
  using standard NS (SNS) and the BPR method, for a range of
  $\theta_\ast$ values in the univariate Gaussian likelihood
  example. The `true' value of the evidence in each case is also
  given, as estimated using standard quadrature techniques.}
\smallskip
\centering
\begin{tabular}{c|ccr|cr}
\hline
$\theta_\ast$ & True & BPR mean & SNS mean & BPR s.d. & SNS s.d. \\
\hline
5 & $-22.04$ & $-21.99$ & $-22.03$ & 0.18 & 0.04\\
10 & $-24.38$ & $-24.28$ & $-24.81$ & 0.21 & 1.30 \\
15 & $-28.27$ & $-28.10$ & $-86.78$ &  0.24 & 103.76\\
20 & $-33.73$ & $-33.66$ & $-222.50$ & 0.27& 297.00 \\
25 & $-40.73$ & $-40.64$ & $-463.27$ & 0.27& 434.38 \\
30 & $-49.30$ & $-49.29$ & $-1319.58$ & 0.28& 1281.99 \\
35 & $-59.43$ & $-59.38$ & $-2530.09$ &  0.29& 1631.83 \\
40 & $-71.11$ & $-71.00$ & $-4309.03$ &  0.30& 2714.91 \\
45 & $-84.35$ & $-84.29$ & $-6963.08$ & 0.30 & 3327.04 \\
50 & $-99.15$ & $-99.07$ & $-10604.46$ & 0.31 & 3816.67 \\
\hline
\end{tabular}
\label{tab:1DEvid}
\end{table*}

\revise{Table \ref{tab:1DmaxSample} compares the number of `effective' posterior samples against the number of likelihood evaluations for $N_{\rm live} = 100$ in the same Gaussian example. One sees that both $N_{\rm eff}$ and $N_{\rm like}$ are comparable in the case $\theta_\ast = 5$, where the true value lies within the range of Gaussian prior $\theta \sim \mathcal{N} (0, 4^2)$. Both ratios decrease as the prior becomes increasingly unrepresentative, but the ratio for standard NS reduces almost to zero, with only one `effective' sample in the extreme case $\theta_\ast = 50$. By contrast, BPR maintains a steady `effective' posterior sample number and ratio across all cases. }

\begin{table}[h]
\caption{Comparison of the `effective' posterior samples between BPR and the standard MultiNest. All results are averaged over 10 realisations. $\theta_\ast$ represents the true value of $\theta$. $N_{\rm eff}$ is the number of `effective' posterior samples, $N_{\rm like}$ is the total number of likelihood evaluations. Ratio equals to $N_{\rm eff}$ divided by $N_{\rm like}$.}
\smallskip
\centering
\begin{tabular}{c|ccc|ccc}
\hline
$\theta_\ast$ & BPR $N_{\rm eff}$ & BPR $N_{\rm like}$ & BPR ratio & NS $N_{\rm eff}$ & NS $N_{\rm like}$ & NS ratio   \\
\hline
5 & $307$ & $857$ &   $35.8\%$ & $278$ & $742$ & $37.5\%$\\
\hline
10 & $371$ & $1233$ &   $30.1\%$ & $279$ & $1008$ & $27.7\%$\\
\hline
20 & $456$ & $2268$ &   $20.1\%$ & $356$ & $2576$ & $13.8\%$\\
\hline
30 & $451$ & $1918$ &   $23.5\%$ & $331$ & $4513$ & $7.3\%$\\
\hline
40 & $434$ & $1979$ &   $21.9\%$ & $298$ & $4361$ & $6.8\%$\\
\hline
50 & $495$ & $2275$ &   $21.8\%$ & $1$ & $5559$ & $0.017\%$\\
\hline
\end{tabular}
\label{tab:1DmaxSample}
\end{table}

We conclude our discussion of this first numerical
  example by performing a comparison between BPR and our original PR
  method \citep{chen2018improving}, in order to illustrate some of the
  advantages of the former, as outlined in
  Section~\ref{Sec:BPRmethod}. In particular, we consider the extreme
  example of the univariate Gaussian likelihood with $\mtheta_\ast =
  40$, again using $N_{\rm live} = 100$ live points (note that this is
  somewhat fewer than $N_{\rm live} =2000$ used in the analysis of the
  same example first presented in \cite{chen2018improving}). For
  simplicity, we adopt a simple linear `annealing schedule' for the
  original PR method, reducing $\beta$ from unity (which corresponds
  to standard NS) in steps of 0.1. Since the dispersion of the RMSE is
  quite large for $N_{\rm live}=100$, especially for $\beta \geq 0.6$
  where the NS process may fail completely, it is difficult to
  identify convergence of the statistical inferences as $\beta$ is
  reduced. Consequently, it is necessary to repeat the analysis 10
  times for each value of $\beta$, in order for the convergence to be
  reliably determined. The resulting RMSE of $\hat{\mtheta}$ and the
  number of likelihood evaluations at each value of $\beta$, both
  averaged over the 10 analyses, are listed in
  Table~\ref{tab:1DPRcompare}. One sees that convergence occurs for
  $\beta \leq 0.2$, which is consistent with the upper limit $\beta_+
  \approx 0.15$ of the effective marginalised posterior
  $\tilde{\mP}(\beta)$ plotted in Figure~\ref{fig:1DComb}.  Turning to
  the average number of likelihood evaluations, one first notes that
  $N_{\rm like}$ diverges from the overall decreasing trend for $\beta
  \geq 0.7$; this occurs since the NS algorithm erroneously terminates
  early for large $\beta$ values in this example of an extreme
  unrepresentative prior.  For $\beta \leq 0.6$, $N_{\rm like}$ then
  decreases monotonically, before converging for $\beta \leq 0.2$ to a
  value consistent with $N_{\rm like}$ required in the BPR method.
  Nonetheless, the total number of likelihood evaluations required by
  the original PR method across the 10 analyses at each $\beta$ value
  is more than two orders of magntiude greater than that required by
  the BPR method, which demonstrates the considerable improved
  effectiveness of the latter.

\begin{table}
\caption{The RMSE of $\hat{\mtheta}$ and the number
    of likelihood evaluations at each considered value of $\beta$ in a
    linear `annealing schedule' for the original PR method with
    $N_{\rm live} = 100$ applied to the extreme example of the
    univariate Gaussian likelihood with $\mtheta_\ast = 40$.  All
    results are averaged over 10 realisations of the data.}
\smallskip
\centering
\begin{tabular}{c|c|c}
\hline
$\beta$ value & $\hat{\mtheta}$ RMSE & $N_{\rm like}$ \\
\hline
PR(0.9) & $38.5533$ & $2312$  \\
\hline
PR(0.8) & $26.5783$ & $2687$ \\
\hline
PR(0.7) & $18.4579$  & $2558$\\
\hline
PR(0.6) & $4.3242$ & $5413$ \\
\hline
PR(0.5) & $2.8731$ & $4414$ \\
\hline
PR(0.4) & $2.3016$ & $4339$  \\
\hline
PR(0.3) & $1.8950$ & $4005$ \\
\hline
PR(0.2) & $0.0087$ & $2482$  \\
\hline
PR(0.1) & $0.0088$ & $2421$  \\
\hline
BPR ($\beta_+ \approx 0.15$) & $0.0090$ & $2304$ 
\end{tabular}
\label{tab:1DPRcompare}
\end{table}

\subsection{Bivariate Laplace likelihood example}

We now consider an example in which the bivariate Gaussian likelihood
used in Section~\ref{sec:bgl} is replaced with a bivariate likelihood
based on the Laplace distribution in order to investigate the
performance of the BPR method in the presence of fatter tails and a
cusp at the likelihood peak. In particular, we consider a bivariate
likelihood that is the product of two identical univariate Laplace
likelihoods of the form
\begin{align}
\mL(\mtheta) = \prod_{n=1}^{N} 
\left\{ \frac{1}{2b} \exp\left(-\frac{|\mtheta - m_n|}{b}\right)
\right\},
\end{align}
where the location parameter is represented by the $n$th measurement
$m_n$ and $b$ is the scale parameter, which is analogous to
$\sigma_\xi$ in the Gaussian distribution in (\ref{eqn:likelihood}).
In this example, we take the scale parameter value to be $b=0.1$ to
produce a narrow, steep Laplace distribution. One should note that a
bivariate Laplace distribution constructed in this way is not
circularly symmetric and differs from the usual form adopted for 
multivariate Laplace distributions \citep{kotz2001laplace}.
Once again, we consider just $N = 1$ measurement to facilitate a
straightforward comparison with the bivariate Gaussian likelihood 
example in the previous section.

For the sake of brevity, we consider only one of the priors used in
the previous section, namely an uncorrelated Gaussian prior centred on
the origin, with $\sigma_{\theta_1}=4$ and $\sigma_{\theta_2} = 4$. We
do, however, consider 8 different centerings of the likelihood, for
which the true values are $\mTheta_{\ast} =
(\theta_1,\theta_2)^{\top}$, where $\theta_1 = \theta_2 = 5, 10, 15,
20, 25, 30, 35$ and $40$, respectively, so that the peak of the
likelihood moves progressively further from the origin.  All other
settings are identical to those used for the bivariate Gaussian
bivariate likelihood example in the previous section; in particular,
we perform the analysis over 10 realisations of the data for each of
the 8 likelihood centres considered.

\begin{figure*}[!ht]
\centering
\subfigure[Estimation of $\beta_{+}$]{
\includegraphics[width = 0.46\linewidth]{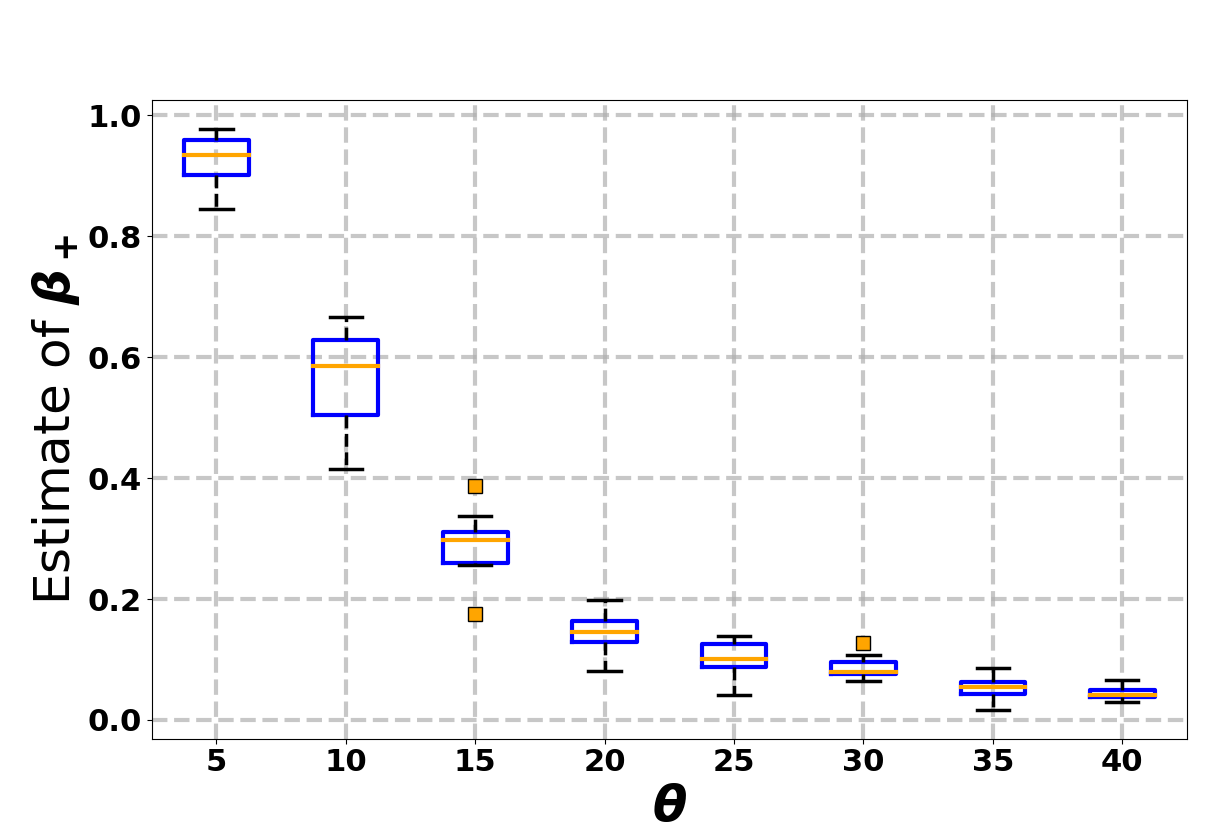}}\qquad
\subfigure[RMSE of $\mtheta$ estimation]{
\includegraphics[width = 0.46\linewidth]{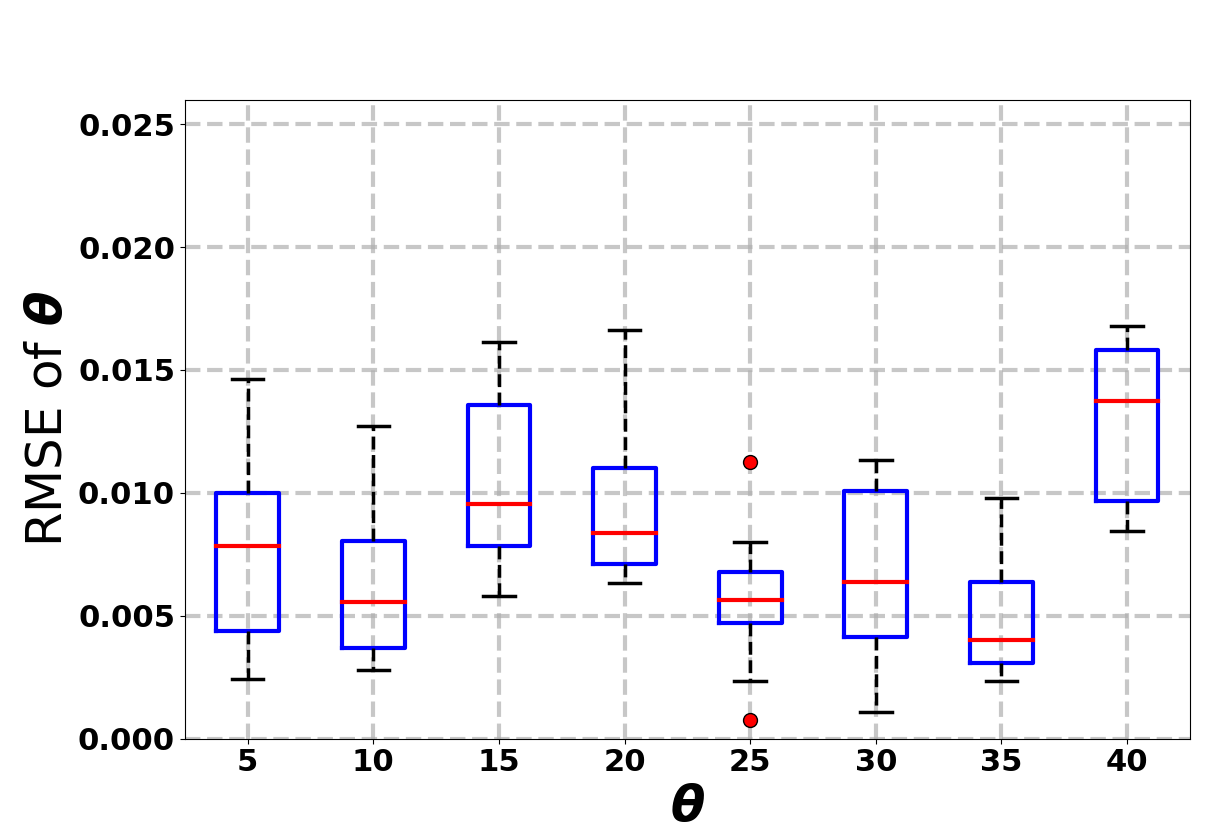}}\\
\subfigure[Error of log-evidence estimates]{
\includegraphics[width = 0.46\linewidth]{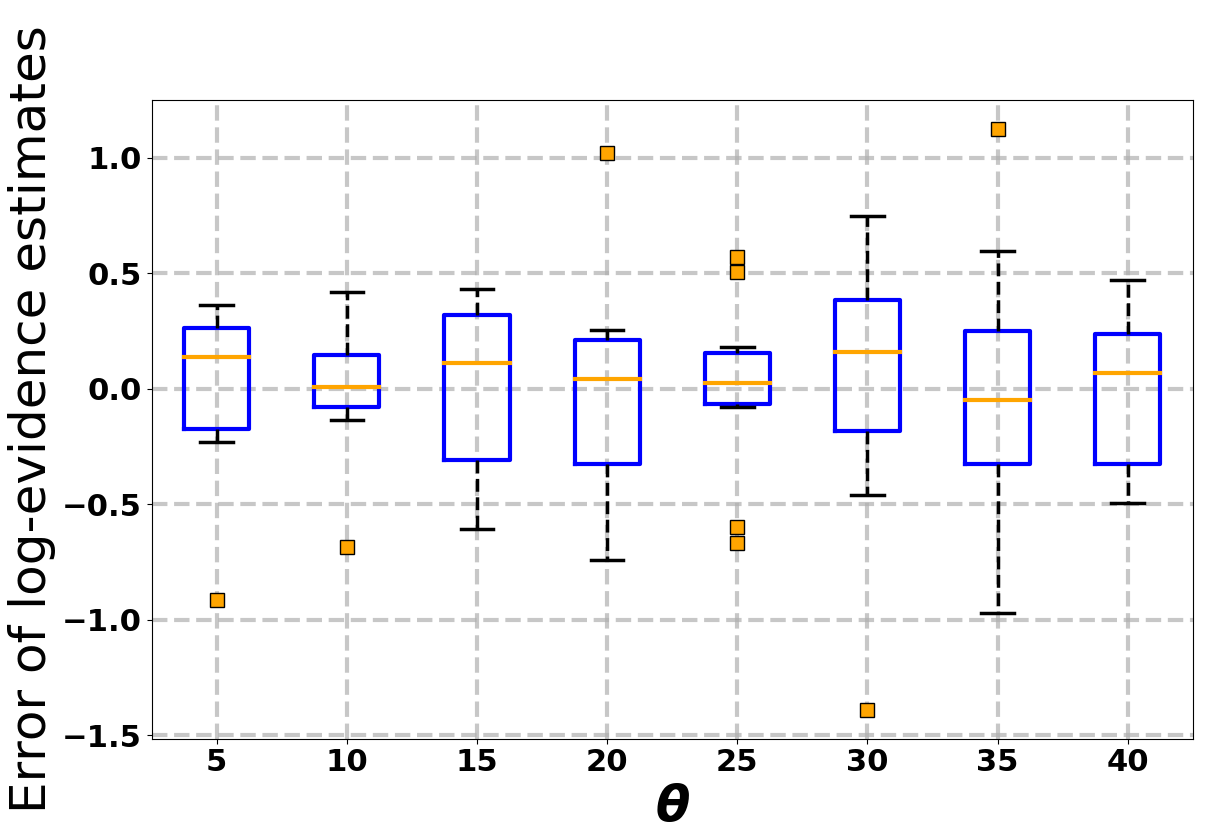}}\qquad
\subfigure[Number of likelihood evaluations]{
\includegraphics[width = 0.46\linewidth]{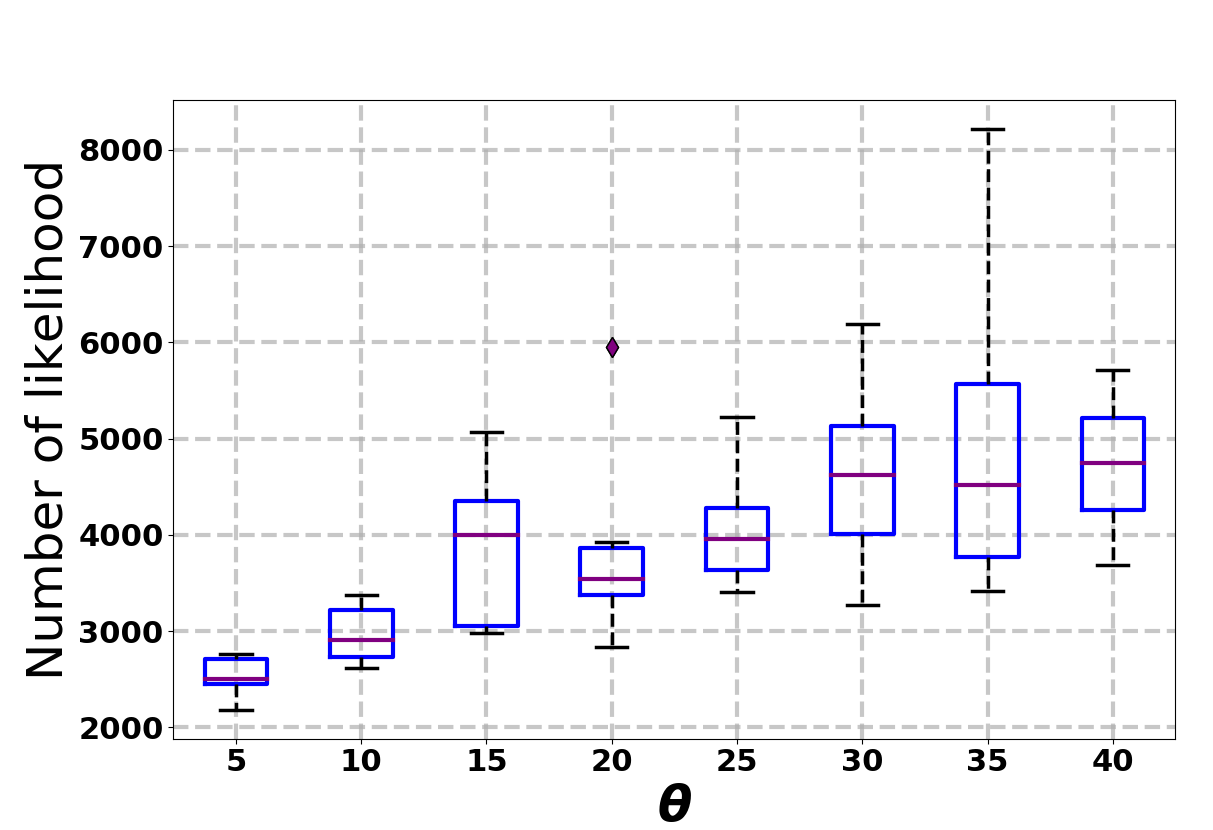}}
\caption{Boxplots for estimated parameter values
    using the BPR method applied to the bivariate Laplace
    likelihood example, with $\mTheta_{\ast}$ ranging from $(5,5)$ to
    $(40,40)$, and an uncorrelated Gaussian prior centred on the
    origin with $\sigma_{\theta_1}=4$ and $\sigma_{\theta_2} = 4$.
    The extent of each box denotes the first and third quartiles,
    $Q_1$ and $Q_3$, of the 10 realisations of the data considered,
    and the error bars denote the `minimum' and `maximum' values, defined as
    $Q_1-\tfrac{3}{2}(Q_3-Q_1)$ and $Q_3+\tfrac{3}{2}(Q_3-Q_1)$,
    respectively.  The line in each box represents median value (or
    second quartile $Q_2$) and the circles and squares denote
    any `outliers' that lie beyond the `minimum' and `maximum'.}
\label{fig:nonGauss}
\end{figure*}

Figure~\ref{fig:nonGauss} summarises the resulting performance of the
BPR method in an alternative format to that used in the previous
section, which makes use of boxplots.  Figure~\ref{fig:nonGauss}(a)
illustrates the estimation of $\beta_+$ (the upper limit of the
marginal posterior on $\beta$) as a function of the true value
$\mTheta_{\ast}$ assumed. As expected, one sees a monotonic decline in
the estimated $\beta_+$ value as $\mTheta_{\ast}$ increases,
indicating that the prior becomes increasingly unrepresentative.
Moreover, one sees that the range of $\beta_+$ estimates over the 10
realisations of the data is stable for each of the 8 cases considered.
Figure~\ref{fig:nonGauss}(b) shows an analogous plot for the RMSE of
the estimate of $\mTheta_{\ast}$, which is calculated as the average
of the RMSE of $\mtheta_1$ and $\theta_2$. The RMSE values typically
lie in the range 0.005--0.01, which indicates that the BPR method
provides an accurate estimate in all cases.  Figure~\ref{fig:nonGauss}
(c) shows the error in the estimate of the log-evidence, which is
centred on zero in all cases, showing that evidence estimates are
unbiassed. Moreover, the accuracy of the log-evidence estimates is
typically $\sim 0.3$, but with the occasionally outlier having an
estimation error up to $\sim 1$; this is broadly consistent with the
accuracies of the log-evidence estimates for the bivariate Gaussian
likelihood example listed in Table~\ref{tab:2DCorrEvid}. Finally,
Figure~\ref{fig:nonGauss} (d) shows the number of likelihood
evaluations $N_{\rm like}$ required for MultiNest to converge using
the BPR method, which gradually increases with $\mTheta_{\ast}$ as
is expected for progressively more unrepresentative
priors. Nonetheless, the rate of increase in $N_{\rm like}$ is
relatively slow, showing that the BPR method is computationally
feasible, even for extreme unrepresentative priors in the presence of
likelihoods with fat tails and a cusp at its peak.

\subsection{More results for the bivariate multi-modal likelihoods example} 
\label{AP: BM}

\begin{figure*}
\centering
\subfigure[Heat-map, varying mode distance]{
\includegraphics[width = 0.9\linewidth]{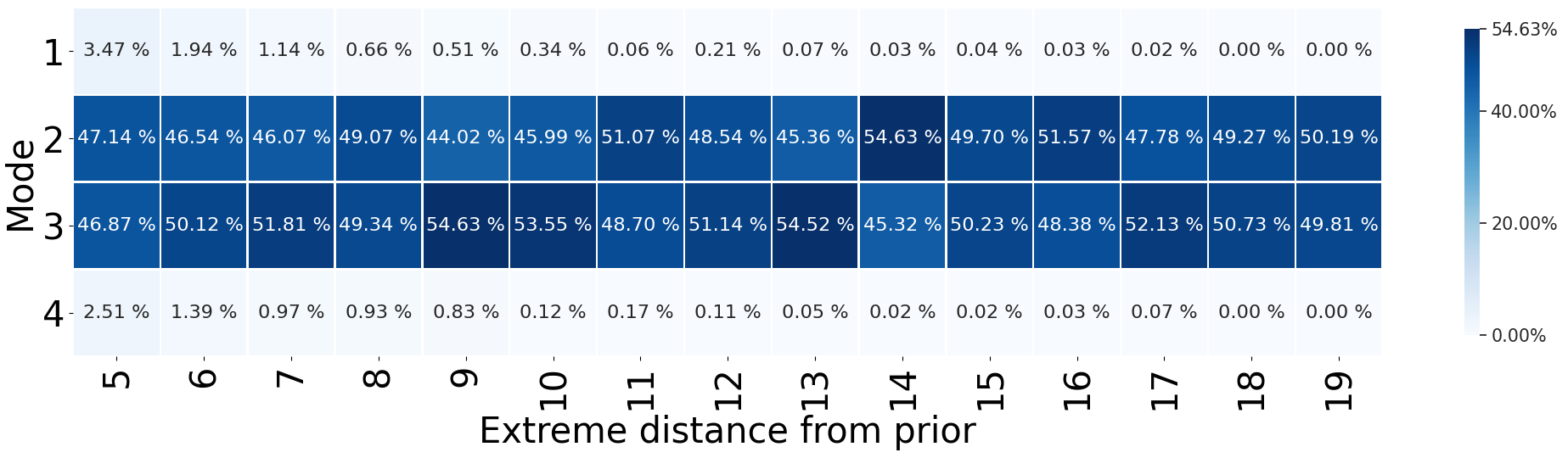}}
\subfigure[Heat-map, varying $N_{\rm live}$]{
\includegraphics[width = 0.9\linewidth]{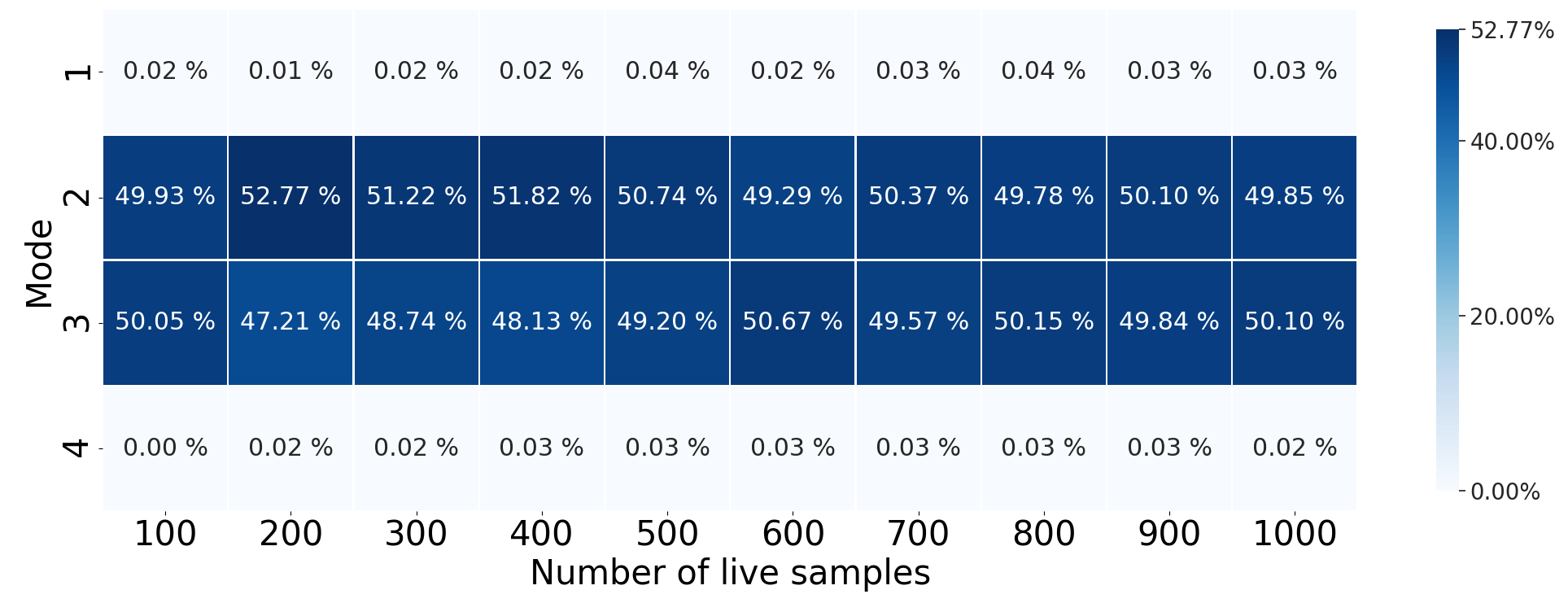}}

\caption{The heatmap of the {\em asymmetric arrangement} bivariate multi-modal case. The panel (e) shows the effect as a function of the distance of the geometric centre of the 4 modes from the origin for $N_{\rm live}=100$, and the panels (f) show the effects of increasing $N_{\rm live}$ for fixed mode centres, with the
    geometric centre of the modes centred at $\theta_1=\theta_2=15$.}
\label{fig:4modesGaussianNonSymHeatmap}
\end{figure*}

In Figure \ref{fig:4modesGaussianNonSymHeatmap}, one sees that the percentage of
equally weighted posterior samples is far larger for modes 2 and 3
than for modes 1 and 4, as expected. From panel (a), the percentage of
samples in modes 1 and 4 drops gradually, again as expected, from a
few percent, when the extreme distance from the origin is 5 units,
down to zero for 18 units and beyond, for the default value of $N_{\rm
  live} = 100$. From panel (b), one sees that the percentage of
samples in modes 1 and 4 is relatively insensitive to $N_{\rm live}$,
with around 0.02\% of samples lying in each mode.

\end{document}